\documentclass[11pt]{article}
\usepackage{amsmath,amssymb}
\usepackage[T1]{fontenc}
\usepackage{lmodern}
\font\dm=cmr9

\numberwithin{equation}{section}
\def\ba{{\bar a}}
\def\hi{\wh\imath}
\def\wi{\wt\imath}
\let\iy\infty
\let\op\oplus
\let\ot\otimes
\let\ov\overline
\let\os\overrightarrow
\let\pa\partial
\let\q\quad
\def\qh#1{\quad\hbox{#1}\quad}
\let\td\tilde
\let\tm\times
\let\ul\underline
\let\wh\widehat
\let\wt\widetilde
\let\a\alpha
\let\b\beta
\let\d\delta
\let\D\varDelta
\let\ve\varepsilon
\let\g\gamma
\let\G\varGamma
\let\k\kappa
\let\la\lambda
\let\La\varLambda
\let\n\nabla
\let\o\omega
\let\O\varOmega
\let\F\varPhi
\let\vf\varphi
\let\Ps\varPsi
\let\PS\varPsi
\let\si\sigma
\let\Si\varSigma
\let\t\theta
\let\T\varTheta
\let\z\zeta
\def\SU{\text{SU}}
\def\U{\text{U}}
\def\(#1){{(#1)}}
\def\[#1]{{[#1]}}
\def\Ft#1#2{\F^{#1}_{\td #2}}
\def\gv{\nad{gauge}}
\def\crt{_{\rm crt}}
\def\pl{_{\rm pl}}
\def\m{{\mu\nu}}
\def\cD{\mathcal D}
\def\cL{\mathcal L}
\def\cLY{\mathcal L_{\rm YM}}
\def\cM{\mathcal M}
\def\cZ{\mathcal Z}
\def\C{\mathbb C}
\def\R{\mathbb R}
\def\fg{\mathfrak g}
\def\fh{\mathfrak h}
\def\fm{\mathfrak m}
\def\tA{{\tilde A}}
\def\tB{{\tilde B}}
\def\tC{{\tilde C}}
\def\dg #1,#2,#3,{{#1{}_{#2}}^{\!#3}}
\def\gd #1,#2,#3,{{#1{}^{#2}}_{\!#3}}
\def\gdg #1,#2,#3,#4,{{{#1{}^{#2}}_{\!#3}}{}^{\!#4}}
\def\dgd #1,#2,#3,#4,{{{#1{}_{#2}}^{\!#3}}{}_{\!#4}}
\def\nad#1#2{\overset{\rm #1}{#2}}
\def\nadd#1#2{\overset{#1}{#2}}
\def\nads#1{\overset{\star}{#1}}
\def\falg{{\mathrel{\lower5pt\hbox{${\scriptstyle\sim}$}\hskip-5pt g}}}
\def\fal#1{{\mathrel{\lower4pt\hbox{${\scriptstyle\sim}$}\hskip-7.5pt #1}}}
\def\tl{\mathopen{\hbox{\dm[}}}
\def\tp{\mathclose{\hbox{\dm]}}}
\newdimen\krop
\setbox0=\hbox{$\scriptstyle\a$}
\krop=\wd0
\def\cdt{\hbox to\krop{$\scriptstyle\hfil\cdot\hfil$}}
\def\lw#1 {\lower#1pt\hbox\bgroup$\scriptstyle}
\def\eg{$\egroup}
\def\hor{\operatorname{hor}}
\def\vol{\operatorname{vol}}
\def\ver{\operatorname{ver}}
\def\id{\operatorname{id}}
\def\ad#1{\operatorname{ad}_{#1}}
\def\Ad{\operatorname{Ad}}
\def\SO{{\rm SO}}

\let\ti\textit
\def\beq#1 #2\e{\begin{equation}\label{#1}#2\end{equation}}
\def\bea#1 #2\e{\begin{align}\label{#1}#2\end{align}}
\def\bml#1 #2\e{\begin{multline}\label{#1}#2\end{multline}}
\def\bg#1 #2\e{\begin{gather}\label{#1}#2\end{gather}}
\let\bal\aligned \let\eal\endaligned
\let\bga\gathered \let\ega\endgathered
\def\bma{\left(\begin{array}{c|c}} \def\ema{\end{array}\right)}
\def\bca{\begin{cases}}
\def\eca{\end{cases}}
\def\bit{\begin{itemize}}
\def\eit{\end{itemize}}

\let\nn\nonumber
\def\lb#1 {\label{#1}}
\let\er\eqref

\def\up#1{\uppercase{#1}}
\def\E{\expandafter\up}
\def\ap{approximat}
\def\cf{coefficient}
\def\cm{composition}
\def\cfn{confinement}
\def\cn{connection}
\def\cv{conservation}
\def\ct{constant}
\def\cd{coordinate}
\def\co{cosmolog}
\def\ci{covariant}
\def\cvt{curvature}
\def\dv{derivative}
\def\di{di\-men\-sion}
\def\elm{electromagneti}
\def\el{element}
\def\e{equation}
\def\fw{following}
\def\f{function}
\def\fn{fundamental}
\def\gn{generalization}
\def\gr{gravitation}
\def\Hm{Higgs' mechanism}
\def\hc{hypercomplex}
\def\ia{interaction}
\def\iv{invarian}
\def\KWK{Kerner--Wong--Kopczy\'nski }
\def\KC{Killing--Cartan }
\def\LC{Levi-Civita }
\def\Li{Lie algebra}
\def\MR{Moffat--Ricci}
\def\mo{morphi}
\def\nA{non-Abelian}
\def\JT{Nonsymmetric Jordan--Thiry Theory}
\def\nos{nonsymmetric}
\def\NK{Nonsymmetric Kaluza--Klein Theory}
\def\pc{particle}
\def\Pl{Planck}
\def\pt{potential}
\def\pp{principle}
\def\rp{represent}
\def\sf{satisf}
\def\so{solution}
\def\spt{space-time}
\def\sn{spontaneous}
\def\sc{structur}
\def\st{such that }
\def\s{symmetr}
\def\tf{transformation}
\def\un{unification}
\def\v{variation}
\def\wrt{with respect to }
\def\YM{Yang--Mills}
\let\TM\texttrademark

\author{M. W. Kalinowski\\
Pracownia Bioinformatyki, Instytut Medycyny
Do\'swiadczalnej i~Klinicznej PAN,\\
ul. Pawi\'nskiego 5, 02-106 Warszawa, Poland\\
e-mail: markwkal@bioexploratorium.pl, mkalinowski@imdik.pan.pl}
\title{The Nonsymmetric Kaluza--Klein Theory and Modern Physics\\
A Novel Approach}
\begin{document}
\maketitle

\rightline{\it To the memory of my teacher Professor Stanis\l aw Szpikowski\hskip30pt}

\vskip20pt
\begin{abstract}
We consider in the paper the \NK\ finding a condition for a color \cfn\ in
the theory. We consider also a Kerner--Wong--Kopczy\'nski \e\ in this theory.
The \NK\ with a spontaneous \s y breaking and Higgs' mechanism is examined.
We find a mass spectrum for a broken gauge bosons and Higgs' \pc s. We derive
a \gn\ of Kerner--Wong--Kopczy\'nski \e\ in the presence of Higgs' field.
A~new term in the \e\ is a \gn\ of a Lorentz force term for a Higgs' field.
We consider also a bosonic part of GSW (Glashow--Salam--Weinberg) model in
our theory, getting masses for $W$, $Z$ bosons and for a Higgs' boson
agreed with an experiment. We consider Kerner--Wong--Kopczy\'nski \e\
in GSW model obtaining some additional charges coupled to Higgs' field.
\end{abstract}

\section*{Introduction}

In this paper we consider the \NK\ in a \nA\ case and the \NK\ with \Hm\ and
spontaneous \s y breaking in a new setting. The paper gives a comprehensive
review of a subject with many new features which are shortly summarized at the
end of the introduction. Moreover, it cannot be considered as a review paper
because it contains new achievements in this rapidly developing subject.

The subject of the paper is specialized of course, but it could be very
interesting for a wide audience because geometrization and unification of
\fn\ physical \ia s are very interesting. This idea gives a justification for
some phenomenological theories which are completely arbitrary. There is no
physics without mathematics, especially without geometry---differential
geometry. Even
Maxwell--Lorentz electrodynamics happens \ti{post factum} geometrized in
fibre bundle formalism. In the case of ordinary Kaluza--Klein Theory the
geometrization and \un\ have been achieved. Unfortunately, without
``interference effects''.
We consider in the paper some additional versions
of the \NK. In particular, except of a real version we consider also \E\nos\
Hermitian Theory in two realizations, complex and hypercomplex.
They are natural (Hermitian) metrization of a fiber bundle over a \spt.
The \nos\ Kaluza--Klein (Jordan--Thiry) Theory (a real version) has
been developed in the past (see Refs \cite1--\cite5). The theory unifies \gr
al theory described by NGT (\E\nos\ \E\gr al Theory, see Ref.~\cite6) and
\YM' fields (also \elm c field). In the case of the \JT\ this theory includes
scalar field. The \NK\ can be obtained from the \JT\ by simply putting this
scalar field to zero. In this way it is a limit of the \JT.

The \JT\ has several physical applications in \co y, e.g.: (1)~\co ical \ct,
(2)~inflation, (3)~quintessence, and some possible relations to the dark
matter problem. There is also a possibility to apply this theory to an
anomalous acceleration problem of Pioneer 10/11 (see Refs~\cite7,~\cite{*6}).

In this paper a scalar field $\Psi=0$ ($\rho=1$). Moreover, the extension to
Jordan--Thiry Theory in any \nos\ version is still possible and will be done
elsewhere. The scalar field can play a role as a dark matter---quintessence
with weak \ia s with ordinary matter. On the classical level, this is only
a \gr al \ia\ with the possibility to change a strength of \gr al \ia\ via
a change of \gr al \ct. On a quantum level due to an excitation of a quantum
vacuum a very weak non\gr al \ia\ with ordinary matter is possible, i.e.\
a scattering of scalarons with ordinary matter particles and also a
scattering of skewons with those particles.

The theory unifies gravity with gauge fields in a nontrivial
way via geometrical \un s of two fundamental \iv ce principles in Physics:
(1)~the \cd\ \iv ce \pp, (2)~the gauge \iv ce \pp. \E\un\ on the level of \iv
ce \pp s is more important and deeper
than on the level of interactions for from \iv ce
\pp s we get \cv\ laws (via the Noether theorem). In some sense Kaluza--Klein
theory unifies the energy-momentum \cv\ law with the \cv\ of a color
(isotopic) charge (an electric charge in an \elm c case).

Let us notice that an idea of geometrization and simultaneously \un\ of \fn\
\ia s is quite old. GR is 100 years old and Kaluza--Klein Theory is almost
100~years old. Both ideas: a geometrization of physical \ia s and a \un\
are well established contemporarily.

This \un\ has been achieved in higher than four-\di al world, i.e.\
$(n+4)$-\di al, where $n=\dim G$, $G$~is a gauge group for a \YM' field,
which is a semisimple Lie group (\nA). In an \elm c case we have $G=\U(1)$ and
a \un\ is in 5-\di al world (see also~\cite8). The \un\ has been achieved
via a natural \nos\ metrization of a fiber bundle. This metrization is
right-\iv t \wrt an action of a group~$G$. We present also an Hermitian
metrization of a fiber bundle in two versions: complex and \hc. The \cn\
on a fiber bundle of frames over a manifold~$P$ (a~bundle manifold) is
compatible with a metric tensor (\nos\ or Hermitian in complex or \hc\
version). In the case of $G=\U(1)$ the geometrical structure is bi\iv t
\wrt an action of~$\U(1)$, in a general \nA\ case this is only right-\iv t.

In the paper we do not mention some ``modern'' Kaluza--Klein developments for
the reason described in Conclusion of Ref.~\cite8 which we do not repeat here.

Let us notice the following fact. We use a notion of a \nos\ metric as an
abuse of nomination for a metric is always \s ic. This will not cause any
misunderstanding. It is similar to an abuse of nomination in the case of
Minkowski metric in Special Relativity for a metric is always positive
definite.

The \un\ is nontrivial for we
can get some additional effects unknown in conventional theories of gravity
and gauge fields (\YM' or \elm c field). All of these effects, which we call
\ti{interference effects} between gravity and gauge fields are testable in
\pp\ in experiment or an observation. The formalism of this \un\ has been
described in Refs \cite1--\cite5,~\cite8 (without Hermitian versions).

The theory considered here is \nA\ and even if there are some formulations
similar to those from Ref.~\cite8 one should remember that the theory
described in Ref.~\cite8 is an Abelian theory with $\U(1)$ group. The
difference is profound not only because a higher level of mathematical
calculations but also because of completely new features which appear in
a \nA\ theory. If we can use similar formulations as in Ref.~\cite8 it means
that a geometrical language is correct to describe a physical reality.

It is possible to extend the \E\nos\ (\nA) Kaluza--Klein Theory to the case
of a spontaneous \s y breaking and \Hm\ (see Ref.~\cite1) by a nontrivial
combination of Kaluza \pp\ (Kaluza miracle) with \di al reduction procedure.
This consists in an extension of a base manifold of a principal fiber bundle
from $E$ (a~\spt) to $V=E\times M$, where $M=G/G_0$ is a manifold of
classical vacuum states.

In this paper we consider a condition for a color \cfn\ in the theory.
We solve the constraints in the case of \nA\ \NK\ getting an exact form of an
induction tensor for \YM' fields in the theory. We find a formula for a \nA\ charge
in the theory in comparison to 4-momentum in \gr\ theory. We derive the Lagrangian for
\YM' field and an energy-momentum tensor in terms of $H^a_\m$ only. We
consider also \NK\ with Higgs' fields and spontaneous \s y breaking. We solve
constraints in the theory getting Lagrangian for \YM' field, kinetic energy
Lagrangian for a Higgs field and Higgs \pt\ in
terms of gauge fields and Higgs fields only. We
derive pattern of masses for a massive intermediate bosons and Higgs' \pc s.
We derive also a \gn\ of \KWK \e\ for a test \pc. In such an \e\ there is a
new charge for a test \pc\ which couples a Higgs' field to the \pc. This is
similar to a Lorentz force term in an \elm c case. This term is also similar
to a new term coupled a \YM' field to a test \pc\ via a color (isotopic)
charge in ordinary \KWK \e\ (see Ref.~\cite3).

The \NK\ is an example of the geometrization of \fn\ \ia\ (described by \YM'
and Higgs' fields) and \gr\ according to the Einstein program for
geometrization of \gr al and \elm c \ia s. It means an example to create a
Unified Field Theory. In the Einstein program we have to do with \elm sm and
gravity only. Now we have to do with more degrees of freedom, unknown in
Einstein times, i.e.\ GSW (Glashow--Salam--Weinberg) model, QCD, Higgs'
fields, GUT (Grand Unified Theories). Moreover, the program seems to be the
same.

We can paraphrase the definition from Ref.~\cite9: \ti{Unified Field Theory:
any theory which attemps to express \gr al theory and \fn\ \ia s theories
within a single unified framework. Usually an attempt to generalize
Einstein's general theory of relativity alone to a theory of gravity and
classical theories describing \fn\ \ia s}. In our case this single unified
framework is a multi\di al analogue of geometry from Einstein Unified Field
Theory (treated as generalized gravity) defined on principal fiber bundles
with base manifolds: $E$ or $E\times V$ and structural groups $G$ or~$H$.
Thus the definition from an old dictionary (paraphrased by us) is still valid.

Summing up, \NK\ connects old ideas of unitary field theories (unified field
theories, see Refs \cite{10,11} for a review) with modern applications. This
is a geometrization and \un\ of a bosonic part of four \fn\ \ia s.

The paper has been divided into four sections. In the first section we give
some \el s of geometry used in the paper. In the second section we give
some elements of the \E\nos\ (\nA) Kaluza--Klein Theory in some new setting.
We give also a condition for the dielectric \cfn\ of a color charge.
We consider in details a \nA\ charge, color charge in static situations.
We consider two versions of the \NK: 1.~the real version and 2.~the Hermitian
version (complex and \hc). We shortly present the second version. In the
third section we give some elements of \NK\ with \sn\ \s y breaking and \Hm.
In this section we consider also two versions of the \NK\ (real and Hermitian,
complex and \hc).
We derive a pattern of masses for broken intermediate bosons and Higgs'
bosons. We write down a \gn\ of \KWK \e\ in this case, getting a coupling of
Higgs' field to a test \pc. In other words, we derive an analogue of a
Lorentz force term for a Higgs' field.

In the fourth section a bosonic part of GSW (Glashow--Salam--Weinberg) model,
according to Manton (6-\di al model) has been extended to the \NK. We get a
realistic pattern of masses of~$W^\pm$, $Z^0$ and Higgs' boson. In particular, we
get a mass of a Higgs boson agreed with an experiment, which is impossible in
a pure Manton model. We have as before the $G2$ exceptional group as a
unification group with a bare Weinberg angle ${\t_W=30^\circ}$ ($\sin^2\t_W=0.25$). We apply here
the Hermitian version of the \NK, i.e.\ \E\nos\ Hermitian Kaluza--Klein
Theory. In the simplest case with $\xi=0$ and $g_\m=\eta_\m$ (in Minkowski
space) we calculate a small deviation $\d$ of a bare Weinberg angle (equal
to~$\frac\pi6$) as a 1-loop and 2-loop corrections using a $\D r$
(or~$\D R$) theory known in literature.
In Appendix~A we give some details of calculations concerning \so s
of constraints in the theory.

In Appendix B we give some \el s of Manton model in a connection to our
approach. In Appendix~C we consider
the \KWK \e\ in GSW model. We derive some explicit influence of new charges
coupled to Higgs' field (from the SM model) on a movement of a test \pc. The
existence of those new charges and their influence on a test \pc\ movement
can be tested in experiment. In Appendix~D we give formulas for \ia s
between gravity and Higgs' field and \YM' fields in Hermitian Kaluza--Klein
Theory in an application to bosonic part of GSW model. These are ``interference
effects'' between \nos\ gravity and GSW model in a unified theory. They can
be considered as effects of \un. In Appendix~E we calculate a correction~$\d$
to a Weinberg angle (equal to~$\frac\pi6$) as radiation corrections to a bare
angle using $\D r$~theory.

In Conclusions we give some prospects for further research, in particular,
how to treat fermions in the \NK\ with a \sn\ \s y breaking and we give a sketch
of a program of quantization of the \NK.

Summing up, the paper contains many novel features (without repetition of
heavy calculations from Refs \cite1--\cite5, \cite8:
\begin{enumerate}
\item Hermitian versions (complex and \hc) in the case of $\U(1)$ and a
general \nA\ semisimple group~$G$ also in the case with \sn\ symmetry breaking
and \Hm.
\item Solutions of constraints appearing in the theory (also in all considered
versions).
\item Detailed calculations of a classical dielectric model of \cfn\ of color
(a~\nA\ gauge charge).
\item Spectrum of masses for broken gauge bosons and scalar (Higgs') particles
in a general case.
\item An application to bosonic part of GSW model, where we get masses for
$W^\pm$, $Z^0$ and Higgs' boson agreed with experiment. In the last case this
is possible only for an Hermitian complex version on~$S^2$ and invokes some
new research connecting the theory to K\"ahlerian structures.
\item A \KWK \e\ in GSW model with some additional charges coupled a test
\pc\ in a motion to Higgs' field (this one from the Standard Model).
\item Additional (non-classical) \ia\ of a Higgs' field (from the Standard
Model) with gravity (described by NGT) and also additional Higgs' phenomena
in~SM.
\item A possibility to tune a \co ical \ct\ to the value obtained from
observational data.
\end{enumerate}

For we have not any traces of GUT or supersymmetry from LHC results we do not
consider extensions of our Kaluza--Klein Theory in these directions. Thus we
stop (temporarily) on 20-\di al \un\ of electro-weak \ia s (a~bosonic part)
and \nos\ gravity (NGT) and on 12-\di al \un s of strong \ia s (a~bosonic
part of~QCD) with \nos\ gravity (NGT). The inclusion of fermions is under
consideration and the work is in progress together with an approach to
quantization.

From technical point of view we get also some additional results:
\begin{enumerate}
\setcounter{enumi}8
\item An exact formula (a covariant one) for $\gd L,a,\m,$ and $L^{a\m}$
(an induction tensor).
\item A Lagrangian for a \YM's field in terms of $\gd H ,a,\m,$ and $g_\m$
only.
\item An exact formula (a covariant one) for a torsion in higher \di\
$\gd Q,a,\m,(\G)$ with an interpretation as a polarization of gauge field
induced by $g_\m$ and~$\ell_{ab}$.
\item In the case of \sn\ \s y breaking and \Hm, i.e.\ for a Kaluza--Klein
Theory with a \di al reduction we get analogous formulas for $\gd L,a,\td n\td b,$,
$\gd L,a,\mu\td b,$ in terms of a Higgs' field $\F^a_{\td a}$ and a covariant
\dv\ $\gv{\n_\mu}\F^a_{\td a}$ of the field. Those formulas are covariant. We
get also similar interpretations of \ti{exact formulas} for torsion in higher
\di s.
\end{enumerate}

Let us notice that we consider geodetic \e s \wrt \LC \cn\ generated
by a \s ic part of any \nos\ tensor on~$P$ as \e s of motion from a
variational principle. 

\section{Elements of geometry}
Let us now describe the notation and definitions of geometric
quantities used in the paper. We use a smooth principal bundle which is an
ordered sequence
\beq{Ap.0}
\ul P=(P,F,G,E,\pi),
\e
where $P$ is a total bundle manifold, $F$ is typical fibre, $G$, a Lie group,
is a structural group, $E$~is a base manifold and $\pi$ is a projection. In our
case $G=\U(1)$, $E$~is a \spt, $\pi:P\to E$.
We have a map $\vf:P\times G\to P$ defining an
action of~$G$ on~$P$. Let $a,b\in G$ and $\ve$~be a unit element of the
group~$G$, then $\vf(a)\circ \vf(b)=\vf(ba)$, $\vf(\ve)=\id$, where $\vf(a)p
=\vf(p,a)$. Moreover, $\pi\circ\vf(a)=\pi$. For any open set $U\subset E$ we
have a local trivialization $U\times G\simeq \pi^{-1}(U)$. For any $x\in E$,
$\pi^{-1}(\{x\})=F_x \simeq G$, $F_x$ is a fibre over~$x$ and is equal to~$F$.
In our case we suppose $G=F$, i.e.\ a Lie group $G$ is a typical fibre.
$\o$~is a 1-form
of \cn\ on~$P$ with values in the algebra of~$G$, $\mathfrak G$.
Let $\vf'(a)$ be a
tangent map to $\vf(a)$ whereas $\vf^\ast(a)$ is the contragradient
to~$\vf'(a)$ at a point~$a$. The form $\o$ is a form of ad-type, i.e.
\beq{Ap.1}
\vf^\ast(a)\o=\ad{a^{-1}}'\o,
\e
where $\ad{a^{-1}}'$ is a tangent map to the internal automorphism of the
group~$G$
\beq{Ap.2}
\ad a(b)=aba^{-1}.
\e
We may introduce the distribution (field) of linear elements $H_r$, $r\in P$,
where $H_r\subset T_r(P)$ is a subspace of the space tangent to~$P$ at a
point~$r$ and
\beq{Ap.4}
v\in H_r \iff \o_r(v)=0.
\e
So
\beq{Ap.5}
T_r(P)=V_r\oplus H_r,
\e
where $H_r$ is called a subspace of \ti{horizontal\/} vectors and $V_r$ of
\ti{vertical\/} vectors.
For vertical vectors $v\in V_r$ we have $\pi'(v)=0$. This means that $v$ is
tangent to the fibres.

Let
\beq{Ap.6}
v=\hor(v)+\ver(v),\quad \hor(v)\in H,\ \ver(v)\in V_r.
\e
It is proved that the distribution $H_r$ is equal to choosing a \cn~$\o$. We
use the operation $\hor$ for forms, i.e.
\beq{Ap.7}
(\hor\b)(X,Y)=\b(\hor X,\hor Y),
\e
where $X,Y\in T(P)$.

The 2-form of a curvature is defined as follows
\beq{Ap.8}
\O=\hor d\o=D\o,
\e
where $D$ means an exterior covariant \dv\ \wrt $\o$. This form is also of
ad-type.

For $\O$ the structural Cartant \e\ is valid
\beq{Ap.9}
\O=d\o+\tfrac12[\o,\o],
\e
where
\beq{Ap.10}
[\o,\o](X,Y)=[\o(X),\o(Y)].
\e
Bianchi's identity for $\o$ is as follows
\beq{Ap.11}
D\O=\hor d\O=0.
\e
The map $f:E\supset U\to P$ such that $f\circ \pi=\id$ is called a
\ti{section} ($U$ is an open set).

From physical point of view it means choosing a gauge. A~covariant \dv\
on~$P$ is defined as follows
\beq{Ap.12}
D\Ps=\hor d\Ps.
\e
This \dv\ is called a \ti{gauge \dv}. $\Ps$ can be a spinor field on~$P$.

In this paper we use also a linear \cn\ on manifolds $E$ and $P$, using the
formalism of differential forms. So the basic quantity is a one-form of the
\cn\ $\gd\o,A,B,$. The 2-form of curvature is as follows
\beq{Ap.13}
\gd\O,A,B,=d\gd\o,A,B,+\gd\o,A,C, \wedge \gd\o,C,B,
\e
and the two-form of torsion is
\beq{Ap.14}
\T^A=D\t^A,
\e
where $\t^A$ are basic forms and $D$ means exterior covariant \dv\ \wrt \cn\
$\gd\o,A,B,$. The following relations are established \cn s with generally met
symbols
\beq{Ap.15}
\bal
\gd\o,A,B,&=\gd\G,A,BC,\t^C\\
\T^A&=\tfrac12\gd Q,A,BC,\t^B\wedge \t^C\\
\gd Q,A,BC,&=\gd\G,A,BC,-\gd\G,A,CB,\\
\gd\O,A,B,&=\tfrac12 \gd R,A,BCD,\t^C \wedge \t^D,
\eal
\e
where $\gd\G,A,BC,$ are \cf s of \cn\ (they do not have to be \s\ in indices
$B$ and~$C$), $\gd R,A,BCD,$ is a tensor of a curvature, $\gd Q,A,BC,$ is a
tensor of a torsion in a holonomic frame. Covariant exterior derivation \wrt $\gd\o,A,B,$ is given
by the formula
\beq{Ap.16}
\bal
D\Xi^A&=d\Xi^A+\gd\o,A,C,\wedge \Xi^C\\
D\gd\Si,A,B,&=
d\gd\Si,A,B,+\gd\o,A,C,\wedge \gd\Si,C,B,-\gd\o,C,B,\wedge \gd\Si,A,C,.
\eal
\e
The forms of a curvature $\gd\O,A,B,$ and torsion $\T^A$ obey Bianchi's
identities
\beq{Ap.17}
\bal
{}&D\gd\O,A,B,=0\\
&D\T^A=\gd\O,A,B,\wedge \t^B.
\eal
\e
All quantities introduced here can be found in Ref.~\cite{13}.

In this paper we use a formalism of a fibre bundle over a \spt~$E$ with an
\elm c \cn~$\a$ and traditional formalism of differential geometry for linear
\cn s on~$E$ and~$P$. In order to simplify the notation we do not use fibre
bundle formalism of frames over $E$ and~$P$. A~vocabulary connected geometrical
quantities and gauge fields (Yang--Mills fields) can be found in
Ref.~\cite{12}.

In Ref.~\cite{Wu} we have also a similar vocabulary (see Table~I, Translation
of terminology). Moreover, we consider a little different terminology. First
of all we distinguished between a gauge \pt\ and a \cn\ on a fibre bundle. In
our terminology a gauge \pt\ $A_\mu \ov\t{}^\mu$ is in a particular gauge $e$
(a~section of a bundle), i.e.
\beq{Ap.18}
A_\mu \ov\t{}^\mu=e^\ast\o
\e
where $A_\mu \ov\t{}^\mu$ is a 1-form defined on $E$ with values in a Lie
algebra $\mathfrak G$ of~$G$. In the case of a strength of a gauge field we have
similarly
\beq{Ap.19}
\tfrac12 F_\m \ov\t{}^\mu \wedge \ov\t{}^\nu=e^\ast\O
\e
where $F_\m \ov\t{}^\mu \wedge \ov\t{}^\nu$ is a 2-form defined on~$E$ with
values in a Lie algebra $\mathfrak G$ of~$G$.

Using generators of a Lie algebra $\mathfrak G$ of $G$ we get
\beq{Ap.20}
A=\gd A,a,\mu, \ov\t{}^\mu X_a=e^\ast \o \quad\hbox{and}\quad
F=\tfrac12\gd F,a,\m,\ov\t{}^\mu \wedge \ov\t{}^\nu X_a=e^\ast \O
\e
where
\beq{Ap.21}
[X_a,X_b]=\gd C,c,ab,X_c, \quad a,b,c=1,2,\dots,n, \ n=\dim G(=\dim \mathfrak G),
\e
are generators of $\mathfrak G$, $\gd C,c,ab,$ are structure \ct s of a Lie
algebra of~$G$, $\mathfrak G$, $[\cdot,\cdot]$ is a commutator of Lie algebra
elements.

In this paper we are using Latin lower case letters for 3-\di al space indices. Here
we are using Latin lower case letters as Lie algebra indices. It does not
result in any misunderstanding.
\beq{Ap.22}
\gd F,a,\m,=\pa_\mu\gd A,a,\nu,-\pa_\nu\gd A,a,\mu,+\gd C,a,bc,\gd A,b,\mu,
\gd A,c,\nu,.
\e
In the case of an \elm c \cn\ $\a$ the field strength~$F$ does not depend on
gauge (i.e.\ on a section of a~bundle).

Finally it is convenient to connect our approach using gauge \pt s $\gd
A,a,\mu,$ with usually met (see Ref.~\cite{Pok}) matrix valued gauge
quantities $A_\mu$ and $F_\m$. It is easy to see how to do it if we consider
Lie algebra generators $X_a$ as matrices. Usually one supposes that $X_a$ are
matrices of an adjoint representation of a Lie algebra~$\mathfrak G$, $T^a$
with a normalization condition
\beq{Ap.23}
{\rm Tr}(\{T^a,T^b\})=2\d^{ab},
\e
where $\{\cdot,\cdot\}$ means anticommutator in an adjoint representation.

In this way
\bea{Ap.24}
A_\mu&=\gd A,a,\mu, T^a,\\
F_\m&=\gd F,a,\m, T^a. \label{Ap.25}
\e
One can easily see that if we take
\beq{Ap.26}
F_\m=\pa_\mu A_\nu - \pa_\nu A_\mu + [A_\mu,A_\nu]
\e
from Ref.~\cite{Pok} we get
\beq{Ap.26a}
F_\m=(\gd F,a,\m,)T^a,
\e
where $\gd F,a,\m,$ is given by \eqref{Ap.22}. From the other side if we take
a section $f$, $f:U\to P$, $U\subset E$, and corresponding to it
\bea{Ap.27}
\ov A=\gd\ov A,a,\mu, \ov\t{}^\mu X_a&=f^\ast \o\\
\ov F=\tfrac12\gd\ov F,a,\m, \ov\t{}^\mu \wedge \ov\t{}^\nu X_a&=f^\ast \O
\label{Ap.28}
\e
and consider both sections $e$ and $f$ we get transformation from $\gd
A,a,\mu,$ to $\gd\ov A{},a,\mu,$ and from $\gd F,a,\m,$ to $\gd\ov F{},a,\m,$ in
the following way. For every $x\in U\subset E$ there is an element $g(x)\in
G$ such that
\beq{Ap.29}
f(x)=e(x)g(x)=\vf(e(x),g(x)).
\e
Due to \eqref{Ap.1} one gets
\bea{Ap.30}
\ov A(x)&=\ad{g^{-1}(x)}'A(x)+{g^{-1}(x)}\,dg(x)\\
\ov F(x)&=\ad{g^{-1}(x)}'F(x) \label{Ap.31}
\e
where $\ov A(x),\ov F(x)$ are defined by \eqref{Ap.27}--\eqref{Ap.28} and
$A(x),F(x)$ by \eqref{Ap.20}. The formulae \eqref{Ap.30}--\eqref{Ap.31} give a
geometrical meaning of a gauge transformation (see Ref.~\cite{12}). In an
\elm c case $G=\U(1)$ we have similarly, if we change a local
section from $e$ to~$f$ we get
$$
f(x)=\vf(e(x), \exp(i\chi(x)))  \quad (f:U\supset E\to P)
$$
and $\ov A=A+d\chi$.

Moreover,
in the traditional approach (see Ref.~\cite{Pok}) one gets
\bea{Ap.32}
\ov A_\mu(x)&=U(x)^{-1}A_\mu(x)U(x)+U^{-1}(x)\pa_\mu U(x)\\
\ov F_\m(x)&=U^{-1}(x)F_\m U(x), \label{Ap.33}
\e
where $U(x)$ is the matrix of an adjoint representation of a Lie group $G$.

For an action of a group $G$ on $P$ is via \eqref{Ap.1}, $g(x)$ is exactly a
matrix of an adjoint representation of~$G$. In this way
\eqref{Ap.30}--\eqref{Ap.31} and \eqref{Ap.32}--\eqref{Ap.33} are equivalent.

Let us notice that usually a Lagrangian of a gauge field (Yang--Mills field)
is written as
\beq{Ap.34}
\cL_{\rm YM} \sim {\rm Tr}(F_\m F^\m)
\e
where $F_\m$ is given by \eqref{Ap.25}--\eqref{Ap.26}. It is easy to see that
one gets
\beq{Ap.35}
\cL_{\rm YM} \sim h_{ab}\gd F,a,\m, F^{b\m}
\e
where
\beq{Ap.36}
h_{ab}=\gd C,d,ac, \gd C,c,bd,
\e
is a Cartan--Killing tensor for a Lie algebra $\mathfrak G$, if we remember
that $X_a$ in adjoint representation are given by structure \ct s $\gd
C,c,ab,$.

Moreover, in Refs \cite{1,3} we use the notation
\beq{Ap.39}
\O=\tfrac12 \gd H,a,\m,\t^\mu \wedge \t^\nu X_a.
\e
In this language
\beq{Ap.40}
\cL_{\rm YM}=\tfrac1{8\pi} h_{ab}\gd H,a,\m, H^{b\m}.
\e
It is easy to see that
\beq{Ap.41}
e^\ast (\gd H,a,\m,\t^\mu \wedge \t^\nu X_a)=\gd F,a,\m,{\ov\t}^\mu
\wedge {\ov \t}^\nu X_a.
\e
Thus \eqref{Ap.40} is equivalent to \eqref{Ap.35} and to \eqref{Ap.34}.
\eqref{Ap.34} is invariant to a change of a gauge. \eqref{Ap.40} is invariant
\wrt the action of a group~$G$ on~$P$.

Let us notice that $h_{ab}\gd F,a,\m, F^{b\m}=h_{ab}\gd H,a,\m, \gd H,b,\m,$,
even $\gd H,a,\m,$ is defined on~$P$ and $\gd F,a,\m,$ on~$E$. In the
non-Abelian case it is more natural to use $\gd H,a,\m,$ in place of $\gd
F,a,\m,$.

Eventually we connect the general fibre bundle formalism and Cartan calculus
with a formalism of linear \cn s on $E$, $P$ and $E\tm G/G_0$.

Let $M$ be an $m$-\di al pseudo-Riemannian manifold with metric~$g$ of
arbitrary signature. Let $T(M)$ be a tangent bundle and $O(M,g)$ the principal
fiber bundle of frames (orthogonal frames) over~$M$. The \sc e group $O(M,g)$
is a group $GL(m,R)$ or the subgroup of $GL(m,R)$ $O(m-p,p)$ which leaves
the metric \iv t. Let $\varPi$ be the projection of $O(M,g)$ onto~$M$. Let
$X$ be a tangent vector at~$x$ in $O(M,g)$. The canonical or soldering form
$\wt\t$ is an $\R^m$-valued form on $O(M,g)$ whose $A$-th component $\wt\t{}^A$
at $x$ of~$X$ is the $A$-th component of $\varPi'(X)$ in the frame~$x$. The \cn\
form $\wt\o =\gd \o,A,B,\gd X,B,A,$ is a 1-form on $O(M,g)$ which takes its
values in the \Li\ $gl(m,R)$ of $Gl(m,R)$ or in $o(m-p,p)$ of $O(m-p,p)$ and
\sf ies the \sc e \e
\beq *
d\wt\o+\frac12[\wt \o,\wt \o]=\wt\O= {\rm\wt Hor}\,d\wt\o
\e
where $\rm\wt Hor$ is understood in the sense of $\wt \o$ and $\wt\O=
\gd\wt\O,A,B,\gd X,A,B,$ is a $gl(m,R)$ ($o(m-p,p)$)-valued 2-form of the
\cvt. $\gd X,A,B,$ are generators of a \Li\ $gl(m,R)$ or $o(m-p,p)$. We can
write Eq.~\er{*} using $\R^{2m}$-valued forms and commutation relations of
the \Li\ $gl(m,R)$ ($o(m,m-p)$)
\beq{**}
\gd\wt\O,A,B,=\gd d\wt \o,A,B, + \gd\wt \o,A,C, \land \gd\wt\o,C,B,.
\e
Taking any local section of $O(M,g)$, $e$, one can get the \cf s of the
\cn, \cvt, basic forms and torsion
\beq{***}
\bal
&e^*\gd\wt \o,A,B,=\gd\o,A,B,\\
&e^*(\gd\wt\O,A,B,)=\gd\wt\O,A,B,\\
&e^*\wt\t{}^A=\t^A\\
&e^*\wt\T{}^A=\T^A.
\eal
\e
The forms on the right-hand side of \e s \er{***} are different in
Eqs \er{Ap.13}--\er{Ap.14}. We call this formalism a linear (affine) metric,
Riemannian--Levi-Civita, Einstein) \cn s on~$M$.

In our theory it is necessary to consider at least four principal bundles:
a principal fiber bundle~$P$ over~$E$ with a \sc al group~$G$ (a~gauge group),
\cn~$\o$ and a projection~$\pi$, an operator of a horizontality $\hor$, a~principal fiber
bundle~$P'$ of frames over $(E,g)$ with a \cn\ $\gd\wt\o,\a,\b,
\gd X,\b,\a,=\o'$, a~\sc al group $GL(4,R)$ ($O(1,3)$), an operator of
horizontality $\ov{\hor}$, and a projection~$\ov\pi$, a~principal fiber bundle $P''$ of frames over $(P,\g)$
(a~metrized fiber bundle~$P$) with a \sc al group $GL(n+4,R)$ ($O(n+3,1)$),
a~\cn\ $\gd\wt\o,A,B,\gd X,B,A,=\wt\o$ and with an operator of horizontality
$\ov{\hor}{}'$, a projection $\ov\pi{}'$ and a principal fiber bundle of frames over~$G$ with a
projection~$\varPi''$, operator of horizontality $(\hor)''$, a~\cn~$\wh\o$
and a \sc al group $Gl(n,R)$. In the case with a \sn\ \s y breaking we need
even more principal bundles of frames, i.e.\ a~principal bundle of frames
over $E\tm G/G_0$ with additional \cn~$\ov{\ov\o}$, a~projection~$\ov{\ov\varPi}$, an
operator of horizontality $\ov{\ov{\hor}}$. In more complicated situation we can also
consider a bundle over $G/G_0$ with \sc al group $GL(n_1,R)$. Moreover, in
order to simplify considerations, we use the formalism of linear \cn\ \cf s
on manifold $(E,g)$, $(P,\g)$, and a principal fiber bundle formalism for~$P$,
i.e.\ a~principal fiber bundle over~$E$ with a \sc al group~$G$, a~gauge
group. In the case with a \sn\ \s y breaking we have also an additional fiber
bundle with a \sc al group~$H$ over $E\tm G/G_0$. I~believe this is a way to
make the formalism more natural and readable. We use tensor formalism with
many kinds of indices which make some formulas very long. Moreover, they
are more readable for a non-expert.

\section{Elements of the \NK\ in general \nA\ case and dielectric model
of a color \cfn}
Let $\ul P$ be a principal fiber bundle over a \spt\ $E$ with a structural
group~$G$ which is a semisimple Lie group. On a \spt~$E$ we define a \nos\
tensor $g_\m=g_\(\m)+g_\[\m]$ \st
\beq2.1
\bal
g&=\det(g_\m)\ne0\\
\wt g&=\det(g_\(\m))\ne0.
\eal
\e
$g_\[\m]$ is called as usual a skewon field (e.g.\ in NGT, see Refs
\cite{6,8}).
We define on $E$ a \nos\ \cn\ compatible with $g_\m$ \st
\beq2.2
\ov Dg_{\a\b}=g_{\a\d}\gd \ov Q,\d,\b\g,(\ov\G)\ov \t{}^\g
\e
where $\ov D$ is an exterior covariant \dv\ for a \cn\ $\gd\ov\o,\a,\b,=
\gd\ov \G,\a,\b\g,\ov \t{}^\g$ and $\gd\ov Q,\a,\b\d,$ is its torsion. We suppose also
\beq2.3
\gd\ov Q,\a,\b\a,(\ov \G)=0.
\e
We introduce on $E$ a second \cn
\beq2.4
\gd\ov W,\a,\b,=\gd\ov W,\a,\b\g,\ov \t{}^\g
\e
\st
\bg2.5
\gd \ov W,\a,\b,=\gd \ov\o,\a,\b,-\tfrac23\,\gd \d,\a,\b,\ov W\\
\ov W=\ov W_\g\ov \t{}^\g=\tfrac12(\gd \ov W,\si,\g\si,-\gd \ov W,\si,\si\g,)
\ov \t{}^\g. \lb2.6
\e

Now we turn to \nos\ metrization of a bundle $\ul P$. We define a \nos\ tensor
$\g$ on a bundle manifold $P$ \st
\beq2.6a
\g=\pi^* g\op \ell_{ab}\t^\a\ot \t^b
\e
where $\pi$ is a projection from $P$ to $E$. On $\ul P$ we define a \cn~$\o$
(a~1-form with values in a Lie algebra $\fg$ of~$G$). In this way we can
introduce on~$P$ (a~bundle manifold) a frame  $\t^A=(\pi^*(\ov \t{}^\a),
\t^a)$ \st
$$
\t^a=\la\o^a,\q \o=\o^a X_a, \q a=5,6,\dots,n+4, \q
n=\dim G=\dim\fg, \q \la={\rm const.}
$$
Thus our \nos\ tensor looks like
\bg2.7
\g=\g_{AB} \t^A\ot\t^B, \q A,B=1,2,\dots,n+4,\\
\ell_{ab}=h_{ab}+\mu k_{ab}, \lb2.8
\e
where $h_{ab}$ is a bi\iv t Killing--Cartan tensor on~$G$ and $k_{ab}$ is a
right-\iv t skew-\s ic tensor on~$G$, $\mu={\rm const}$.

We have
\beq2.9
\bal
h_{ab}&=\gd C,c,ad,\gd C,d,bc,=h_{ab}\\
k_{ab}&=-k_{ba}
\eal
\e
Thus we can write
\bea2.10
\ov \g(X,Y)=\ov g(\pi'X,\pi'Y)+\la^2h(\o(X),\o(Y))\\
\ul \g(X,Y)=\ul g(\pi'X,\pi'Y)+\la^2k(\o(X),\o(Y)) \lb2.11
\e
($\gd C,a,bc,$ are structural \ct s of the Lie algebra $\fg$).

$\ov \g$ is the \s ic part of $\g$ and $\ul \g$ is the anti\s ic part of~$\g$.
We have as usual
\beq2.12
[X_a,X_b]=\gd C,c,ab,X_c
\e
and
\beq2.13
\O=\frac12 \gd H,a,\m,\t^\mu \land \t^\nu
\e
is a curvature of the \cn\ $\o$,
\beq2.14
\O=d\o+\frac12[\o,\o].
\e
The frame $\t^A$ on $P$ is partially nonholonomic. We have
\beq2.15
d\t^a=\frac\la2\Bigl(\gd H,a,\m,\t^\mu\land \t^\nu - \frac1{\la^2}\,
\gd C,a,bc,\t^b\land\t^c\Bigr)\ne0
\e
even if the bundle $\ul P$ is trivial, i.e.\ for $\O=0$. This is different than
in an \elm c case (see Ref.~\cite3). Our \nos\ metrization of a
principal fiber bundle gives us a right-\iv t structure on~$P$ \wrt an action
of a group~$G$ on~$P$ (see Ref.~\cite3 for more details). Having $P$ \nos
ally metrized one defines two \cn s on~$P$ right-\iv t \wrt an action of a
group $G$ on~$P$. We have
\beq2.16
\g_{AB}=\left(\begin{array}{c|c}
g_{\a\b}&0\\
\hline
0&\ell_{ab}\end{array}\right)
\e
in our left horizontal frame $\t^A$.
\bg2.17
D\g_{AB}=\g_{AD}\gd Q,D,BC,(\G)\t^C\\
\gd Q,D,BD,(\G)=0 \label{2.18}
\e
where $D$ is an exterior covariant \dv\ \wrt a \cn\ $\gd \o,A,B,=\gd \G,A,BC,
\t^C$ on~$P$ and $\gd Q,A,BC,(\G)$ its torsion. One can solve
Eqs~\er{2.17}--\er{2.18} getting the following results
\beq2.19
\gd \o,A,B,=\left(
\begin{array}{c|c}
\pi^*(\gd \ov\o,\a,\b,)-\ell_{db}g^{\mu\a}\gd L,d,\mu\b,\t^b&\gd L,a,\b\g,\t^\g\\
\hline
\ell_{bd}g^{\a\b}(2\gd H,d,\g\b,-\gd L,d,\g\b,)\t^\g & \gd \wt\o,a,b,
\end{array}\right)
\e
where $g^{\mu\a}$  is an inverse tensor of $g_{\a\b}$
\beq2.20
g_{\a\b}g^{\g\b}=g_{\b\a}g^{\b\g}=\d^\g_\a,
\e
$\gd L,d,\g\b,=-\gd L,a,\b\g,$ is an Ad-type tensor on~$P$ \st
\beq2.21
\ell_{dc}g_{\mu\b}g^{\g\mu}\gd L,d,\g\a,+\ell_{cd}g_{\a\mu}g^{\mu\g}
\gd L,d,\b\g,=2\ell_{cd}g_{\a\mu}g^{\mu\g}\gd H,d,\b\g,,
\e
$\gd\wt\o,a,b,=\gd \wt\G,a,bc,\t^c$ is a \cn\ on an internal space (typical
fiber) compatible with a metric $\ell_{ab}$ \st
\bg2.22
\ell_{db}\gd \wt\G,d,ac,+\ell_{ad}\gd \wt\G,d,cb,=-\ell_{db}\gd C,d,ac,\\
\gd \wt\G,a,ba,=0, \q \gd \wt\G,a,bc,=\gd -\wt\G,a,cb, \lb2.23
\e
and of course $\gd \wt Q,a,ba,(\wt\G)=0$ where $\gd \wt Q,a,bc,(\G)$ is a
torsion of the \cn~$\gd \wt\o,a,b,$.

We also introduce an inverse tensor of $g_\(\a\b)$
\beq2.24
g_\(\a\b)\wt g{}^\(\a\g)=\d^\g_\b.
\e
We introduce a second \cn\ on~$P$ defined as
\beq2.25
\gd W,A,B,=\gd \o,A,B,-\frac4{3(n+2)}\,\gd\d,A,B,\ov W.
\e
$\ov W$ is a horizontal one form
\bg2.26
\ov W=\hor\ov W\\
\ov W=\ov W_\nu\t^\nu =\tfrac12(\gd\ov W,\si,\nu\si, - \gd \ov W,\si,\si\nu,).
\lb2.27
\e

In this way we define on $P$ all analogues of four-\di al quantities from NGT
(see Refs \cite{6,14,15,16}). It means, $(n+4)$-\di al analogues from Moffat
theory of \gr, i.e.\ two \cn s and a \nos\ metric $\g_{AB}$. Those quantities
are right-\iv t \wrt an action of a group~$G$ on~$P$. One can calculate a
scalar curvature of a \cn\ $\gd W,A,B,$ getting the following result (see
Refs \cite{1,3}):
\beq2.28
R(W)=\ov R(\ov W)-\frac{\la^2}4 \bigl(2\ell_{cd}H^cH^d - \ell_{cd}L^{c\m}
\gd H,d,\m,\bigr) + \wt R(\wt \G)
\e
where
\beq2.29
R(W)=\g^{AB}\bigl(\gd R,C,ABC,(W)+\tfrac12 \, \gd R,C,CAB,(W)\bigr)
\e
is a Moffat--Ricci curvature scalar for the \cn~$\gd W,A,B,$,
$\ov R(\ov W)$ is a Moffat--Ricci curvature scalar for the \cn~$\gd \ov W,\a,\b,$,
and $\wt R(\wt \G)$ is a Moffat--Ricci curvature scalar for the \cn~$\gd \wt\o,a,b,$,
\bg2.30
H^a=g^\[\m]\gd H,a,\m,\\
L^{a\m}=g^{\a\mu}g^{\b\nu}\gd L,a,\a\b,. \lb2.31
\e
Usually in ordinary (\s ic) Kaluza--Klein Theory one has
$\la=2\frac{\sqrt{G_N}}{c^2}$, where $G_N$ is a Newtonian \gr al \ct\ and
$c$~is the speed of light. In our system of units $G_N=c=1$ and $\la=2$. This
is the same as in \NK\ in an \elm ic case (see Refs \cite{4,8}). In the \nA\
Kaluza--Klein Theory which unifies GR and \YM\ field theory we have a \YM\
lagrangian and a \co ical term. Here we have
\beq2.32
\cLY=-\frac1{8\pi}\,\ell_{cd}\bigl(2H^cH^d-L^{c\m}\gd H,d,\m,\bigr)
\e
and $\wt R(\wt\G)$ plays a role of a \co ical term.

It is easy to see that $\cLY$ is \iv t \wrt an action of a group~$G$ on~$P$
(it is gauge \iv t). $\wt R(\wt \G)$ is also gauge \iv t.

$L^{c\m}$ plays a role of an induction tensor of \YM\ field (a~gauge field).

According to Refs \cite{1,3} we have
\beq2.33
\gd Q,a,\m,(\G)=2(\gd H,a,\m,-\gd L,a,\m,)
\e
where $\gd Q,A,BC,(\G)$ is a torsion of a \cn~$\G$. Writing $\gd L,a,\m,$ in
the form
\beq2.34
\gd L,a,\m,=\gd H,a,\m, - 4\pi \gd M,a,\m,
\e
we get
\beq2.35
\gd Q,a,\m,(\G)=8\pi \gd M,a,\m,.
\e
One can solve Eq.~\er{2.21} getting the result (see Appendix~A):
\bml2.36
\gd L,n,\o\mu,=\gd H,n,\o\mu, +\mu h^{na}k_{ad}\gd H,d,\o\mu,
+\bigl(\gd H,n,\a\o, \wt g{}^\(\a\d) g_\[\d\mu] - \gd H,n,\a\mu,
\wt g{}^\(\a\d)g_\[\d\o]\bigr)\\
{}-2\mu h^{na}k_{ad}\wt g{}^\(\d\tau)\wt g{}^\(\a\b)\gd H,d,\d\a, g_\[\tau\o]
g_\[\b\mu]
-2\mu h^{na}k_{ad}\wt g{}^\(\d\b)\wt g{}^\(\a\tau)
\gd H,d,\b\tl \o,g_{\mu\tp\tau}g_\[\d\a]\\
{}+2\mu^2 h^{na}h^{bc}k_{ac}k_{bd} \wt g{}^\(\a\b)\gd H,d,\a\tl\o,
g_{[\mu\tp\b]}.
\e
In this way we get that
\beq2.37
\gd L,a,\o\mu,=-\gd L,a,\mu\o,
\e
and simultaneously $\gd Q,a,\m,(\G)$ has a physical interpretation as a
polarization tensor of \YM\ theory (a~difference between an induction tensor
and a gauge field strength). Moreover, it seems from Eq.~\er{2.32} that
$L^{a\m}$ plays the role of an induction tensor. Thus one can get
\bml2.38
\cLY=\frac1{8\pi}\biggl(h_{nk}H^{k\o\mu}\gd H,n,\o\mu,
-2h_{cd}H^cH^d+ 2h_{nk}H^{k\o\mu}\gd H,n,\d\o,g_\[\a\mu]\wt g{}^\(\a\d)\\
{}+\mu\Bigl[2k_{nk}H^{k\o\mu}\gd H,n,\d\o,\wt g{}^\(\d\a)g_\[\a\mu]
-2k_{kd}H^{k\o\mu}\gd H,d,\d\a,\wt g{}^\(\d\b)\wt g{}^\(\a\rho)
g_\[\b\o]g_\[\rho\mu]\\
{}-k_{kd}H^{k\o\mu}\gd H,d,\eta\o,\wt g{}^\(\eta\b)\wt g{}^\(\a\rho)
g_\[\mu\a]g_\[\b\rho]
+k_{kd}H^{k\o\mu}\gd H,d,\eta\o,\wt g{}^\(\eta\d)\wt g{}^\(\a\rho)
g_\[\d\b]g_\[\o\d]\Bigr]\\
{}+\mu^2\Bigl[k_{nk}\gd k,n,d,H^{k\o\mu}\gd H,d,\eta\mu,\wt g{}^\(\rho\b)
\wt g{}^\(\eta\a)g_\[\o\b]g_\[\a\rho]
-2k_{nk}\gd k,n,d,H^{k\o\mu}\gd H,d,\d\a,\wt g{}^\(\d\eta)\wt g{}^\(\a\rho)
g_\[\eta\o]g_\[\rho\mu]\\
{}-k_{nk}\gd k,n,d,H^{k\o\mu}\gd H,d,\eta\o,\wt g{}^\(\rho\a)
\wt g{}^\(\eta\b)g_\[\mu\a]g_\[\b\rho]
+\dg k,k,b,k_{bd}H^{k\o\mu}\gd H,d,\a\o,\wt g{}^\(\a\b)g_\[\mu\a]\\
{}
-\dg k,k,b,k_{bd}H^{k\o\mu}\gd H,d,\a\mu,\wt g{}^\(\a\b)g_\[\o\b]
+\gd k,p,n,k_{pk}H^{k\o\mu}\gd H,n,\o\mu,\Bigr]\\
{}+\mu^3\Bigl[k_{nk}k^{nb}k_{bd}H^{k\o\mu}\gd H,d,\a\o,\wt g{}^\(\a\b)g_\[\mu\b]
-k_{nk}k^{nb}k_{bd}H^{k\o\mu}\gd H,d,\a\mu,\wt g{}^\(\a\b)g_\[\o\b]\Bigr]
\biggr).
\e
Eq.~\er{2.38} is written in term of $\gd H,a,\m,$ only. Moreover, the form
$\cLY$, i.e.\ Eq.~\er{2.32}, is more convenient for theoretical
considerations. One can say the same for $\gd L,a,\m,$. One gets
\begin{multline}\label{2.34*}
\gd Q,n,\o\mu,(\G)=2\Bigl(-\mu h^{na}k_{ad}\gd H,d,\o\mu,
-\bigl(\gd H,n,\a\o,\wt g{}^\(\a\d)g_\[\d\mu]
-\gd H,n,\a\mu, \wt g{}^\(\a\d)g_\[\d\o]\bigr)\\
{}+2\mu h^{na}k_{ad}\wt g{}^\(\d\tau)\wt g{}^\(\a\b)\gd H,d,\d\a,g_\[\tau\o]g_\[\b\mu]
+2\mu h^{na}k_{ad}\wt g{}^\(\d\b)\wt g{}^\(\a\tau) \gd H,d,\b[\o,g_{\mu]\tau}g_\[\d\a]\\
{}-2\mu^2 h^{na}h^{bc}k_{ac}k_{bd}\wt g{}^\(\a\b)\gd H,d,\a[\o,g_{|\mu|\b]}\Bigr)
\tag {\ref{2.33}*}
\end{multline}

Let us introduce the following notation:
\bg2.39
\gd H,a,\m,=\left(\begin{matrix}
0 & -B^a_3 & B^a_2 & -E^a_1 \\
B^a_3 & 0 & -B^a_1 & -E^a_2 \\
-B^a_2 & B^a_1 & 0 & -E^a_3 \\
E^a_1 & E^a_2 & E^a_3 & 0 \end{matrix}\right)\\
L^{a\m}=\left(\begin{matrix}
0 & -H^{3a} & H^{2a} & -D^{1a} \\
H^{3a} & 0 & -H^{1a} & -D^{2a} \\
-H^{2a} & H^{1a} & 0 & -D^{3a} \\
D^{1a} & D^{2a} & D^{3a} & 0 \end{matrix}\right). \lb2.40
\e
In this way we write $\gd H,a,\m,$ in terms of $\os E{}^\a=(E^a_\ba)=
(E^a_1,E^a_2,E^a_3)$, $\ba=1,2,3$, and $\os B{}^a = (B^a_\ba)=(B^a_1,B^a_2,
B^a_3)$, $L^{a\mu}$ in terms of $\os D{}^a = (D^{\ba a})=(D^{1a},D^{2a},
D^{3a})$ and $\os H{}^a = (H^{\ba a})=(H^{1a},H^{2a},H^{3a})$.

In this way
\beq2.41
\bal
E^a_\ba&=\gd H,a,4a, \q D^{\ba a}=L^{a4a},\\
\os B{}^a&=-(\gd H,a,23,, \gd H,a,31,, \gd H,a,12,)\\
\os H{}^a&=-(L^{a23},L^{a31},L^{a12})
\eal
\e
or
\bea2.42
\gd B,a,\ba,&=-\tfrac12 \, \dg \ve,\ba,\bar b\bar c,\gd H,a,\bar b\bar c,\quad
\gd H,a,\bar c\bar m,=-\dg \ve,\bar c\bar m,\bar e, \gd B,a,\bar e,
\\
H^{\ba a}&=-\tfrac12\,\gd \ve,\ba,\bar b\bar c,L^{a\bar b\bar c},\q
L^{a\bar c\bar m}=-\gd \ve,\bar c\bar m,\bar e, H^{\bar ea}
\lb2.43
\e
where $\ve_{\ba\bar b\bar c}$, $\ba,\bar b,\bar c=1,2,3$, is a usual 3-\di al
anti\s ic symbol, $\ve_{123}=1$ and it is unimportant for it if its indices
are in up or down position. We keep these indices in up or down position only
for convenience.

One gets
\beq2.44
D^{n\bar e}=\gdg\ov B,n,f,\bar d\bar e,\gd B,f,\bar d, + \gdg\ov A,n,f,\bar
v\bar e,\gd E,f,\bar v,
\e
where
\bml 2.45
\gdg \ov B,n,d,\bar p\bar e, = \dg\ve,\bar m\bar z,\bar p,\Bigl(
g^{\bar z4}g^{\bar m\bar e}\gd \d,n,d,
+\mu \gd k,n,d, g^{\bar z4}g^{\bar m\bar e}
+g^{\mu \bar e}g_\[\d\mu]g^{\bar z4} \wt g{}^\(\bar m\d)\gd \d,n,d,
-g^{\o4}g_\[\d\mu]g^{\bar m\bar e}\wt g{}^\(\bar z4)\gd \d,n,d,\\
{}+\mu \gd k,n,d, g^{\mu\bar e}\wt g{}^\(\a\tau)g_\[\mu\tau]g_\[\d\a]
g^{\bar z4}\wt g{}^\(\d\bar m)
+\mu \gd k,n,d, g^{\o4}\wt g{}^\(\a\tau)g_\[\mu\tau]g_\[\d\a]
g^{\bar m\bar e}\wt g{}^\(\d\bar z)
-\mu^2\gd k,n,c,\gd k,c,d,g^{\mu\tau}g_\[\mu\b]g^{\bar z4}\wt g{}^\(m\b)\\
{}-\mu^2\gd k,n,c,\gd k,c,d,g^{\o4}g_\[\o\b]g^{\bar m\bar e}\wt g{}^\(\bar z\b)
-2\mu\gd k,n,d,g^{\o4}g^{\mu\tau}g_\[\tau\o]g_\[\b\mu]\wt g{}^\(\bar z\tau)
\wt g{}^\(\bar m\b)\Bigr),
\e
\bml2.46
\gdg \ov A,n,d,\bar m\bar e, = g^{44}g^{\bar m\bar e}\gd \d,n,d,
-\mu\gd k,n,d,g^{\bar m4}g^{4\bar e}
+\mu \gd k,n,d,g^{44}g^{\bar m\bar e}+g^{\mu\bar e}g_\[\d\mu]
\wt g{}^\(4\d)g^{\bar m4}\gd \d,n,d,
- g^{\mu\bar e}g_\[\d\mu]g^{44}\wt g{}^\(\bar m\d)\gd \d,n,d,\\
{}-g^{\o4}g_\[\d\o]\wt g{}^\(4\d)\wt g{}^{\bar m\bar e}\gd \d,n,d,
+g^{\o4}g_\[\d\o]g^{4\bar e}\wt g{}^\(\bar m\d)\gd \d,n,d,
-\mu \gd k,n,d,g^{\mu\bar e}\wt g^\(\a\tau)
g_\[\mu\tau]g_\[\d\a]g^{\bar m\bar e}\wt g{}^\(\d\mu)\\
{}+\mu\gd k,n,d, g^{\mu\d}\wt g^\(\a\tau)g_\[\mu\tau]g_\[\d\a]g^{44}\wt g^\(\d\bar m)
+\mu\gd k,n,d, g^{\o4}\wt g{}^\(\a\tau)g_\[\mu\tau]g_\[\d\a]g^{\bar m\bar e}
\wt g{}^\(\d\mu)\\
{}-\mu\gd k,n,d, g^{\o4}\wt g{}^\(\a\tau)g_\[\mu\tau]g_\[\d\a]g^{4\bar e}
\wt g{}^\(\d\bar m)
-\mu^2 \gd k,n,c,\gd k,c,d,g^{\mu\bar e}g_\[\mu\b]g^{\bar m4}\wt g^\(\b4)
-\mu^2 \gd k,n,c,\gd k,c,d,g^{\mu\bar e}g_\[\mu\b]g^{44}\wt g^\(\bar m\b)\\
{}-\mu^2 \gd k,n,c,\gd k,c,d,g_\[\o\b]g^{\bar m\bar e}\wt g{}^\(4\b)
\wt g{}^\(4\o)
+\mu^2 \gd k,n,c,\gd k,c,d,g^{\o4}g_\[\o\b]g^{4\bar e}\wt g{}^\(\bar m\b)\\
{}+2\mu \gd k,n,d, g^{\o4}g^{\mu\bar e}g_\[\tau \o]g_\[\b\mu]\wt g{}^\(\bar
m\tau) \wt g{}^\(4\b)
+2\mu \gd k,n,d, g^{\o4}g^{\mu\bar e}g_\[\tau \o]g_\[\b\mu]\wt g{}^\(4
\tau) \wt g{}^\(\bar m\b)
-\gd \d,n,d, g^{\bar m4}g^{4\bar e},
\e
\beq2.47
H^{nd}=\gdg \ov C,n,f,\bar d\bar p, \gd B,f,\bar p, + \gdg\ov D,n,f,d\bar v,
\gd E,f,\bar v,,
\e
where
\bml2.48
\gdg \ov C,n,d,\bar p\bar f, = \frac12 \,\gd\ve ,\bar p,\bar e\bar k,
\dg \ve,\bar w\bar m,\bar f, \Bigl(
g^{\bar w\bar k}g^{\bar m\bar e}\gd\d,n,d,
+\mu \gd k,n,d,g^{\bar w\bar k}g^{\bar m\bar e}
-g^{\bar w\bar k}g^{\mu\bar e}g_\[\d\mu]\wt g{}^\(\bar m\d)\gd \d,n,d,\\
{}+\mu \gd k,n,d, g^{\bar w\bar k}g^{\mu\bar e}g_\[\d\mu]\wt g{}^\(\bar m\d)
\gd\d,n,d,
-g^{\o\bar k}g_\[\d\o]g^{\bar m\bar e}\wt g{}^\(\bar w\d)\gd \d,n,d,\\
{}-2\mu \gd k,n,d, g^{\o\bar k}g^{\mu\bar e}g_\[\tau\o]g_\[\b\mu]
\wt g{}^\(\bar w\b)\wt g{}^\(\bar m\tau)
+\mu \gd k,n,d,g^{\o\bar k}\wt g^\(\a\tau)g_\[\o\tau]g_\[\b\mu]
g^{\bar m\bar e}\wt g{}^\(\bar w\b)\Bigr),
\e
\bml2.49
\gdg \ov D,n,d,\bar p\bar m,=\frac12\,\gd \ve,\bar p,\bar e\bar k,\Bigl(
g^{4\bar k}g^{\bar m\bar e}\gd\d,n,d, - g^{\bar m\bar k}g^{4\bar e}\gd\d,n,d,
-\mu\gd k,n,d,g^{\bar m\bar k}g^{4\bar e}
+\mu\gd k,n,d,g^{\bar e\bar k}g^{4\bar k}
+g^{\bar m\bar k}g^{\mu\bar e}g_\[\d\mu]\wt g{}^\(4\d)\gd\d,n,d,\\
{}-g^{4\bar k}g^{\mu\bar e}\wt g{}^\(\bar m\d)g_\[\d\mu]\gd\d,n,d,
-g^{\o\bar k}g_\[\d\o]g^{\bar m\bar e}\wt g{}^\(4\d)\gd\d,n,d,
+2\mu\gd k,n,d, g^{\o\bar k}g^{\mu\bar e}g_\[\tau\o]g_\[\b\mu]
\wt g{}^\(\bar m\b)\wt g{}^\(4\tau)\\
{}-2\mu \gd k,n,d,g^{\o\bar k}g^{\mu\bar e}g_\[\tau\o]g_\[\b\mu]
\wt g{}^\(4\b)\wt g{}^\(\bar m\tau)
+\mu\gd k,n,d,g^{\mu\bar e}g_\[\d\a]g_\[\mu\tau]\wt g{}^\(\a\tau)
g^{4\bar k}\wt g{}^\(\d\bar m)\\
{}-\mu \gd k,n,d,g^{\mu\bar e}g_\[\d\a]g_\[\mu\tau]\wt g^\(\a\tau)
g^{\bar m\bar k}\wt g{}^\(4\d)
+\mu\gd k,n,d,g^{\o\bar k}\wt g{}^\(\a\tau)g_\[\o\tau]g_\[\d\a]
g^{\bar m\bar e}\wt g{}^\(4\d)
+\mu^2\gd k,n,c,\gd k,c,d,g^{\mu\bar e}g_\[\mu\b]g^{\bar m\bar k}\wt g{}^\(4\b)\\
{}-\mu^2\gd k,n,c,\gd k,c,d,g^{\mu\bar e}g_\[\mu\b]g^{4\bar k}\wt g{}^\(\bar m\b)
-\mu^2\gd k,n,c,\gd k,c,d,g^{\o\bar k}g_\[\o\b]g^{\bar m\bar e}\wt g{}^\(4\b)
+\mu^2\gd k,n,c,\gd k,c,d,g^{\o\bar k}g_\[\o\b]g^{4\bar e}\wt g{}^\(\bar m\b).
\Bigr)
\e

The \cfn\ condition in this theory means
\beq2.50
D^{\bar aa}=0
\e
with $E^{\bar aa}\ne0$ and can be satisfied by special arrangement of the
\nos\ tensor $g_\m$. This generalizes a notion of a charge \cfn\ from
Ref.~\cite8 and can be considered as a color \cfn\ in the case of $G=\SU(3)_c$
(QCD).

In this case \gr\ behaves as a medium which generalizes a notion of
bianisotropic medium in \elm c theory to \nA\ \YM\ field. This is a
dielectric model of \cfn.

It is easy to see that if $g_\[\m]=k_{ab}=0$ we get
$$
\gd L,a,\m,=\gd H,a,\m,.
$$
We have identities concerning $\gd H,a,\m,$ and $\gd L,a,\m,$ coming from
Eq.~\er{2.21}:
\bg2.51
g^\[\m]\gd L,a,\m, = h^{ac}\ell_{cp}\gd H,p,\m,g^\[\m]\\
\ell_{dc}g^{\si\nu}g^{\a\mu}\gd L,d,\si\a,\gd H,c,\m,
+\ell_{cd}g^{\mu\si}g^{\nu\b}\gd L,d,\b\si,\gd H,c,\m, =
2\ell_{cd}g^{\mu\si}g^{\nu\b}\gd H,d,\b\si,\gd H,c,\m, \lb2.52 \\
\ell_{dc}g^{\a\o}g^{\b\mu}\gd L,d,\a\b,\gd L,c,\o\mu,
= \ell_{cd}g^{\a\o}g^{\b\mu}\gd L,d,\a\b,\gd L,c,\o\mu,. \lb2.53
\e

The problem of a \cfn\ emerged in QCD on a quantum level. Moreover, up to now
we have not any realistic explanation of this problem. QCD is a quantum field
theory obtained via quantization procedure
from classical \YM' field theory in a perturbative regime.
The \cfn\ is a strictly non-perturbative effect. The natural way to solve the
problem is to pose it on a classical level and afterwards to quantize the new
theory  (classical) using non-perturbative methods to get a quantum model of
the \cfn. The theory is of course highly nonlinear. Nonperturbative
quantization of nonlinear theories including gravity can be achieved by using
canonical quantization as in~GR (Ashtekar--Lewandowski approach) or using
nonlocal approach (as we shortly described in Conclusions). Even strings
models need quantization.

There is no dielectric
classical model of \cfn\ in a \s ic theory, i.e.\ with $g_\[\m]=0$, zero
skewon field.

Let us give some remarks on a \cfn.
According to modern ideas (see \cite{x1}, \cite{a}, \cite{y}, \cite{z}) the \cfn\ of
color could be connected to dielectricity of the vacuum (dielectric model of
\cfn). Due to the so-called antiscreening mechanism, the effective dielectric
\ct\ is equal to zero. This means that the energy of an isolated charge goes to
infinity. There are also so-called classical-dielectric models of \cfn\ (see
Refs \cite{u1}, \cite{u2}). The \cfn\ is induced by a special kind of dielectricity
of the vacuum, \st $\os E\ne0$ and $\os D=0$ ($\os E{}^a\ne0$, $\os D{}^a=0$).
In this case we do not have a distribution of a charge. This is similar to the electric
type of Meissner effect.

It is easy to see that in our case (the \NK) the dielectricity is induced by the
\nos\ tensors $g_\m$ and $\ell_{ab}$. If $g_\[\m]=0$, $\os D=\os E$, $\os B=\os H$
(in an \elm c case see Ref.~\cite8).
The \gr al field described by the \nos\ tensor $g_\m$ behaves as a medium for an \elm c
field (or \YM' field). In this way the skewon field $g_\[\m]$ plays a double role:
\bit
\item[1)] additional \gr al \ia\ from NGT,
\item[2)] a strong \ia\ field connected to the \cfn\ problem.
\eit
In other words we can say that we get a \cfn\ from higher \di s due to a torsion
in higher \di s.

In Refs \cite{w}, \cite{v} one can find some ideas of nonlocal field theory with
an application to \cfn\ problem which can be connected to dielectric model
of \cfn.

There is a body of works on classical models of \cfn\ for Abelian and \nA\ gauge
fields (see Refs \cite{p1}, \cite{p2}, \cite{p3}) which are not directly connected
to our approach. Moreover, it is worth to mention that an idea of a \cfn\ in QCD
for $\SU(3)$ group (see Ref.~\cite{p4}) can be applied for electrodynamics in
order to get a \cfn\ of plasma in thermonuclear fusion.

We do not confuse here
the ``\cfn'' problem in strong \ia s (i.e.\ the fact that quarks are
permanently bound in hadrons and never manifest themselves as free \pc s,
unlike leptons) and the ``\cfn'' problem in thermonuclear fusion. We turn
only an attention of a reader that some ideas from strong \ia\ ``\cfn''
problem can be applied to a thermonuclear fusion problem. Moreover, the ideas
are really far away from our idea of dielectric model of \cfn\ (see Refs
\cite{p1}--\cite{p4}).

An important problem is to find an exact \so\ with axial \s y for the \NK\ with
fermion sources for $G=\SU(3)$. This could offer us a model of a hadron with
a \cfn\ condition ($\os D{}^a=0$, $\os E{}^a\ne0$). The axially \s ic, stationary
case seems to be very interesting from more general point of view. We have in
General Relativity very peculiar properties of stationary, axially \s ic \so s
of the Einstein--Maxwell \e s. These \so s describe the \gr al and \elm c fields
of a rotating charged mass. Thus we get the magnetic field component.
Asymptotically (these \so s are asymptotically flat) the magnetic field behaves
as a dipole field. We can calculate the gyromagnetic ratio at infinity, i.e.\
the ratio of the magnetic dipole moment and the angular momentum moment.
It is worth noticing that we get the anomalous gyromagnetic ratio, i.e.\
the gyromagnetic ratio for an electron (for a charged Dirac \pc). We cannot
interpret the Kerr--Newman \so\ as a model of a fermion for we have a
singularity. In the \NK\ we can expect completely nonsingular \so s. We can
also expect the asymptotic behavior of Einstein--Maxwell theory. Thus it seems that
we probably get the \so s with an anomalous gyromagnetic ratio. Such a \so\
could be treated as a (classical) model of $\frac12$-spin \pc.

In a \nA\ case (for $G=\SU(3)_c\tm \U(1)_{\rm em}$) the \so\ of field \e s
could offer us a model
of a charged barion (i.e.\ proton), where the skewon field $g_\[\m]$ induces
a \cfn\ of color. Such \so s should be considered also for a zero charge and
without and with fermion sources. Let us mention that fermion fields (quarks
fields) are coupled to the Riemannian part (a \LC \cn\ induced by $g_\(\a\b)$
metric) of the \cn\ $\gd \ov\o,\a,\b,$ on~$E$ (i.e.\ $\gd\wt{\ov \o},\a,\b,)$.

Let us come back to our presentation of the \NK.
One can easily calculate $\wt R(\wt \G)$ (see Appendix~A) getting
\beq2.54
\wt R(\wt \G)=-\frac14\,\ell^{ab}h_{ab}
\e
(it is a \co ical \ct). In the \NK\ (in the \nA\ case) we consider a special
nonholonomic frame. Moreover, we can consider a different frame which is
still nonholonomic, moreover, it looks more classical. Let us take a section
$e:E\to P$ and attach to it a frame $v^a$, $a=5,6,\dots,n+4$, selecting
$x^\mu={\rm const}$ on a fiber in such a way that $e$~is given by the
condition $e^*v^a=0$ and the \fn\ fields $\z_a$ \st $v^a(\z_b)=\gd \d,a,b,$
\sf y $[\z_a,\z_b]=\frac1\la\,\gd C,c,ab,\z_c$. Thus we have
\beq2.55
\o=\frac1\la\,v^aX_a + \pi^*(\gd A,a,\mu,\ov\t{}^\mu)X_a,
\e
where
\beq2.56
e^*\o=A=\gd A,a,\mu,\ov\t{}^\mu X_a.
\e
In this frame a tensor $\g$ takes a form
\beq2.57
\g_{AB}=\left(\begin{array}{c|c}
g_{\a\b}+\la^2\ell_{ab}\gd A,a,\a,\gd A,b,\b,& \la \ell_{cb}\gd A,c,\a,\\
\hline
\la\ell_{ac}\gd A,c,\b, & \ell_{ab}
\end{array}\right).
\e
This frame is also unholonomic.
One gets
\beq2.58
dv^a=-\frac1{2\la}\,\gd C,a,bc,v^b\land v^c.
\e
In this way a \nA\ gauge field four-\pt\ is a part of our theory. We present
here a model of a color \cfn. This model is a dielectric model of a \cfn. It
is a classical model of \cfn. We
know that a \cfn\ of a color is a nonperturbative effect. Moreover, our
theory is nonlinear and contains a gravity. Thus we should quantize the
theory using Ashtekar--Lewandowski method (see Ref.~\cite{17}) or using different
methods described in Conclusions, which we mentioned already above.

In our theory test \pc s move along geodesic \e s induced by Levi-Civita \cn\
induced by a \s ic part of a metric~$\g$, i.e.~$\g_\(AB)$. This \cn\ has a
form
\beq2.59
\gd \wt \o{},A,B,=\left(\begin{array}{c|c}
\pi^*(\gd\wt{\ov\o}{},\a,\b,)- h_{db}\wt g{}^\(\mu \a)\gd H,d,\mu\b,\t^b &
\gd H,a,\b\g,\t^\g \\
\hline
h_{bd}\wt g{}^\(\a\b)\gd H,d,\g\b,\t^\g & \gd \wt{\wt\o}{},a,b,
\end{array}\right)
\e
where $\gd \wt{\wt\o}{},a,b,$ means a \cn\ (Levi-Civita one) induced by a
Killing--Cartan metric on~$G$.

One can write a geodetic \e\ on $P$:
\beq2.60
u^A\wt\n_Au^B=0,
\e
where $u^B(\tau)$ is a tangent vector to geodetic and $\wt\n_A$ means a
covariant \dv\ \wrt a \LC \cn\ $\gd \wt\o{},A,B,$ (Eq.~\er{2.59}). One gets
\beq2.61
\bga
\frac{\wt{\ov D} u^\a}{d\tau} - 2u^bh_{ba}\wt g{}^\(\a\b)\gd H,a,\b\g,
u^\g=0\\
\frac{du^b}{d\tau}=0.
\ega
\e
This \e\ is written on $P$. We have a normalization of a
four-velocity~$u^\a$, $g_\(\a\b)u^\a u^\b=1$. The second \e\ gives us
a constancy of a color charge of a test \pc. We can identify
\beq2.62
q^b=2mu^b.
\e
Moreover, if we take a section $e:E\to P$ we get
\beq2.63
\bga
\frac{\wt{\ov D}u^\a}{d\tau}+\frac{Q^c}{m_0}\,u^\b g^{\a\d}h_{cd}\gd F,d,\b\d,=0\\
\frac{dQ^a}{d\tau}-\gd C,a,cb,Q^c \gd A,b,\nu,u^\nu=0
\ega
\e
where
\beq2.64
\bga
e^*\O=\frac12 \gd F,d,\b\d,\ov\t{}^\b \land \ov\t{}^\d X_d\\
e^*(q^bX_b)=Q^bX_b.
\ega
\e

Eq.\ \er{2.63} is called a Wong \e\ in the case of $G=\SU(2)$ (see
Ref.~\cite{00}). Moreover, for the first time the \e\ has been derived by
R.~Kerner (see Ref.~\cite{+0}) in general case (an arbitrary group~$G$). In
\E\nA\ Kaluza--Klein Theory
W.~Kopczy\'nski derived this \e\ on a principal bundle~$\ul P$ and
afterwards projected it on~$E$ (see Ref.~\cite{0+0}), i.e.\ in the
form~\er{2.61}. (This was of course Kaluza--Klein Theory with a \s ic metric.)

In our theory an action is given by
\beq2.65
S=\int_U d^{n+4}x\, R(W)\sqrt{\det \g_{AB}}
\e
where $U=V\times G$, $V\subset E$.

The Palatini \v al \pp\ adopted here is along the main theoretical stream.
Even more unconventional approach is advocated by J.~Pleba\'nski, where we vary
not only \wrt a metric and a \cn, but also \wrt a~\cvt. We do not apply the
mentioned formalism. The Palatini \v al \pp\ is really interesting if applied
to Kaluza--Klein Theory in a \nos\ version.

From the Palatini \v al \pp\ (\wrt $\gd \ov W{},\a,\b,$, $g_\m$, $\o$)
\beq2.66
0=\d S=\vol(G)\d\int_V d^4x \Bigl(\ov R(\ov W)+\ov R(\wt \G)
+\ell_{cd}(2H^cH^d - L^{c\m}\gd H,d,\m,)\Bigr)\sqrt{-g}
\e
one gets field \e s
\bg2.67
\ov R_{\a\b}(\ov W)-\frac12\,g_{\a\b}\ov R(\ov W)=
8\pi\gv{T_{\a\b}}+\La g_{\a\b}\\
\gd \falg{},\[\m],{,},=0 \lb2.68 \\
g_{\m,\si}-g_{\z\nu}\gd\ov\G{},\z,\mu\si, - g_{\mu\z}\gd \ov\G{},\z,\si\nu,=0
\lb2.69 \\
\gv{\n_\mu}\bigl(\ell_{ab}\fal L{}^{a\a\mu}\bigr)=
2g^\[\a\b]\gv{\n_\b}\bigl(h_{ab}\falg{}^\[\m]\gd H,a,\m,\bigr) \lb2.70
\e
where
\beq2.71
\gv {T_{\a\b}}=-\frac{\ell_{ab}}{4\pi}\Bigl(g_{\g\b}g^{\tau\z}g^{\ve\g}
\gd L,a,\z\a,\gd L,b,\tau\ve, - 2g^\[\m]\gd H,(a,\m,\gd H,b),\a\b,
-\frac14 \,g_{\a\b}\bigl(L^{a\m}\gd H,b,\m,-2H^aH^b\bigr)\Bigr).
\e
The skew-\s ic part of the metric induces a current
\beq2.72
\gd J,\a,a,=\frac1{2\pi} \falg{}^\[\a\b]\gv{\n_\b}(h_{ab}g^\[\m]\gd H,b,\m,).
\e
This current vanishes if
$$
g_\[\m]=0.
$$
One can easily see here that if $D^{a\bar a}=0$ we have zero color charge
distribution on the right-hand side of Eq.~\er{2.70}. Moreover, a color
charge is in general gauge-dependent.

We have here the same problem to define a \nA\ charge as to define an energy
in General Relativity. We cannot define an energy or a charge at a \spt\ point.
However, in \nA\ gauge field  theory the situation is even more severe.
According to Ref.~\cite{Q}, the most important difference between theories
of \YM' type and \gr\ is that the underlying bundle of the latter---the
bundle of linear frames---is ``concrete'', has more \sc e than ``abstract''
bundles occurring in other gauge theories.

A \fn\ difference between these two theories (in NGT there is the same
problem as in~GR) happens if we consider asymptotic behavior (at large
distances) of static fields. A~gauge \tf\ $U$ of $A\to\ov A$, $U:E\to G$,
\bea2.74n
\ov A_\mu(x)&=U^{-1}(x)A_\mu(x)U(x) + U^{-1}(x)\pa_\mu U(x)\\
\ov F_\m(x)&=U^{-1}(x)F_\m(x)U(x) \lb2.75n
\e
is compatible with a static $A$ iff
\beq2.76n
U(r,\psi,\vf)=U_0(\psi,\vf)\Bigl(1+\frac{\ov u(\psi,\vf)}r + \ldots\Bigr)
\e
($U$ does not depend on time), where $\psi$ and $\vf$ are defined as usual
on~$S^2$ ($r,\psi,\vf$ are spherical \cd s).
$U_0 : S^2\to G$, $\ov u(\psi,\vf)$ is a real \f. From Eqs \er{2.74n}--\er{2.76n}
one gets\break ($F_\m\ov\t{}^\mu \wedge \ov\t{}^\nu=e^*(H_\m\t^\mu\wedge \t^\nu)$,
$H_\m=\gd H,a,\m,X_a$)
\beq2.77n
\ov F_\m(r,\psi,\vf)=U_0^{-1}F_\m U_0 + O(r^{-3}).
\e
In the case of \gr al fields we have
\beq2.78n
g_\m=\eta_\m+O(r^{-1}),
\e
$\eta_\m$ is a Minkowski tensor, $g_\m$ is \nos,
\beq2.79n
\ov\G\simeq O(r^{-2}),
\e
$\ov\G$ is a $\gd \ov\o,\a,\b,$ \cn\ in static configurations.

One gets also for $G^{a\a\mu}$
\beq2.80n
\ov G{}^{\a\mu}=U_0^{-1}G^{\a\mu}U_0+O(r^{-2})
\e
where $G^{a\a\mu}=g^{\b\a}g^\m \gd G,a,\b\nu,$, $\gd G,a,\m,\ov \t{}^\mu
\wedge\ov\t{}^\nu= e^*(h^{\a d}\ell_{db}\gd\fal L,d,\a\mu,\t^\a \wedge\t^\mu)$
and $G^{\a\mu}=G^{a\a\mu}X_a$.

We define a \LC symbol and a dual Cartan basis
\bg2.81n
\ov\eta_{\a\b\g\d},\ \ov\eta_{1234}=\sqrt{-g}\\
\ov\eta_{\a\b\g}=\ov\t{}^\d \ov\eta_{\a\b\g\d} \lb2.82n \\
\ov\eta_{\a\b}=\tfrac12 \,\ov\t{}^\g \wedge \ov\eta_{\a\b\g} \lb2.83n \\
\ov\eta_{\a}=\tfrac13\, \ov\t{}^\b \wedge \ov\eta_{\a\b} \lb2.84n \\
\ov\eta = \tfrac14 \,\ov\t{}^\a \wedge \ov\eta_\a. \lb2.85n
\e
Eq. \er{2.70} can be rewritten after taking a section $e$
\beq2.86n
\pa_\mu(G^{a\a\mu})=4\pi J^{\a a}-\gd C,a,dc, A^c_\mu G^{d\a\mu}
\e
where we raise Latin indices by a \KC\ tensor $h^{ab}$.

We rewrite Eq.~\er{2.86n} using dual forms
\beq2.87n
\wh D\nads G=4\pi\nads J
\e
or in the Gauss form
\beq2.88n
d\nads G=4\pi\nads J - [A,\nads G]
\e
where $A=A_\mu \ov\t{}^\mu$, $\nads J=J^{\a a}X_a \ov\eta_\a$, $\star$ means
a Hodge star and $\nads J$ and~$\nads G$ mean dual forms for $J^\a$ and~$G^{\a\mu}$
$$
\nads G=G^{\a\mu}\ov\eta _{\a\b}.
$$
$\wh D$ means an exterior \ci\ \dv\ \wrt a gauge \cn~$\o$ on a fiber bundle $\ul P$
(in a section~$e$), $d$~is an ordinary exterior \dv\ of~$E$. $\ov\eta_\a,
\ov\eta_{\a\b}$ mean a dual Cartan base.

In this way a total \nA\ charge
\beq2.89n
\frac1{4\pi} \oint \nads G
\e
is ill-defined.

A \cv\ law for a \nA\ charge can be written
\beq2.90n
d\Bigl(\nads J-\frac1{4\pi}[A,\nads G]\Bigr)=0
\e
(see Ref.~\cite{Q}).

One gets
\beq2.91n
J^4 _a= \frac1{2\pi}\falg^\[4\bar b] \gv{\n_{\bar b}}(h_{ab}g^\[\m] \gd H,b,\m,)
\e
and
\beq2.92n
\pa_{\bar m}(\ell_{ab}D^{a\bar m}) + \gd C,d,be, \gd A,e,\bar m,
D^{a\bar m}\ell_{ab} = J^4_a.
\e
Our \cfn\ condition ($D^{a\bar m}=0$) is considered in a static (or stationary)
limit. In this way $J^4_a=0$ and a fact that a total \nA\ charge is ill-defined
does not concern us. In the case with some external sources, i.e.\ quark
fields (fermion colored fields) $J^4_a$ will compensate a charge caused by
fermions and a total color charge distribution remains zero. In this situation
we can develop a color \cfn\ program on the level of exact \so s with fermion
sources which we mentioned before.

In the case of \gr al field (see Ref.~\cite{Q}) an analogue of a \nA\ charge
\er{2.89n} is not ill-defined in static situation for an asymptotically flat
\spt\ (also in the case of \nos\ \gr\ field).

Let us consider a \tf\ of a \cn\ $\gd\ov\o,\a,\b,$, ($\gd\ov\G,\a,\b\g,$),
i.e.
\beq2.93n
\ov\o{}'=U^{-1}(x)\ov\o U(x)+U^{-1}\,dU
\e
where
\beq2.94n
\o=\gd\o,\a,\b, \gd X,\b,\a,
\e
where $\gd X,\b,\a,$ are generators of a $GL(4,R)$ group and $U(x)$ has
the form \er{2.76n}. Now of course the group~$G$ is $GL(4,R)$ group. In order
to have the asymptotic behavior \er{2.79n} in static configurations, $U_0$
must be a \ct\ Lorentz \tf, i.e.\ it belongs to $\SO(1,3)\subset GL(4,R)$. Introducing
pseudotensor of an energy-momentum for a \gr al field we can define
a conserved four-momentum from a \cv\ law
\beq2.95n
d(\nads T_\mu + \nads t_\mu)=0
\e
where
\beq2.96n
T_\mu = T_\m \ov\t{}^\nu, \q t_\mu=t_\m \ov\t{}^\nu,
\e
$\nads T_\mu$, $\nads t_\mu$ are dual Hodge forms to $T_\mu$ and $t_\mu$.

$T_\m$ is an energy-momentum tensor for a matter (\YM' field) which is \nos\
in general.
In some future extensions we include also an energy-momentum tensor for
fermion (quark) fields. $t_\m$ is a pseudotensor of an energy-momentum for
a \gr al field, defined in such a way that Eq.~\er{2.95n} is equivalent to
\gr al field \e s via Bianchi identity. In this way we can define super\pt s
$V_\mu$ \st
\beq2.97n
dV_\mu=4\pi (\nads T_\mu+\nads t_\mu)
\e
and a conserved 4-momentum
\beq2.98n
P_\mu = \frac1{4\pi} \oint V_\mu.
\e
$P_\mu$ is well defined in static situation (under condition \er{2.78n}
and~\er{2.79n}).

Using \er{2.36} one easily gets
\bml2.99n
\nads G=h^{ed}g^{\a\mu}g^{\nu\o}e^*\Bigl(\ell_{dn}\Bigl(
\gd H,n,\o\mu, - \mu h^{na}k_{ad}\gd H,d,\o\mu,\\
{}+\bigl(\gd H,n,\a\o,\wt g{}^\(\a\d) g_\[\d\mu] - \gd H,n,\a\mu,
\wt g{}^\(\a\d)g_\[\d\o]\bigr)
-2\mu h^{na}k_{ad}\wt g{}^\(\d\tau) \wt g{}^\(\a\b)\gd H,d,\d\a,g_\[\tau\o]g_\[\b\mu]\\
{}-2\mu h^{na}k_{ad}\wt g^\(\d\b)\wt g^\(\a\tau)\gd H,d,\b[\o,g_{\mu]\tau}g_\[\d\a]
+2\mu^2h^{na}h^{bc}k_{ac}k_{bd}\wt g{}^\(\a\b)\gd H,d,\a[\o,g_{|\mu|\b]}
\Bigr) \eta_{\m}X_e\Bigr),
\e
where $\eta_\m=\pi^*(\ov\eta_\m)$.

One also derives
\beq2.100n
V'_\mu=V_\nu U^\nu_{0\mu}+O(r^{-3})
\e
where $U_0=(U^\nu_{0\mu})\in\SO(1,3)$.

If we want to consider fermions (quarks) in the theory we should add
a Lagrangian of fermions (quarks)
\beq2.101n
\cL_{\rm fermions}=\sqrt{-\wt g}\sum_f (i\ov\psi_{jf}(\g^\mu D_\mu)_{jk}
-m_f\d_{jk})\psi_{kf}
\e
where
\beq2.102n
(D_\mu)_{jk}=\wt{\ov \n}_\mu \d_{jk}+\ov g \gd A,a,\mu,(X_a)_{jk}
\e
is a \ci\ \dv\ of a spinor field \wrt Riemannian part of a \cn\ $\gd\ov\o,\a,\b,$
($\gd\wt{\ov\o},\a,\b,$) generated by $g_\(\a\b)$ and a gauge field at once
($\wt g=\det(g_\(\a\b))$).
\beq2.103n
\wt{\ov\n}_\mu=\pa_\mu +\dg\si,\a,\b,\gd\wt{\ov \G},\a,\b\mu,, \q
\si_{\a\b}=\tfrac18[\g_\a,\g_\b],
\e
$\ov g$ is a coupling \ct, $m_f$ is the mass of a quark of flavour~$f$.
$X_a$ is a generator of the Lie algebra of a group~$G$ (equal to $\SU(3)$)
in a \fn\ \rp ation and repeated indices are summed over.

One derives a color current for fermions (quarks) getting
\beq2.104n
\gd J,q\mu,a,= \frac{i\ov g}{4\pi} \sqrt{-\wt g} \sum_f \ov\psi_{jf}\g^\mu
(X_a)_{jk}\psi_{kf}
\e
and a color charge distribution
\beq2.105n
\gd J,q4,a, = \frac{i\ov g}{4\pi} \sqrt{-\wt g} \sum_f \ov\psi_{jf}\g^4
(X_a)_{jk}\psi_{kf}.
\e
The color \cfn\ condition reads now
\beq2.106n
\gd J,q4,a, + \gd J,4,a,=0
\e
or
\beq2.107n
\frac{i\ov g}{2} \sqrt{-\wt g} \sum_f \ov\psi_{jf}\g^4(X_a)_{jk}\psi_{kf}
+\falg^\[4\bar b] e^*\Bigl(\gv{\n_{\bar b}}(h_{ab}g^\[\m] \gd H,a,\m,)\Bigr)=0.
\e
Spinor fields (quark fields) are defined in the same gauge $e$.
In this way we can write
\beq2.108n
e^*\Bigl(\frac{i\ov g}{2} \sqrt{-\wt g} \sum_f \ov\PS_{jf}\g^4(X_a)_{jk}\PS_{kf}
+\falg^\[4\bar b] \gv{\n_{\bar b}}(h_{ab}g^\[\m] \gd H,a,\m,)\Bigr)=0,
\e
i.e.\ in any ``gauge'', $\PS_{jf},\ov\PS_{jf}$ are spinor fields on $P$. This means that
\beq2.109n
\frac{i\ov g}{2} \sqrt{-\wt g} \sum_f \ov\PS_{jf}\g^4(X_a)_{jk}\PS_{kf}
+\falg^\[4\bar b] \gv{\n_{\bar b}}(h_{ab}g^\[\m] \gd H,a,\m,)=0.
\e
Eq.~\er{2.109n} means a dielectric color \cfn\ condition in a presence of
fermion (quark) sources.

It is interesting to express $\os E{}^a$ and $\os B{}^a$ fields in terms of
$\os D{}^a$ and $\os H{}^a$ fields. From Eq.~\er{2.21} one gets
\beq2.110n
\tfrac12(\ell^{ce}\ell_{dc}g_{\nu\o}g_{\mu\b} + \gd\d,e,d, g_{\b\mu}g_{\o\nu})L^{d\m}
=\gd H,e,\b\o,.
\e
From Eq. \er{2.110n} we obtain
\bg2.111
\gd E,e,\bar w,=\tfrac12 \Bigl(\gd\ve,\bar n\bar m,\bar l, \bigl(\ell^{ce}\ell_{dc}
g_{\bar n\bar w}g_{\bar m4} + \gd\d,e,d,g_{4\bar n}g_{\bar w\bar n}\bigr)H^{\bar ld}\hskip40pt \nn\\
\hskip40pt{}+\bigl(\ell^{ce}\ell_{dc}(g_{\bar m\bar w}g_{44} - g_{4\bar w}g_{\bar m4})
+\gd \d,e,d,(g_{44}g_{\bar w\bar m}-g_{4\bar m}g_{\bar w4})\bigr)D^{d\bar m}\Bigr)\lb2.111n \\
\gd B,e,\bar a, = \dg\ve,\bar a,\bar w\bar b, \Bigl[\gd \ve,\bar n\bar m,\bar l,
\bigl(\ell^{ce}\ell_{dc} g_{\bar n\bar w}g_{\bar m\bar b} +
\gd\d,e,d, g_{\bar b\bar m}g_{\bar w\bar n}\bigr)H^{\bar ld}\hskip40pt \nn\\
\hskip40pt{}+\bigl(\ell^{ce}\ell_{dc}(g_{\bar n\bar w}g_{4\bar b}-g_{4\bar w}g_{\bar n\bar b})
+\gd\d,e,d,(g_{\bar b4}g_{\bar w\bar n}-g_{\bar b\bar n}g_{\bar w4})\bigr)D^{\bar nd}\Bigr]
\lb2.112n
\e

It is interesting to ask how to construct a \NK\ with a group $G=G_0\otimes \U(1)_{\rm em}$
with an obvious application $G_0=\SU(3)_c$.

The simplest choice is to suppose that a \nos\ right-\iv t tensor on~$G$ has the form
\beq2.113n
\left(\begin{matrix} \ell_{ab}&\ &0 \\ 0&&-1
\end{matrix}\right)
\e
where $\ell_{ab}=h_{ab}+\mu k_{ab}$ is a \nos\ right-\iv t tensor on~$G_0$.

In this case we can consider also a fermion sources (quarks) adding to the
Lagrangian a Lagrangian of fermion (quark) fields. We considered such a situation
before. Moreover, now we should add also a coupling with an \elm c field, i.e.
\beq2.114n
iq\sqrt{-\wt g} \sum_f \ov\psi_{fj} \g^\mu \psi_{fj}A_\mu
\e
where $q$ is an \el ary charge, $q_f$ is a charge of a quark measured in~$q$,
$A_\mu$~is a four-\pt\ of an \elm c field and repeated indices are summed over.

In this way we get an electric current as a source of Maxwell \e s in our theory:
\beq2.115n
J^\a_{n+5} = \frac1{2\pi}\,\falg^\[\a\mu]\pa_\mu \bigl(g^\[\nu\b] F_{\nu\b}\bigr)
+\frac{q\sqrt{-\wt g}}{4\pi} \sum_fq_f \sum_j \ov\psi_{fj}\g^\mu \psi_{fj}
\e
a density of a charge is
\beq2.116n
J^4_{n+5} = \frac1{2\pi} \falg^\[4\bar m]\pa_{\bar m}(g^\[\m]F_{\nu\b}) +
\frac{q\sqrt{-\wt g}}{4\pi} \sum_f q_f \sum_j \ov\psi_{fj}\g^4\psi_{fj}.
\e

If we want a \cfn\ of a charge we should have $\os D=0$ (see Ref.~\cite8).
This means that even $\os E\ne0$, $J^4_{n+5}=0$, for $G=\SU(3)_c$ $n=8$.
Let us consider Eqs \er{2.110n}--\er{2.112n} for $G=\U(1)$, i.e.\ in an
\elm c case. Let us notice that in the formulas below Latin small cases
$a,b,c=1,2,3$ correspond to space indices as in Ref.~\cite8. One gets
\bg2.117n
\tfrac12(g_{\nu\o}g_{\mu\b}+g_{\b\mu}g_{\o\nu})H^\m = F_{\b\o}\\
E_w = \tfrac12\bigl(\gd\ve,nm,e, (g_{nw}g_{m4}+g_{4m}g_{wn})H^e
+(g_{nw}g_{44}-g_{4w}g_{m4}-g_{44}g_{wm}-g_{4m}g_{w4})D^n\bigr) \lb2.118n \\
B_a=\dg\ve,a,wb,\bigl(\gd\ve,nm,e, (g_{nw}g_{mb}+g_{bm}g_{wn})H^e
+(g_{nw}g_{4b}-g_{4w}g_{mb}+g_{b4}g_{wn}-g_{bn}g_{w4})D^n\bigr) \lb2.119n
\e

Thus we get
\bea2.120n
E_a&=\ov K_{ae}H^e + \ov L_{an}D^n\\
B_a&=K_{ae}H^e + L_{an}D^n \lb2.121n
\e
where
\bea2.122n
\ov K_{we}&=\tfrac12\,\gd \ve,nm,e, (g_{nw}g_{m4}+g_{4m}g_{wn}) \\
\ov L_{an}&=\tfrac12\,(g_{nw}g_{44}-g_{4w}g_{m4}-g_{44}g_{wm}-g_{4m}g_{w4}) \lb2.123n \\
K_{ae}&=\gd \ve,wb,a, \gd \ve,mn,e, (g_{nw}g_{mb}+g_{bm}g_{wm}) \lb2.124n \\
L_{an}&=\gd\ve,wb,a,(g_{nw}g_{4b}-g_{4w}g_{mb}-g_{b4}g_{wn}-g_{bn}g_{w4}) \lb2.125n
\e

Let us give the following remark on \cfn\ condition $\os D{}^a=0$. This condition
should be \sf ied outside a hadron, which in our model is a \so\ of field \e\ for the
\NK\ with or without fermion (quark) sources with $G=\SU(3)_c$ or $G=\SU(3)_c
\ot \U(1)_{\rm em}$. In this way we consider a soliton model of hadrons. Inside a
hadron, i.e.\ a \so\ of field \e\ this condition is not \sf ied. The \so\ should be
static (or stationary) with spherical \s y or axial \s y.

Eq.~\er{2.67} can be rewritten in a different shape
\beq2.73
\ov R_{\a\b}(\ov W)=8\pi\gv {T_{\a\b}} - \La g_{\a\b}.
\e
From Eq.~\er{2.73} one gets
\bea2.74
\ov R_\(\a\b)(\ov\G)&=8\pi\gv {T_\(\a\b)} - \La g_\(\a\b)\\
\ov R_{\tl[\a\b],\g\tp}(\ov\G)&=8\pi\gv {T_{\tl[\a\b],\g\tp}} -
\La g_{\tl[\a\b],\g\tp} \lb2.75
\e
We use of course a fact that a trace of $\gv {T_{\a\b}}$ is zero
\beq2.76
\gv {T_{\a\b}}g^{\a\b}=0.
\e
Eq.~\er{2.70} can be rewritten in the form
\beq2.77
\gv{\ov\n_\mu}\bigl(\ell_{ab}L^{a\a\mu}\bigr)  =
2g^\[\a\b]\gv{\ov \n_\b}\bigl(h_{ab}g^\[\m]\gd H,a,\m,\bigr),
\e
$\gv{\n_\mu}$ means a gauge \dv\ \wrt a \cn~$\o$, $\gv{\ov\n_\mu}$ means a
covariant \dv\ \wrt \cn~$\o$ (a~gauge \dv) and a \cn\ $\gd \ov\o,\a,\b,$
on~$E$ at once,
\beq2.78
\fal L^{a\m}=\sqrt{-g}\,L^{a\m}, \q
\falg^\[\m]=\sqrt{-g}\,g^\m.
\e

Let us consider a different approach to NGT coming from Einstein Unified
Field Theory. This is a so called Hermitian-\E\nos\ Theory (see Refs
\cite{15,*4}). In this theory we have a \fn\ tensor $g_\m$ as before.
Moreover, now this tensor is complex and Hermitian
\beq2.80
g_{\nu\mu}^\ast = g_\m.
\e
(In this case the star $*$ means the complex conjugation. Do not mix up
it with the Hodge star.)
In such a way one gets
\beq2.81
g_\m=g_\(\m)+g_\[\m]
\e
and $g_\[\m]$ is pure imaginary
\beq2.82
g_\[\m]=ip_\m
\e
where $p_\m$ is a real anti\s ic tensor
\beq2.83
p_\m=-p_{\nu\mu}.
\e
In the theory we have two \cn s as before (in the real version)
$\gd\ov\o,\a,\b,=\gd\ov\G,\a,\b\g,\ov\t{}^\g$ and $\gd\ov W,\a,\b, =
\gd\ov W,\a,\b\g,
\ov \t$. The first \cn\ is an Hermitian \cn\ (in holonomic system of \cd s)
\beq2.84
\gd\ov\G,*\a,\b\g, = \gd\ov\G,\a,\g\b,
\e
and
\beq2.85
\gd\ov\G,\a,\[\b\a],=0
\e
or
\beq2.85a
\gd\ov Q,\a,\b\a,(\ov\G)=0
\e
where $\gd\ov Q,\a,\b\g,(\ov\G)$ is a torsion of the \cn\ $\gd\ov\o,\a,\b,$.
The second \cn\ is not Hermitian
\beq2.86
\gd\ov W,\a,\b,=\gd\ov\o,\a,\b, - \frac23\,\gd\d,\a,\b,\ov W
\e
and the form $\ov W$ is pure imaginary.

The Ricci tensor is defined as before (\MR) tensor. This tensor is Hermitian.
The inverse tensor of $g_\m$, $g^\m$ is also Hermitian. It is easy to prove
that in a nonholonomic system of \cd s we have in place of Eq.~\er{2.84}
\beq2.87
\gd\ov\G,*\nu,\mu\o, = \gd\ov\G,\nu,\o\mu,+(\gd\wt{\ov\G},\nu,\mu\o,
-\gd\wt{\ov\G},\nu,\o\mu,)
\e
where $\gd\wt{\ov\G},\nu,\o\mu,$ is a \LC \cn\ generated by the \s ic part
of $g_\m$, $g_\(\m)$. The \cn\ $\gd\ov\o,\a,\b,$ \sf ies Eqs
\er{2.2}--\er{2.3}.

Now we construct a \NK\ exactly as before (see Refs \cite{3,4}). In the
5-\di al (\elm c) case we have in place of the \nos\ tensor $g_\m$ an
Hermitian tensor $g^*_\m=g_{\nu\mu}$. Thus we get \E\nos\ Hermitian
Kaluza--Klein Theory. It is an Hermitian metrization of a fiber bundle.
This is a natural Hermitian metrization of a fiber bundle (5-\di al case).

All the formulas are the same. Moreover, we should remember that $g_\[\m]$ is
pure imaginary. The \cn\ $\gd\o,A,B,$ on~$\ul P$ is Hermitian in holonomic
system of \cd s. Moreover, in our lift-horizontal basis it \sf ies a
different condition
\beq2.88
\gd\G,*N,MW, = \gd\G,N,MW, + (\gd\wt\G,N,MW,-\gd\wt\G,N,WM,)
\e
where $\gd\wt\o,N,M,=\gd\wt\G,N,MW,\t^W$ is a \LC \cn\ generated by
$\g_\(AB)$, where $N,W,M,A,B=1,2,3,4,5$. If a frame is holonomic, we get a
condition to be Hermitian from Eq.~\er{2.88}.

In the case of an exact \so\ from Ref.~\cite8 we have
\beq2.89
g_\m=\left(\begin{matrix}
-\a & 0   & 0 & \o \\
0   & -r^2 & 0 & 0 \\
0   & 0    &-r^2\sin^2\t & 0 \\
-\o & 0    & 0 & \g
\end{matrix}\right)
\e
where
$$
\o=\frac{i\ell^2}{r^2} \q\hbox{and}\q \g=\Bigl(1-\frac{\ell^4}{r^4}\Bigr)\a^{-1}
$$
and all the remaining formulae are the same as in Ref.~\cite8 (Eqs (4.1),
(4.3), (4.5), (4.6) of Ref.~\cite8). The energy-momentum tensor for an \elm c
field is also Hermitian
\beq2.90
\nad{\rm em}T{}^*_{\a\b}=\nad{\rm em}T_{\b\a}.
\e

Let us come to the \NK\ in a general \nA\ case. In this way we consider an
Hermitian \nos\ tensor on~$\ul P$ (as in this section) $\g_{AB}$. But now our
\ct~$\mu$ is a pure imaginary \ct, e.g.\ $\mu=i\ov\mu$, where $\ov\mu$ is a
real number. Our tensor is right \iv t \wrt the group action on~$\ul P$ and
Hermitian,
\beq2.91
\g^*_{AB}=\g_{BA}.
\e
In all the formulae derived here it is enough to put $i\ov\mu$ in place
of~$\mu$. The \cn\ on~$\ul P$, $\gd\o,A,B,$ \sf ies condition \er{2.88}, but now
$A,B=1,2,\dots,n+4$,
\bg2.92
\ell_{ab}=h_{ab}+\mu k_{ab}=h_{ab}+i\ov\mu k_{ab}, \\
\ell^*_{ab}=\ell_{ba} \lb2.93
\e
In this way we consider a natural Hermitian metrization of a fiber bundle
in a general \nA\ case (for a semisimple gauge group).
All energy-momentum tensors are now Hermitian, e.g.
\beq2.94
\gv{T_{\a\b}}{}^*= \gv{T_{\b\a}}.
\e

Let us give the following remark. A~lift horizontal basis in Kaluza--Klein
Theory is nonholonomic. For this a \LC \cn\ \cf s are not Christoffel symbols.
They are not \s ic in lower indices.

One can consider in place of a complex (Hermitian) metric also a hypercomplex
(Hermitian) metric (see Ref.~\cite{x}). Hypercomplex numbers (see
Ref.~\cite{xx}) are defined as
\beq2.96
x=x_1+Ix_2
\e
where $x_1,x_2$ are real numbers and
\beq2.97
I^2=\begin{cases}
-1 &\hbox{for complex numbers,}\\
+1 &\hbox{for \hc\ numbers,}\\
0  &\hbox{for dual or parabolic numbers.}
\end{cases}
\e
\E\hc\ numbers form a ring. They do not form a field. Addition and
multiplication are defined as usual taking into account the fact
that $I^2=1$ ($I\ne 1$). An inverse of a number does not always exist. This
ring contains also divisors of zero. One gets
\beq2.98
\bga
E_{1,2}=\tfrac12(I\pm 1), \q E^2_{1,2}=E_{1,2}\\
E_1\cdot E_2=0.
\ega
\e
In this way
\beq2.99
x=\wt x_1E_1 + \wt x_2E_2 = (x_1+x_2)E_1 + (x_2-x_1)E_2
\e
and the ring of \hc\ numbers is iso\mo c to a simple product of two copies
of real numbers field. In some sense it is a trivial structure in
comparison with the complex numbers field.

Thus if we take
\beq2.100
\wt g_\m = g_\(\m)+I g_\[\m]=E_1 g_\m-E_2g_{\nu\mu}
\e
where
\beq2.101
g_\m=g_\(\m)+g_\[\m]
\e
as in the real version of the theory, we get two disconnected real versions
of the theory for $g_\m$ and transpose $g_{\nu\mu}$. The \nos\ natural
metrization of a fiber bundle in Hermitian (\hc) can be done analogously to
the complex one in 5-\di al case and in general \nA\ (for a semisimple group)
case.

Moreover, all the calculations given by us in the case of a real version can
be repeated remembering that $g_\[\m]$ should be shifted to $Ig_\[\m]$ and
also $k_{ab} \to Ik_{ab}$. In this way we get \E\nos--Hermitian (Hypercomplex)
Kaluza--Klein Theory. Moreover, we can write also in the case of a tensor
$\g_{AB}$
\bg2.102
\wt \g_{AB} = \g_\(AB) + I\g_\[AB] = E_1\g_{AB} - E_2\g_{BA}\\
\g_{AB} = \g_\(AB) + \g_\[AB] \lb 2.103
\e
and we have as before in a 4-\di al case two disconnected real versions.
Thus in the case of Hermitian (Hypercomplex) Kaluza--Klein Theory we are
reduced to a real version. Moreover, from the methodological point of view
it is better to consider a Hermitian approach.

The solution \er{2.89} will now look
\bg2.104
\o=\frac{I\ell^2}{r^2} \qh{and} \g=\Bigl(1+\frac{\ell^4}{r^4}\Bigr)\a^{-1}.
\e
This \so\ can be written also in the form \er{2.101}--\er{2.102}, i.e.
\bml2.105
\wt g_\m =\left(\begin{matrix}
-\a & 0 & 0 & 0 \\
0 & -r^2 & 0 & 0 \\
0 & 0 & -r^2\sin^2\t & 0 \\
0 & 0 & 0 & \g
\end{matrix}\right) + J
\left(\begin{matrix}
0 & 0 & 0 & \frac{l^2}{r^2}\\
0 & 0 & 0 & 0 \\
0 & 0 & 0 & 0 \\
-\frac{l^2}{r^2} & 0 & 0 & 0
\end{matrix}\right)\\
{}=E_1 \left(\begin{matrix}
-\a & 0 & 0 & \frac{l^2}{r^2} \\
0 & -r^2 & 0 & 0 \\
0 & 0 & -r^2\sin^2\t & 0 \\
-\frac{l^2}{r^2} & 0 & 0 & \g
\end{matrix}\right) - E_2
\left(\begin{matrix}
-\a & 0 & 0 & -\frac{l^2}{r^2} \\
0 & -r^2 & 0 & 0 \\
0 & 0 & -r^2\sin^2\t & 0 \\
\frac{l^2}{r^2} & 0 & 0 & \g
\end{matrix}\right)
\e
where $\g$ is given by the second formula of Eq.~\er{2.104}.

\section{\E\sn\ \s y breaking and \Hm\ in the \NK}
In order to incorporate a \sn\ \s y breaking and \Hm\ in our geometrical \un\
of \gr\ and \YM' fields we consider a fiber bundle $\ul P$ over a base manifold
$E\times G/G_0$, where $E$ is a \spt, $G_0\subset G$, $G_0,G$ are semisimple Lie
groups. Thus we are going to combine a Kaluza--Klein theory with a \di al
reduction procedure.

Let $\ul P$ be a principal fiber bundle over $V=E\times M$ with a structural
group~$H$ and with a projection~$\pi$, where $M=G/G_0$ is a homogeneous
space, $G$~is a semisimple Lie group and $G_0$ its semisimple Lie subgroup.
Let us suppose that $(V,\g)$ is a manifold with a \nos\ metric tensor
\beq3.1
\g_{AB}=\g_\(AB)+\g_\[AB].
\e
The signature of the tensor $\g$ is ${(}+{-}-{-},
\underbrace{{}-{-}-\cdots-}_{n_1})$. Let us
introduce a natural frame on~$\ul P$
\beq3.2
\t^{\tilde A}=(\pi^*(\t^A),\t^0=\la\o^a), \q \la={\rm const.}
\e
It is convenient to introduce the following notation. Capital Latin indices
with tilde $\wt A,\wt B,\wt C$ run $1,2,3,\dots,m+4$, $m=\dim H+\dim M
=n+\dim M=n+n_1$, $n_1=\dim M$, $n=\dim H$. Lower Greek indices $\a,\b,\g,\d
=1,2,3,4$ and lower Latin indices
$a,b,c,d=n_1+5,n_2+5,\dots,\break n_1+6,\dots,m+4$. Capital Latin indices
$A,B,C=1,2,\dots,n_1+4$. Lower Latin indices with tilde $\wt a,\wt b,\wt c$
run $5,6,\dots,n_1+4$. The symbol over $\t^A$ and other quantities indicates
that these quantities are defined on~$V$. We have of course
$$
n_1=\dim G-\dim G_0=n_2-(n_2-n_1),
$$
where $\dim G=n_2$, $\dim G_0=n_2-n_1$, $m=n_1+n$.

On the group $H$ we define a bi-\iv t (\s ic) Killing--Cartan tensor
\beq3.3
h(A,B)=h_{ab}A^aB^b.
\e
We suppose $H$ is semisimple, it means $\det(h_{ab})\ne0$. We define a skew-\s
ic right-\iv t tensor on~$H$
$$
k(A,B)=k_{bc}A^bB^c, \q k_{bc}=-k_{cb}.
$$

Let us turn to the \nos\ metrization of~$\ul P$.
\beq3.4
\k(X,Y)=\g(X,Y)+\la^2\ell_{ab}\o^a(X)\o^b(Y)
\e
where
\beq3.5
\ell_{ab}=h_{ab}+\xi k_{ab}
\e
is a \nos\ right-\iv t tensor on~$H$. One gets in a matrix form (in the
natural frame \er{3.2})
\beq3.6
\k_{\tA\tB}=\left(\begin{array}{c|c}
\g_{AB}&0\\ \hline 0&\ell_{ab}
\end{array}\right),
\e
$\det(\ell_{ab})\ne0$, $\xi={\rm const}$ and real, then
\beq3.7
\ell_{ab}\ell^{ac}=\ell_{ba}\ell^{ca}=\gd\d,c,b,.
\e
The signature of the tensor $\k$ is $(+,-{-}-,\underbrace{-\cdots-}_{n_1},
\underbrace{{}-{-}\cdots-}_n)$.
As usual, we have commutation relations for Lie algebra of~$H$, $\fh$
\beq3.8
[X_a,X_b]=\gd C,c,ab,X_c.
\e
This metrization of $\ul P$ is right-\iv t \wrt an action of $H$ on $P$.

Now we should \nos ally metrize $M=G/G_0$. $M$~is a homogeneous space for~$G$
(with left action of group~$G$). Let us suppose that the Lie algebra of~$G$,
$\fg$ has the following reductive decomposition
\beq3.9
\fg=\fg_0 \mathrel{\dot+} \fm
\e
where $\fg_0$ is a Lie algebra of $G_0$ (a~subalgebra of~$\fg$) and $\fm$
(the complement to the subalgebra~$\fg_0$) is $\Ad G_0$ \iv t, $\dot+$
means a direct sum. Such a decomposition might be not unique, but we assume
that one has been chosen. Sometimes one assumes a stronger condition
for~$\fm$, the so called \s y requirement,
\beq3.10
[\fm,\fm]\subset \fg_0.
\e
Let us introduce the following notation for generators of $\fg$:
\beq3.11
Y_i\in\fg, \q Y_{\tilde\imath}\in\fm, \q Y_{\hat a}\in \fg_0.
\e
This is a de\cm\ of a basis of $\fg$ according to \er{3.9}. We define a \s ic
metric on~$M$ using a \KC form on~$G$ in a classical way. We call this tensor
$h_0$.

Let us define a tensor field $h^0(x)$ on $G/G_0$, $x\in G/G_0$, using tensor
field $h$ on~$G$. Moreover, if we suppose that $h$ is a bi\iv t metric on~$G$
(a~\KC tensor) we have a simpler construction.

The complement $\fm$ is a tangent space to the point $\{\ve G_0\}$ of~$M$,
$\ve$~is a unit element of~$G$. We restrict $h$ to the space $\fm$ only. Thus
we have $h^0(\{\ve G_0\})$ at one point of~$M$. Now we propagate
$h^0(\{fG_0\})$ using a left action of the group $G$
$$
h^0(\{fG_0\})=(L_f^{-1})^\ast (h^0(\{\ve G_0\})).
$$
$h^0(\{\ve G_0\})$ is of course $\Ad G_0$ \iv t tensor defined on~$\fm$
and $L_f^\ast h^0=h^0$.

We define on $M$ a skew-\s ic 2-form $k^0$. Now we introduce a natural frame
on~$M$. Let $\gd f,i,jk,$ be structure \ct s of the Lie algebra $\fg$, i.e.
\beq3.12
[Y_j,Y_k]=\gd f,i,jk, Y_i.
\e
$Y_j$ are generators of the \Li~$\fg$. Let us take a local section $\si:V\to
G/G_0$ of a natural bundle $G\mapsto G/G_0$ where $V\subset M=G/G_0$. The
local section $\si$ can be considered as an introduction of a \cd\ system
on~$M$.

Let $\o_{MC}$ be a left-\iv t Maurer--Cartan form and let
\beq3.13
\gd \o,\si,MC,=\si^\ast \o_{MC}.
\e
Using de\cm\ \er{3.9} we have
\beq3.14
\gd \o,\si,MC,=\gd \o,\si,0,+\gd \o,\si,\fm,=
\t^{\hi}Y_{\hi}+\ov t{}^{\td a}Y_{\td a}.
\e
It is easy to see that $\ov \t{}^{\td a}$ is the natural (left-\iv t) frame
on~$M$ and we have
\bea3.15
h^0&=\gd h,0,\td a\td b,\ov\t{}^{\td a}\otimes \ov\t{}^{\td b}\\
k^0&=\gd k,0,\td a\td b,\ov\t{}^{\td a}\land \ov\t{}^{\td b}. \lb3.16
\e
According to our notation $\wt a,\wt b=5,6,\dots,n_1+4$.

Thus we have a \nos\ metric on $M$
\beq3.17
\g_{\td a\td b}=r^2\bigl(\gd h,0,\td a\td b,+\z \gd k,0,\td a\td b,\bigr)
=r^2g_{\td a\td b}.
\e
Thus we are able to write down the \nos\ metric on $V=E\tm M=E\tm G/G_0$
\beq3.18
\g_{AB}=\left(\begin{array}{c|c}
g_{\a\b}&0\\
\hline
0 & r^2g_{\td a\td b}
\end{array}\right)
\e
where
\begin{align*}
g_{\a\b}&=g_\(\a\b)+g_\[\a\b]\\
g_{\td a\td b}&=\gd h,0,\td a\td b,+\z \gd k,0,\td a\td b,\\
\gd k,0,\td a\td b,&=-\gd k,0,\td b\td a,\\
\gd h,0,\td a\td b,&=\gd h,0,\td b\td a,,
\end{align*}
$\a,\b=1,2,3,4$, $\wt a,\wt b=5,6,\dots,n_1+4=\dim M+4=\dim G-\dim G_0+4$.
The frame $\ov\t{}^{\td a}$ is unholonomic:
\beq3.19
d\ov\t{}^{\td a}=\frac12\,\gd \k,\td a,\td b\td c, \ov\t{}^{\td b}\land
\ov \t{}^{\td c}
\e
where $\gd \k,\td a,\td b\td c,$ are \cf s of nonholonomicity and depend  on
the point of the manifold $M=G/G_0$ (they are not \ct\ in general). They
depend on the section~$\si$ and on the \ct s $\gd f,\td a,\td b\td c,$.

We have here three groups $H,G,G_0$. Let us suppose that there exists a
homomorphism $\mu$ between $G_0$ and~$H$,
\beq3.20
\mu:G_0 \to H
\e
\st a centralizer of $\mu(G_0)$ in $H$, $C^\mu$ is isomorphic to~$G$.
$C^\mu$, a centralizer of $\mu(G_0)$ in $H$, is a set of all \el s of~$H$
which commute with \el s of $\mu(G_0)$, which is a subgroup of~$H$. This
means that $H$~has the following structure, $C^\mu=G$.
\beq3.21
\mu(G_0)\otimes G\subset H.
\e
If $\mu$ is a iso\mo sm between $G_0$ and $\mu(G_0)$ one gets
\beq3.22
G_0\otimes G\subset H.
\e
Let us denote by $\mu'$ a tangent map to $\mu$ at a unit \el. Thus $\mu'$ is
a differential of~$\mu$ acting on the \Li\ \el s. Let us suppose that the
\cn~$\o$ on the fiber bundle $\ul P$ is \iv t under group action of~$G$ on the
manifold $V=E\tm G/G_0$. According to Refs \cite{13,18,19,20} this means the
following.

Let $e$ be a local section of $\ul P$, $e:V\subset U\to P$ and $A=e^*\o$. Then
for every $g\in G$ there exists a gauge \tf\ $\rho_g$ \st
\beq3.23
f^*(g)A=\Ad_{\rho_g^{-1}}A+\rho_g^{-1}\,dg_g,
\e
$f^*$ means a pull-back of the action $f$ of the group $G$ on the
manifold~$V$. According to Refs \cite{18,19,20,21,22,23}
(see also Refs \cite{**,*1,*2}) we are able to write
a general form for such an~$\o$. Following Ref.~\cite{20} we have
\beq3.24
\o=\wt \o_E +\mu'\circ \gd \o,\si,0,+\F\circ \gd \o,\si,\fm,.
\e
(An action of a group $G$ on $V=E\tm G/G_0$ means left multiplication on a
homogeneous space $M=G/G_0$.)
where $\gd \o,\si,0,+\gd \o,\si,\fm,=\gd \o,\si,MC,$ are components of the
pull-back of the Maurer--Cartan form from the de\cm~\er{3.14}, $\wt\o_E$ is a
\cn\ defined on a fiber bundle $Q$ over a \spt~$E$ with structural
group~$C^\mu$ and a projection~$\pi_E$. Moreover, $C^\mu=G$ and  $\wt\o_E$ is
a 1-form with values in the \Li~$\fg$. This \cn\ describes an ordinary \YM'
field gauge group $G=C^\mu$ on the \spt~$E$. $\F$~is a \f\ on~$E$ with values
in the space $\wt S$ of linear maps
\beq3.25
\F:\fm \to \fh
\e
\sf ying
\beq3.26
\F([X_0,X])=[\mu'X_0,\F(X)], \q X_0\in\fg_0.
\e
Thus
\beq3.27
\bal
\wt\o_E=\gd\wt\o,i,E,Y_i, &\q Y_i\in\fg,\\
\gd\o,\si,0,=\t^{\hi}Y_{\hi}, &\q Y_{\hi}\in\fg_0,\\
\gd\o,\si,\fm,=\ov\t{}^{\td a}Y_{\td a}, &\q Y_{\td a}\in\fm.
\eal
\e

Let us write condition \er{3.24} in the base of left-\iv t form
$\t^{\hi},\ov\t{}^{\td a}$, which span respectively dual spaces
to~$\fg_0$ and~$\fm$ (see Refs \cite{24,25}). It is easy to see that
\beq3.28
\F\circ \gd\o,\si,\fm,=\gd \F,a,\td a,(x)\ov\t{}^{\td a}X_a, \q X_a\in\fh
\e
and
\beq3.29
\mu'=\gd\mu,a,\hi, \t^{\hi}X_a.
\e
From \er{3.26} one gets
\beq3.30
\F_{\td b}^c (x)f_{\hi\td a}^{\td b}=
\mu^a_{\hi}\F^b_{\td a}(x)\gd C,c,ab,
\e
where $f^{\td b}_{\hi\td a}$ are structure \ct s of the \Li~$\fg$ and
$\gd C,c,ab,$ are structure \ct s of the \Li~$\fh$. Eq.~\er{3.30} is a
constraint on the scalar field $\F^a_{\td a}(x)$. For a curvature of~$\o$ one
gets
\bml3.30a
\O=\frac12\,\gd H,C,AB,\t^A\land \t^BX_C=
\frac12\,\gd\wt H,i,\m,\t^\mu\land\t^\nu \a^c_iX_c
+\gv{\n_\mu} \F^c_{\td a}\t^\mu\land \t^{\td a}X_c\\
{}+\frac12\,\gd C,c,ab,\F^a_{\td a}\F^b_{\td b}\t^{\td a}\land \t^{\td b}
X_c - \frac12\,\F^c_{\td d}f^{\td d}_{\td a\td b}\t^{\td a}\land\t^{\td b}X_c.
\e
Thus we have
\bg3.31
\gd H,c,\m,=\gd \a,c,i,\gd \wt H,i,\m,\\
\gd H,c,\mu\td a,=\gv{\n_\mu} \F^c_{\td a}=-\gd H,c,\td a\mu, \lb3.32 \\
\gd H,c,\td a\td b,=\gd C,c,ab,\cdot\F^a_{\td a}\F^b_{\td b} - \mu^c_{\hi}
f^{\hi}_{\td a\td b} - \F^c_{\td d}\gd f,\td d,\td a\td b, \lb3.33
\e
where $\gv{\n_\mu}$ means gauge \dv\ \wrt the \cn\ $\wt\o_E$ defined on a
bundle~$Q$ over a \spt~$E$ with a \sc al group~$G$
\beq3.34
Y_i=\a^c_i X_c.
\e
$\gd \wt H,i,\m,$ is the curvature of the \cn\ $\ov\o_E$ in the base
$\{Y_i\}$, generators of the \Li\ of the Lie group~$G$, $\fg$, $\a^c_i$ is
the matrix which connects $\{Y_i\}$ with $\{X_c\}$. Now we would like to
remind that indices $a,b,c$ refer to the \Li~$\fh$, $\wt a,\wt b,\wt c$ to
the space~$\fm$ (tangent space to~$M$), $\wh \imath,\wh \jmath,\wh k$ to the
\Li~$\fg_0$ and $i,j,k$ to the \Li\ of the group $G$, $\fg$. The matrix
$\a^c_i$ establishes a direct relation between generators of the \Li\ of the
subgroup of the group~$H$ iso\mo c to the group~$G$.

Let us come back to a construction of the \NK\ on a manifold~$P$. We should
define \cn s. First of all, we should define a \cn\ compatible with a \nos\
tensor $\g_{AB}$, Eq.~\er{3.18},
\bg3.35
\gd \ov\o,A,B,=\gd \ov\G,A,BC,\t^C\\
\ov D\g_{AB}=\g_{AD}\gd\ov Q,D,BC,(\ov\G)\t^C \lb3.36 \\
\gd \ov Q,D,BD,(\ov \G)=0 \nn
\e
where $\ov D$ is the exterior covariant \dv\ \wrt $\gd \ov\o,A,B,$ and $\gd
\ov Q,D,BC,(\ov\G)$ its torsion.

Using \er{3.18} one easily finds that the \cn\ \er{3.35} has the following
shape
\beq3.37
\gd \ov\o,A,B,=\left(\begin{array}{c|c}
\pi^*_E(\gd \ov\o,\a,\b,) & 0 \\ \hline
0 & \gd \wh{\bar\o},\td a,\td b,
\end{array}\right)
\e
where $\gd \ov\o,\a,\b,=\gd \ov \G,\a,\b\g,\ov\t{}^\g$ is a \cn\ on the
\spt~$E$ and $\gd\wh{\ov \o},\td a,\td b,=\gd \wh{\ov\G},\td a,\td b\td c,
\ov \t{}^{\td c}$ on the manifold $M=G/G_0$ with the following properties
\bg3.38
\ov Dg_{\a\b}=g_{\a\d}\gd \ov Q,\d,\b\g,(\ov \G)\ov \t{}^\g=0\\
\gd \ov Q,\a,\b\a,(\ov\G)=0 \lb3.39 \\
\wh{\ov D}g_{\td a\td b}=g_{\td a\td d}\gd\wh{\ov Q},\td d,\td b\td c,(\wh{\ov\G}).
\lb3.40 \\
\gd\wh{\ov Q},\td d,\td b\td d,(\wh{\ov\G})=0 \nonumber
\e

$\ov D$ is an exterior covariant \dv\ \wrt a \cn~$\gd \ov\o,\a,\b,$. $\gd \ov
Q,\a,\b\g,$ is a tensor of torsion of a \cn~$\gd \ov\o,\a,\b,$. $\wh{\ov
D}$~is an exterior covariant \dv\ of a \cn~$\gd \wh{\ov\o},\td a,\td b,$ and
$\gd \wh{\ov Q},\td a,\td b\td c,(\wh{\ov\G})$ its torsion.

On a \spt\ $E$ we also define the second affine \cn\ $\gd \ov W,\a,\b,$ \st
\beq3.41
\gd \ov W,\a,\b,= \gd \ov\o,\a,\b, - \frac23\,\gd \d,\a,\b,\ov W,
\e
where
$$
\ov W=\ov W_\g \ov\t{}^\g = \tfrac12(\gd \ov W,\si,\g\si,-\gd \ov W,\si,\g\si,).
$$
We proceed a \nos\ metrization of a principal fiber bundle $\ul P$ according
to \er{3.18}. Thus we define a right-\iv t \cn\ \wrt an action of the
group~$H$ compatible with a tensor $\k_{\tA\tB}$
\bg3.42
D\k_{\tA\tB}=\k_{\tA\td D}\gd Q,\td D,\tB\tC,(\G)\t^{\tC}\\
\gd Q,\td D,\td B\td D,(\G)=0 \nn
\e
where $\gd \o,\tA,\tB,=\gd \G,\tA,\tB\tC,\wt \t{}^\tC$. $D$ is an exterior
covariant \dv\ \wrt the \cn\ $\gd \o,\tA,\tB,$ and $\gd Q,\tA,\tB\tC,$ its
torsion. After some calculations one finds
\beq3.43
\gd \o,\tA,\tB,= \bma
\pi^*(\gd \ov\o,A,B,)-\ell_{db}\g^{MA}\gd L,d,MB,\t^b & \gd L,a,BC,\t^C \\
\hline
\ell_{bd}\g^{AB}(2\gd H,d,CB,-\gd L,d,CB,)\t^C & \gd \wt\o,a,b,
\ema
\e
where
\bg3.44
\gd L,d,MB,=-\gd L,d,BM,\\
\ell_{dc}\g_{MB}\g^{CM}\gd L,d,CA, + \ell_{cd}\g_{AM}\g^{MC}\gd L,d,BC,
=2\ell_{cd}\g_{AM}\g^{MC}\gd H,d,BC,, \lb3.45
\e
$\gd L,d,CA,$ is Ad-type tensor \wrt $H$ (Ad-\ci\ on~$\ul P$)
\bg3.46
\gd \wt\o,a,b,=\gd \wt\G,a,bc,\t^c\\
\ell_{db}\gd \wt\G,d,ac,+\ell_{ad}\gd \wt\G,d,cb,=-\ell_{db}\gd C,d,ac, \lb3.47 \\
\gd \wt\G,d,ac,=-\gd \wt\G,d,ca,, \q \gd \wt\G,d,ad,=0. \lb3.48
\e
We define on $\ul P$ a second \cn
\beq3.49
\gd W,\tA,\tB, = \gd \o,\tA,\tB, - \frac4{3(m+2)}\,\gd \d,\tA,\tB,\ov W.
\e
Thus we have on $P$ all $(m+4)$-\di al analogues of geometrical quantities from
NGT, i.e.
$$
\gd W,\tA,\tB,, \q \gd \o,\tA,\tB, \qh{and} \k_{\tA\tB}.
$$

Let us calculate a Moffat--Ricci curvature scalar for the \cn\ $\gd W,\tA,\tB,$
\beq3.50
R(W)=\k^{\tA\tB}\bigr(\gd R,\tC,\tA\tB\tC,(W)+\tfrac12 \gd R,\tC,\tC\tA\tB,
(W)\bigr)
\e
where $\gd R,\tC,\tC\tA\tB,(W)$ is a curvature tensor for a \cn\ $\gd
W,\tA,\tB,$ and $\k^{\tA\tB}$ is an inverse tensor for $\k_{\tA\tB}$
\beq3.51
\k^{\tA\tC}\k_{\tA\tB} = \k^{\tC\tA}\k_{\tB\tA}=\gd \d,\tC,\tB,.
\e
Using results from Ref.~\cite1 one gets (having in mind some analogies from a
theory with a base space $E$ to the theory with the base space
$V=E\tm M=E\tm G/G_0$)
\beq3.52
R(W)=\ov R(\ov W)+\frac1{r^2}\,R(\wh{\ov \G})+\frac1{\la^2}\,
\wt R(\wt\G) - \frac{\la^2}4\,\ell_{ab}\bigl(2H^aH^b - L^{aMN}\gd H,b,MN,\bigr)
\e
where $\ov R(\ov W)$ is a \MR\ \cvt\ scalar on the \spt~$E$ for a \cn~$\gd\ov
W,\a,\b,$, $R(\wh{\ov \G})$ is a \MR\ \cvt\ scalar for a \cn~$\gd
\wh{\ov\o},\td a,\td b,$ on a homogeneous space $M=G/G_0$, $\wt R(\wt\G)$
is a \MR\ \cvt\ scalar for a \cn~$\gd\wt\o,a,b,$,
\beq3.53
H^a=\g^\[AB]\gd H,a,\[AB],=g^\[\a\b]\gd H,a,\a\b,+\frac1{r^2}\,
g^\[\td a\td b]\gd H,a,\td a\td b,
\e
\bml3.54
L^{aMN}=\g^{AM}\g^{BN}\gd L,a,AB,=\gd \d,M,\mu,\gd \d,N,\g,g^{\a\mu}
g^{\b\g}\gd L,a,\a\b,\\
{}+ \frac1{r^2}\bigl(g^{\a\mu}g^{\td b\td n}\gd L,a,\a\td b,
+g^{\td a\td n}g^{\b\g}\gd L,a,\td a\b,\bigr)\gd \d,M,\mu,\gd \d,N,\td n,
+\frac1{r^4}\,g^{\td a\td m}g^{\td b\td n}\gd L,a,\td a\td b,\gd \d,M,\td m,
\gd \d,N,\td n,.
\e
One finds that
\bml3.55
-\ell_{ab}L^{aMN}\gd H,b,MN,=-\ell_{ab}\Bigr(g^{\a\mu}g^{\b\nu}\gd L,a,\a\b,
\gd H,b,\m,+\frac2{r^2}\,g^{\a\mu}g^{\td b\td n}\gd L,a,\a\td b,\gd
H,b,\mu\td n,+\frac1{r^4}\,g^{\td a\td m}g^{\td b\td n}\gd L,a,\td a\td b,
\gd H,b,\td m\td n,\Bigr)\\
{}=-\ell_{ab}\Bigl(L^{a\m}\gd H,b,\m,+\frac2{r^2}\,g^{\td b\td n}
\gd L,\a\mu,\td b,\gd H,b,\mu\td n,+\frac1{r^4}\,g^{\td a\td m}g^{\td b\td n}
\gd L,a,\td a\td b,\gd H,b,\td m\td n,\Bigr).
\e
We get conditions from Eq.~\er{3.45}
\bea3.56
\ell_{dc}g_{\mu\b}g^{\g\mu}\gd L,d,\g\a,
+ \ell_{cd}g_{\a\mu}g^{\mu\g}\gd L,d,\b\g,
&= 2\ell_{cd}g_{\a\mu}g^{\mu\g}\gd H,d,\b\g, \\
\ell_{dc}g_{\td m\td b}g^{\td c\td m}\gd L,d,\td c\td a,
+ \ell_{cd}g_{\td a\td m}g^{\td m\td c}\gd L,d,\td b\td c,
&= 2\ell_{cd}g_{\td a\td m}g^{\td m\td c}\gd H,d,\td b\td c, \lb3.57 \\
\ell_{dc}g_{\mu\b}g^{\g\mu}\gd L,d,\g\td a,
+ \ell_{cd}g_{\td a\td m}g^{\td m\td c}\gd L,d,\b\td c,
&= 2\ell_{cd}g_{\td a\td m}g^{\td m\td c}\gd H,d,\b\td c, \lb3.58 \\
L^{a\m}&=g^{\a\mu}g^{\b\nu}\gd L,a,\a\b, \lb3.59 \\
\gd L,a\mu,\td b,&=g^{\a\mu}\gd L,a,\mu\td b,. \lb3.60
\e

For $\ell_{ab}H^aH^b=h_{ab}H^aH^b$ we have the following:
\beq3.61
h_{ab}H^aH^b = h_{ab}\gd H,a,0,\gd H,b,0, + \frac2{r^2}\,h_{ab}
\gd H,a,0,\gd H,b,1, + \frac1{r^4}\,h_{ab}\gd H,a,1,\gd H,b,1,
\e
where
\beq3.62
\gd H,a,0,=g^{\a\b}\gd H,a,\a\b,, \q \gd H,a,1,=g^\[\td a\td b]\gd H,a,\td
a\td b,.
\e

Finally, we have for a density of $R(W)$, i.e.
\bml3.63
\sqrt{|\k|}\,R(W) = \sqrt{-g}\,r^{n_1} \sqrt{|\wt g|}\,\sqrt{|\ell|}\,R(W)\\
{}=\sqrt{-g}\,r^{n_1} \sqrt{|\wt g|}\,\sqrt{|\ell|}\,
\biggl(\ov R(\ov W)+\frac{\wt R(\wt \G)}{\la^2} + \frac1{r^2}\,R(\wh{\ov\G})
+\frac{\la^2}4 \, \ell_{ab}\bigl(2\gd H,a,0,\gd H,b,0, - L^{a\m}
\gd H,b,\m,\bigr)\\ {}+ \frac{\la^2}{4r^2} \, \ell_{ab}\bigl(4\gd H,(a,0,
\gd H,b),1, - 2g^{\td b\td n}\gd L,a\mu,\td b,\gd H,b,\mu\td n,\bigr)
+\frac{\la^2}{4r^2}\,\ell_{ab}\bigl(2\gd H,a,1,\gd H,b,1, - g^{\td a\td m}
g^{\td b\td n}\gd L,a,\td a\td b,\gd H,b,\td m\td n,\bigr)\biggr).
\e

We define an integral of action
\beq3.64
S \sim \int_U \sqrt{|\k|}\,R(W)\,d^{m+4}x,
\e
where
$$
U=M \tm G \tm V, \q V\subset E, \q d^{m+4}x=d^4x\,d\mu_H(h)\,dm(y),
$$
$d\mu_H(h)$ is a bi\iv t measure on a group $H$ and $dm(y)$ is a measure
on~$M$ induced by a bi\iv t measure on~$G$. $\ov R(\ov W)$ is a \MR\ \cvt\
scalar for a \cn~$\gd\ov W,\a,\b,$ on~$E$.

Let us consider Eqs \er{3.56}--\er{3.58} modulo \e s \er{3.31}--\er{3.33}.
One gets
\beq3.65
\ell_{ij}g_{\mu\b}g^{\g\mu}\gd \wt L,i,\g\a, + \ell_{ji}g_{\a\mu}g^{\mu\g}
\gd \wt L,i,\b\g, = 2\ell_{ji}g_{\a\mu}g^{\mu\g}\gd \wt H,i,\b\g,
\e
where $\ell_{ij}=\ell_{cd}\gd\a,c,i,\gd \a,d,j,$ is a right-\iv t \nos\
metric on the group~$G$ and
\beq3.66
\gd L,c,\m,=\gd \a,c,i, \gd \wt L,i,\m,.
\e
$\gd \wt L,i,\m,$ plays a role of an induction tensor for the \YM' field with
the gauge group~$G$. $\gd \wt H,i,\m,$~is of course the tensor of strength of
this field. The polarization tensor is defined as usual
\beq3.67
\gd \wt L,i,\m,=\gd \wt H,i,\m, - 4\pi \gd \wt M,i,\m,.
\e
We introduce two $\Ad_G$-type 2-forms with values in the \Li~$\fg$ (of~$G$)
\bea3.68
\wt L&=\tfrac12\,\gd \wt L,i,\m,\t^\mu \land \t^\nu Y_i\\
\wt M&=\tfrac12\,\gd \wt M,i,\m,\t^\mu \land \t^\nu Y_i\lb3.69
\e
and we easily write
\beq3.70
\wt L=\wt\O_E - 4\pi \wt M=\wt\O_E - \tfrac12\,Q
\e
where $\wt Q=\frac12\,\gd \wt Q,i,\m,\t^\mu\land \t^\nu Y_i$, $\gd \wt Q,i,\m,
=\gd \a,i,c, \gd Q,c,\m,$.
$\wt\O_E$ is a 2-form of a \cvt\ of a \cn\ $\wt\o_E$ (Eq.~\er{3.27}) in
Eq.~\er{3.30a} (the first term of this \e).

In this way we get a geometrical interpretation of a \YM' induction tensor in
terms of the \cvt\ tensor and torsion in additional \di s (see Refs
\cite{1,3}). Afterwards we get
\bg3.71
\ell_{cd}g_{\td m\td b}g^{\td c\td m}\gd L,d,\td c\td a,
+\ell_{cd}g_{\td a\td m}g^{\td m \td c}\gd L,d,\td b\td c,
=2\ell_{cd}g_{\td a\td m}g^{\td m\td c}\bigl(\gd C,d,ab,\F^a_{\td b}
\F^b_{\td b} - \gd \mu,d,\hi, \gd f,\hi,\td b\td c, - \F^d_{\td d}
\gd f,\td d,\td b\td c,\bigr),\\
\ell_{cd}g_{\mu\b}g^{\g\mu}\gd L,d,\g\td a,+\ell_{cd}g_{\td a\td m}
g^{\td m\td c}\gd L,d,\b\td c,
=2\ell_{cd}g_{\td a\td m}g^{\td m\td c}\gv{\n_\b}\F^d_{\td c}. \lb3.72
\e

Let us rewrite an action integral
\bg3.73
S=-\frac1{V_1V_2r^{n_1}} \int_U \bigl(R(W)\,d^nx\bigr)\,d^{n_1}x\,d^4x,
\q U=V\tm M\tm H, \q V\subset E,\\
V_1=\int_H \sqrt{|\ell|}\,d^nx \lb3.74 \\
V_2=\int_M \sqrt{|\wt g|}\,d^{n_1}x. \lb3.75
\e
Thus we get
\beq3.76
S=-\int_V \sqrt{-g}\,d^4x \,\cL(\ov W,g,\wt A,\F)
\e
where
\bg aa
\cL(\ov W,g,\wt A,\F) \hskip320pt \nn \\
\hskip20pt{}= \ov R(\ov W) + \frac{\la^2}4 \Bigr(
8\pi \cLY(\wt A)+\frac2{r^2}\,\cL_{\rm kin}(\gv\n \F) + \frac1{r^4}\,V(\F)
-\frac4{r^2}\,\cL_{\rm int}(\F,\wt A)\Bigr)+\la_c \lb3.77 \\
\cLY(\wt A)=-\frac1{8\pi}\,\ell_{ij}\bigl(2\wt H{}^i \wt H{}^j - L^{i\m}
\gd \wt H,j,\m,\bigr) \lb3.78
\e
is the lagrangian for the \YM' field with the gauge group~$G$ (see Eqs
\er{2.32} and~\er{2.38}),
\bml3.79
\cL_{\rm kin}(\gv\n \F)=\frac1{V_2}\int_M \sqrt{|\wt g|}\,d^{n_1}x
\bigl(\ell_{ab}g^{\td b\td n}\gd L,a\mu,\td b,\gv{\n_\mu} \F^b_{\td n}\bigr)\\
{}=\ell_{ab}g^{\a\mu}\,\frac1{V_2}\int_M \sqrt{|\wt g|}\,d^{n_1}x
\bigl(g^{\td b\td n}\gd L,a,\a\td b,\gv{\n_\mu} \F^b_{\td n}\bigr)
\e
is a kinetic part of a lagrangian for a scalar field $\F^a_{\td a}$. It is
quadratic in gauge \dv\ of $\F^a_{\td a}$ and is \iv t \wrt the action of
groups $H$ and~$G$.
\bml3.80
V(\F)=\frac{\ell_{ab}}{V_2} \int_M \sqrt{|\wt g|}\,d^{n_1}x
\Bigl[ 2g^\[\td m\td n]\bigl(\gd C,a,cd,\F^c_{\td m}\F^d_{\td n} -
\gd\mu,a,\hi,\gd f,\hi,\td m\td n,- \F^a_{\td e}\gd f,\td e,\td m\td n,\bigr)\\
{}g^\[\td a\td b]\bigl(\gd C,b,ef,\F^e_{\td a}\F^f_{\td b}
-\gd \mu,b,\wh\jmath,\gd f,\wh\jmath,\td a\td b,
-\Ft ba\gd f,\td d,\td a\td b,\bigr)
-g^{\td a\td m}g^{\td b\td n}\gd L,a,\td a\td b,
\bigl(\gd C,b,cd,\F^c_{\td m}\F^d_{\td n}-\gd \mu,b,\hi,\gd f,\hi,\td m\td n,
-\F^b_{\td e}\gd f,\td e,\td m\td n,\bigr)\Bigr]
\e
is a self-interacting term for a field $\F$. It is \iv t \wrt the action of
the groups $H$ and~$G$. This term is a polynomial of fourth order in $\F$'s
(a~Higgs' field \pt\ term)
\beq3.81
\cL_{\rm int}(\F,\wt A)=h_{ab}\gd \mu,a,i,\wt H{}^i \ul g^\[\td a\td b]
\bigl(\gd C,b,cd,\F^c_{\wt a} \F^d_{\td b} - \gd \mu,b,\hi,\gd f,\hi,\td a\td
b,-\F^b_{\td d}\gd f,\td d,\td a\td b,\bigr)
\e
where
\beq3.82
\ul g^\[\td a\td b]=\frac1{V_2}\int_M \sqrt{|\wt g|}\,d^{n_1}x \,g^\[\td a\td
b]
\e
is the term describing non-minimal coupling between the scalar field $F$ and
the \YM' field. This term is also \iv t \wrt the action of the groups $H$
and~$G$.
\beq3.83
\la_c = \frac1{\la^2}\,\wt R(\wt \G)+\frac1{r^2V_2} \int_M \sqrt{|\wt g|}\,
\wh{\ov R}(\wh{\ov \G})\,d^{n_1}x = \frac1{\la^2}\,\wt R(\wt \G)+ \frac1{r^2}\,\ul{\wt P}.
\e

The condition \er{3.72} can be explicitly solved (see Appendix A). One gets
\bml3.84
\gd L,n,\o\td m,=\gv{\n_\o}\F^n_{\td m}+ \xi \gd k,n,d,\gv{\n_\o}
\F^d_{\td m} - \bigl(\z\gv{\n_\o}\F^n_{\td a}h^{0\td a\td d}k_{0\td d\td m}
+\wt g{}^\(\a\mu)\gv{\n_\a}\F^n_{\td m}g_\[\mu\o]\bigr)\\
{}-2\xi\z\gd k,n,d,\gv{\n_\o}\F^d_{\td d}\wt g{}^\(\d\a)g_\[\a\o]
h^{0\td d\td a}\gd k,0,\td a\td m,+\xi\gd k,n,d,\bigl(\z^2h^{\td d\td a}
\gv{\n_\o}\F^d_{\td a}\gd k,0,\td d\td b,\gd k,0,\td m\td c,h^{0\td c\td b}\\
{}+\gv{\n_\b}\F^d_{\td m}\wt g{}^\(\d\b)g_\[\d\a]g_\[\o\mu]\wt g{}^\(\a\mu)
\bigr)
-\xi^2k^{nb}k_{bd} \bigl(\z\gv{\n_\o}\F^d_{\td a}h^{0\td a\td b}\gd k,0,\td
m\td b, + \wt g{}^\(\a\b)\gv{\n_a} \F^d_{\td m}g_\[\o\b]\bigr)
\e
where
\beq3.85
k^{nb}=h^{na}h^{bp}k_{ap}.
\e

The condition \er{3.71} can be also explicitly solved. One gets
\bml3.86
\gd L,n,\td w\td m,= \gd H,n,\td w\td m, + \mu \gd k,n,d,\gd H,d,\td w\td m,
+\z\bigl(h^{0\td a\td d}\gd H,n,\td a\td w,\gd k,0,\td d\td m,
-h^{0\td a\td d}\gd H,n,\td a\td m,\gd k,0,\td a\td w,\bigr)\\
-2\mu\z^2h^{0\td d\td c}h^{0\td a\td b}\gd H,d,\td d\td a,\gd k,0,\td c\td w,
\gd k,0,\td b\td m,
-2\mu\z\gd k,n,d,h^{0\td a\td p}h^{0\td d\td b}\gd H,d,\td b[\td w,
\gd k,0,\td m]\vphantom{\td b}\td p,\gd k,0,\td d\td a,\\
{}+2\mu^2\z k^{nb}k_{bd}\gd H,d,\td a[\td w,
\gd k,0,\td m]\td p,h^{0\td p\td a}.
\e
In this case a kinetic term for a scalar field takes a form
\bml3.87
\cL_{\rm kin}(\gv\n \F)=\frac1{V_2}\int_M\sqrt{|\wt g|}\,d^{n_1}x
\Bigl[\ell_{nk}g^{\o\mu}g^{\td m\td p}\gv{\n_\mu}\F^k_{\td p}
\Bigl\{\gv{\n_\o}\F^n_{\td m} + \z \gd k,n,d,\gv{\n_\o}\F^d_{\td m}\\
{}-\z \gv{\n_\o}\Ft da h^{0\td a\td q}\gd k,0,\td q\td m,
-\gv{\n_\a}\Ft am \wt g{}^\(\a\mu)g_\[\eta\o]
-2\xi\z\gv{\n_\d}\Ft da\gd k,n,d,\wt g^\(\d\a)g_\[\a\o]h^{o\td d\td q}
\gd k,0,\td q\td m,\\
{}-\xi\bigl(\z^2 \gd k,n,d,\gv{\n_\o}\Ft da h^{0\td b\td q}h^{0\td a\td w}
\gd k,0,\td q\td m,\gd k,0,\td w\td b,
+\gd k,n,d,\gv{\n_\b}\Ft dm\wt g{}^\(\a\nu)\wt g{}^\(\b\rho)g_\[\nu\o]
g_\[\rho\a]\bigr)\\
{}+\xi^2 \bigl(\z k^{nb}k_{bd}\gv{\n_\o}\Ft dah^{0\td a\td q}\gd k,0,\td q\td
m,
+\gv{\n_\a}\Ft dm\wt g{}^\(\a\b)g_\[\b\o]\bigr)\Bigr\}\Bigr].
\e
In the case of $g_\m=\eta_\m$ (a Minkowski \spt) one gets
\bml3.88
\cL_{\rm kin}(\gv\n \F)=\frac1{V_2}\int_M \sqrt{|\wt g|}\,d^{n_1}x\Bigl[
\ell_{nk}g^{\td m\td p}\gv{\n^\o}\Ft kp \Bigl\{\gv{\n_\o}\Ft nm
+\xi \gd k,n,d,\gv{\n_\o}\Ft dm\\
{} - \z\gv{\n_\o}\Ft da\gd k,0\td a,\td m,
-\xi\z^2 \gd k,n,d,\gd k,0\td b,\td m,\gd k,0\td a,\td b,\gv{\n_\o}\Ft da
+\xi^2\z k^{nb}k_{bd}\gd k,0\td a,\td m,\gv{\n_\o}\Ft da\Bigr\}\Bigr]
\e
where $\gv{\n^\o}\Ft kp=\eta^{\o\mu}\gv{\n_\mu}\Ft kp$, $\gd k,0\td a,\td b,
=h^{0\td a\td c}k_{0\td c\td b}$.

The Higgs \pt\ is given by
\bml3.89
V(\F)=\frac1{V_2}\int_M \sqrt{|\wt g|}\,d^{n_1}x\,\Bigl\{
g^{\td w\td p}g^{\td m\td q}\Bigl[h_{nk}\gd H,n,\td w\td m,
+2\z h_{nk}\gd H,n,\td d\td w,\gd k,0\td d,\td m,\\
{}+ \mu\z\Bigl(2k_{nk}\gd H,n,\td d\td w,\gd k,0\td d,\td m,
+\z\bigl(-2k_{kd}\gd H,d,\td d\td n,\gd k,0\td d,\td w,\gd k,0\td a,\td m,
-k_{kd}\gd H,d,\wi\td w,\gd k,0,\td m\td a,k^{0\wi\td a}
+k_{kd}\gd H,d,\wi\td m,k^{0\wi\td a}\gdg k,0,\td w,\td a,\bigr)\\
{}+\z\bigl(k_{nk}\gd k,n,d,\gd H,d,\wi\td m,\gdg k,0,\td w,\td r,\gd k,\wi,\td
r,
-2k_{nk}\gd k,n,d,\gd H,d,\td d\td a,\gd k,0\td d,\td w,\gd k,0\td a,\td m,
-k_{nk}\gd k,n,d,\gd H,d,\wi\td w,\gdg k,0,\td m,\td r,\gd k,0\wi,\td r,
\bigr)\Bigr)\\
{}+\mu^3\z\bigl(k_{nk}k^{nb}k_{bd}\gd H,d,\td a\td w,\gdg k,0,\td m,\td a,
-k_{nk}k^{nb}k_{bd}\gd H,d,\td a\td m,\gdg k,0,\td w,\td a,\bigr)\Bigr]
\gd H,k,\td p\td q,
-2h_{cd}\bigl(\gd H,c,\td p\td q,g^\[\td p\td q]\bigr)
\bigl(\gd H,d,\td a\td b,g^\[\td a\td b]\bigr)\Bigr\}
\e
or
\bml3.90
V(\F)=\frac1{V_2}\int_M \sqrt{|\wt g|}\,d^{n_1}x\,\Bigl(
\dg P,kl,[\td p\td q][\td a\td b],\gd H,k,\td p\td q,\gd H,l,\td a\td b,
-2h_{kl}\bigl(\gd H,k,\td p\td q,g^\[\td p\td q]\bigr)
\bigl(\gd H,l,\td a\td b,g^\[\td a\td b]\bigr)\Bigr)\\
{}=\frac1{V_2}\int_M \sqrt{|\wt g|}\,d^{n_1}x\,\dg Q,sk,[\td c\td d][\td p\td q],
\gd H,s,\td c\td d,\gd H,k,\td p\td q,.
\e
$$
Q_{sk}^{[\td c\td d][\td p\td q]}=Q_{ks}^{[\td p\td q][\td c\td d]}=
-Q_{sk}^{[\td d\td c][\td p\td q]}=-Q_{sk}^{[\td c\td d][\td q\td p]}=
Q_{sk}^{[\td d\td c][\td q\td p]}
$$
where
\bml3.91
\dg P,sk,[\td c\td d][\td p\td q],=g^{[\td c\tl \td p}g^{\td d] \td q\tp}h_{sk}
-2\z h_{sk}\gd k,0[\td d,|\td e|,g^{\td c][\td p}g^{|\td e|\td q]}
+\mu\z\Bigl(-2k_{sk}\gd k,0[\td d,|\td e|,g^{\td c][\td p|\td e|\td q]}\\
{}+\z\bigl(2k_{sk}\gd k,0[\td c,|\td e|,\gd k,0\td d],\td f,
g^{\td e[\td p}g^{|\td f|\td q]}
-k_{sk}\gd k,0,\td e\td a,k^{0[\td d|\td a|}g^{\td c][\td p}g^{|\td e|\td q]}
-k_{sk}k^{0[\td c|\td a|}g^{\td d][\td q}g^{|\td e|\td p]}
\gd k,0,\td e\td a,\bigr)\Bigr)\\
{}+\mu^2\z\Bigl(-k_{bs}\dg k,k,b,\gd k,0[\td d,|\td a|,g^{\td c[\td p}
g^{|\td a|\td q]}
-2\gd k,n,s,k_{nk}\gd k,0[\td c,|\td e|,\gd k,0\td d],\td f,g^{\td e[\td p}
g^{|\td e|\td q]}
+\z\bigr(\gd k,n,s,k_{nk}\gdg k,0,\td a,\td r,\gd k,0[\td c,|\td r|,
g^{\td a\tl \td p}g^{\td d]\td q\tp}\\
{}+\z\bigl(\gd k,n,s,k_{nk}\gdg k,0,\td a,\td r,\gd k,0[\td c,|\td r|,
g^{\td a\tl \td p\td a]\td q\tp}
-2\gd k,n,s,k_{nk}\gd k,0[\td c,|\td e|,\gd k,0\td d],\td f,
g^{\td e[\td p}g^{|\td f|\td q]}\bigr)\bigr)\Bigr)\\
{}+\mu^3\z\bigl(-k_{bs}k^{nb}k_{nk}\gdg k,0,\td e,[\td d,g^{\td c][\td p}
g^{|\td e|\td q]}
-k_{bs}k^{nb}k_{nk}\gdg k,0,\td e,[\td c, g^{|\td e|\tl p}g^{\td d]\td q\tp}
\bigr)
+\mu^2 g^{[\td c\tl \td p}g^{\td a]\td q\tp}\gd k,p,s,k_{pk},
\e
$$
\dg Q,sk,[\td c\td d][\td p\td q],=\dg P,sk,[\td c\td d][\td p\td q],
-2h_{sk}g^{[\td c\td d]}g^{[\td p\td q]}.
$$

Let us do some manipulations concerning physical \di s. The \cn\ $\o$ on the
fiber bundle $P$ has no correct physical \di s. Let us pass in all formulas
from~$\o$ to $\a_s\,\frac1{\hbar c}\,\o$,
\beq3.92
\o \mapsto \a_s\,\frac1{\hbar c}\,\o,
\e
where $\hbar$ is a \Pl\ \ct, $c$ is the velocity of light in the vacuum and
$\a_s$ is a \di less coupling \ct\ for the \YM' field if this field couples
to a matter. For example in the \elm c case $\a_s=\frac1{\sqrt{137}}$. We use
$\a_g=\a_s^2=\frac{g^2}{\hbar c}$ where $g$ is a coupling \ct\ for a gauge
field. The redefinition of~$\o$ is equivalent to a usual treatment in local
section $e:V\supset U\to P$, $e^*\o=\frac{g}{\hbar c}A$.

Let us notice that we do this redefinition for a \cn~$\o$, not only
for~$\o_E$. This means that we treat Higgs' field as a part of \YM' field
(gauge field). This is a part of our geometrical \un\ of fundamental \ia s.
One easily writes an integral of action
\bml3.93
S=-\frac1{r^2}\int \sqrt{-g}\,d^4x\,\Bigl[\ov R(\ov W)\\
{}+\frac{8\pi
\la^2\a_s^2}{4c\hbar} \Bigl(\cLY+\frac1{4\pi r^2}\,\cL_{\rm kin}
-\frac1{8\pi r^2}\,V(\F) - \frac1{2\pi r^2}\,\cL_{\rm int}(\F,\wt A)
\Bigr)+\la_c\Bigr].
\e

If we want to be in line with an ordinary coupling between gravity and matter
we should put
\beq3.94
\frac{8\pi \la^2\a_s^2}{4c\hbar} = \frac{8\pi G_N}{c^4}\,.
\e
One gets
\beq3.95
\la=\frac2{\a_s}\,\ell\pl=\frac2{\sqrt{\a_g}}\,\ell\pl
\e
where $\ell\pl$ is the \Pl\ length $\ell\pl=\sqrt{\frac{G_N\hbar}{c^3}}
\simeq 10^{-33}$~cm.
In this case we have
\beq3.96
\la_c=\Bigl(\frac{\a_s^2}{\ell\pl^2}\,\wt R(\wt \G)+\frac{\wt{\ul P}}{r^2}
\Bigr).
\e

Let us pass to \sn\ \s  y breaking and \Hm\ in our theory. In order to do
this we look for the critical points (the minima) of the \pt\ $V(\F)$.
However, our field \sf ies the constraints
\beq3.97
\Ft cb\gd f,\td b,\hi\td a, - \gd \mu,a,\hi,\Ft ba \gd C,c,ab,=0.
\e
Thus we must look for the critical points of
\beq3.98
V'=V+\gd \psi,\hi\td d,c,\bigl(\Ft cb\gd f,\td b,\hi\td a,
-\gd \mu,a,\hi, \Ft ba \gd C,c,ab,\bigr)
\e
where $\gd \psi,\hi\td d,c,$ is a Lagrange multiplier. Moreover, we should
change \di s of the scalar field $\Ft aa$ in the \pt. It is in the following
exchange form
\beq3.99
\gd H,b,\td a\td b,=\gd C,d,cd,\Ft ca\Ft d b - \gd \mu,b,\hi,\gd f,\hi,\td a
\td b, - \Ft bc \gd f,\td c,\td a\td b,
\e
to
\beq3.100
\gd H,b,\td a\td b,=\a_s\,\frac1{\sqrt{\hbar c}}\,\gd C,d,cd,\Ft ca\Ft db
-\frac1{\a_s}\sqrt{\hbar c}\,\gd \mu,b,\hi,\gd f,\hi,\wt a\wt b,
-\Ft bc\gd f,\td c,\td a\td b,.
\e
It is easy to see that, if
\beq3.101
\gd H,a,\td m\td n,=0
\e
then
\beq3.102
\frac{\d V'}{\d \F}=0
\e
if \er{3.97} is \sf ied.

This was noticed in Refs \cite{20}, \cite1 and it is known in the \s ic
theory. $\gd H,a,\td n\td m,$ is a part of the curvature of $\o$ over
a~manifold~$M$. Thus it means that $\F\crt$ \sf ying Eq.~\er{3.101} is a
``pure gauge''. If the \pt\ $V(\F)$ is positively defined, then we have the
absolute minimum of~$V$
\beq3.103
V(\F\crt^0)=0.
\e
But apart from this \so\ there are some others due to an influence of \nos\
metric on~$H$ and~$M$. The details strongly depend on \ct s $\xi$,~$\z$ and on
groups $G,G_0,H$. There are also some critical which are minima. Moreover, we
expect the second critical point $\F\crt^1\ne \F\crt^0$ \st
$V(\F\crt^1)\ne0$ and
\bg3.104
\gd H,a,\td m\td n,(\F\crt^1)\ne0 \\
\frac{\d V'}{\d\F}(\F\crt^i)=0, \q i=0,1. \lb3.105
\e
This means that $\F\crt^1$ is not a ``pure gauge'' and a gauge configuration
connected to $\F\crt^1$ is not trivial. This indicates that the local minimum
is not a vacuum state. It is a ``false vacuum'' in contradiction to ``true
vacuum'' for the absolute minimum $\F\crt^0$.

Now we answer the question of what is a \s y breaking if we choose one of the
critical values of $\F\crt^0$ (we choose one of the degenerated vacuum states
and the \sn\ breaking of the \s y takes place). In Ref.~\cite{20} it was
shown that if $\gd H,a,\td m\td n,=0$ and Eq.~\er{3.97} is \sf ied then the
\s y is reduced to~$G_0$. In the case of the second minimum (local
minimum---false vacuum) the unbroken \s y will be in general different.

Let us call it $G_0'$ and its \Li~$\fg_0'$. This will be the \s y which
preserves $\F\crt^1$ and the constraint \er{3.97}. It is easy to see that the
\Li\ of this unbroken group preserves $\F\crt^1$ under Ad-action. For
the \s y group $V$ is larger than $G$ (it is~$H$) we expect some scalars
which remain massless after the \s y breaking in both cases (i.e., $i=0,1$,
``true'' and false vacuum case). They became massive only through radiative
corrections. They are often referrred as the pseudo-Goldstone bosons.

Let us pass to the integral of action \er{3.93} in the two vacuum cases
$\F\crt^0$, $\F\crt^1$. Let us expand the Higgs' field $\F^{\td a}_a$ in the
neighbourhood of $(\F\crt^k)^a_{\td b}$, $k=0,1$,
\beq3.106
\Ft ab= (\F\crt^k)_{\td b}^a + (\vf^k)_{\td b}^a
\e
and apply this formula for $e^*(\gv{\n_\mu}\Ft ba)$:
\beq3.107
e^*(\gv{\n_\mu}\Ft ba)=e^*(\gv{\n_\mu}(\vf^k)^b_{\td a})
+\a_s\,\frac1{\hbar c}\bigl((\F\crt^k)_{\td a}^a \gd C,b,ac,\gd \a,c,j, \gd
\wt A,j,\mu, + (\F\crt^k)_{\td b}^a \gd f,\td b,\td a j,\gd\wt A,j,\mu,\bigr)
\e
and for $V(\F)$
\beq3.107a
V(\F)=V(\F\crt^k)+\wt V^k(\vf^k), \q k=0,1,
\e
where $V(\F\crt^k)$ is the value for the critical value of~$\F$ and $\wt
V{}^k(\vf^k)$ is the polynomial of fourth order in $\vf^k$. If we use
Eq.~\er{3.87} we get a mass matrix for vector bosons $\gd\wt A,j,\mu,$ which
strongly depends on~$\F\crt^k$
\beq3.108
N^\m M_{ij}^2 (\F\crt^k) \gd\wt A,i,\mu, \gd\wt A,j,\nu,.
\e
The matrix $N^\m$ depends on $g_\m$ and in the case $g_\m=\eta_\m$ (Minkowski
tensor) we have
\beq3.109
N^\m =\eta^\m
\e
and
\bml3.110
M_{ij}^2(\F\crt^k) = \frac{\a_s^2}{4\pi r^2\hbar c}\,\frac1{V_2}
\int_M \sqrt{|\wt g|}\,d^{n_1}x\,\Bigl\{
\ell_{np}g^{\td m\td p} \gd \nadd{(k)}B{},p,\b(i, \Bigr(\gd \nadd{(k)}B{},
d,\td mj), + \xi\gd k,n,d, \gd \nadd{(k)}B{},d,\td mj), - \z\gd \nadd{(k)}B{},
d,\td aj,\gd k,0\td a,\td m,\\ {}- \xi\z^2 \gd k,n,d,\gd k,0\td b,\td m,
\gd k,0\td a,\td b,\gd \nadd{(k)}B{},d,\td aj), + \xi^2\z k^{nb}k_{bd}
\gd k,0\td a, \td m,\gd \nadd{(k)}B{},d,\td aj),\Bigr)\Bigr\}, \q k=0,1,
\e
where
\beq3.111
\gd \nadd{(k)}B{},b,\td ni, = \bigl[\gd \d,\td m,\td n,\gd C,b,ms,\gd \a,s,i,
+\gd \d,b,m,\gd f,\td m,\td ni,\bigr][\F\crt^k]_{\td m}^m \q
(M_{ij}^2=M_{ji}^2).
\e

In the case of a \s ic theory $\ell_{ab}=h_{ab}$, $g_{\td a\td b}=
\gd h,0,\td a\td b,$ one gets
\beq3.112
M_{ij}^2 = \frac{\a_s^2}{4\pi r^2\hbar c}\,\frac1{V_2} \int_M \sqrt{|\wt g|}\,
d^{n_1}x\,\bigl\{ h_{bn}h^{0\td m\td p}\gd B,b,\td p(i, \gd B,n,\td mj),
\bigr\}.
\e
Let us consider an expression
\beq3.113
(\F\crt^k)^m_{\td n}\gd C,b,ms,\gd\a,s,i, + (\F\crt^k)^b_{\td m}
\gd f,\td m,\td ni,
\e
in order to find its interpretation. One easily notices that it equals to
\beq3.114
\bigl(\bigl[ \Ad_H'(Y_i)+\Ad_G'(Y_i)\bigr]\F\crt^k\bigr)^b_{\td m}
\e
($\Ad_H'$ and $\Ad_G'$ mean the adjoint representation of \Li s of~$H(\fh)$
and $G(\fg)$, respectively).

Thus if $k=0$ \er{3.113} equals zero for $Y_i\in \fg_0$ and if $k=1$
\er{3.113} equals zero for $Y_i\in\fg_0'$. The latest statement comes from
the \iv cy of the vacuum state \wrt the action if the group~$G_0$ for $k=0$
($G_0'$ for $k=1$). Generators of $\fg_0$ ($\fg_0'$) should annihilate vacuum
state. Thus the matrix \el s $M_{ij}^2(\F\crt^k)$ are zero for $i,j$
corresponding to $\fg_0$ ($\fg_0'$).

From the \iv cy of the \pt\ $V$ \wrt the action of the group~$G$ one gets
\beq3.115
\frac{\pa^2V}{\pa\Ft bn \pa\Ft dd}\Bigr|_{\F=\F\crt^k}
\bigl({{{T^{b}}_{\!\td n}}^{\!c}}_{\!\td c}\bigr)_i\gd[\F\crt^k],\td c,c,=0
\e
where
\beq3.116
\bigl({{{T^{b}}_{\!\td n}}^{\!c}}_{\!\td c}\bigr)_i\gd[\F\crt^k],\td c,c,=
\gd[\F\crt^k],m,\td n,\gd C,b,ms,\gd\a,s,i, +\gd[\F\crt^k],b,\td m,
\gd f,\td m,\td ni,.
\e
Eigenvalues of $M_{ij}^2(\F\crt^k)$ are the squares of the masses of the
gauge bosons. The secular \e\ $\det(M^2-m^2I)=0$ gives us a mass spectrum of
massive vector bosons. Thus there is an orthogonal matrix $(A^i_j)=A$ \st
$A^T=A^{-1}$ and
\bg3.117
A^{-1}M^2(\F\crt^k)A=\left|\begin{matrix}
m_1^2(\F\crt^k) & \dots &0 \\
\vdots& \ddots &\vdots \\
0 & \dots &m_{l_k}^2(\F\crt^k) \end{matrix} \right|,\\
l_0=n_1,\ l_1=\dim G-\dim G_0'. \nn
\e
In this way we transform the broken vector fields into massive vector fields
\beq3.118
\gd \wt B,i',\mu,=\sum_{j=1}^{l_k} \gd A,i',j,\gd \wt A,j,\mu,
\e
\st
\beq3.119
\eta^\m \sum_{j=1}^{l_k} m_j^2 (\F\crt^k)\gd \wt B,j,\mu,\gd \wt B,j,\nu,
=\eta^\m M_{ij}^2 (\F\crt^k)\gd \wt A,i,\mu,\gd \wt A,j,\nu,.
\e
Moreover, we should remember the formula \er{3.97} which is a constraint on
Higgs' field. The mass matrix of masses for Higgs' bosons can be obtained in
a similar way,
\beq3.120
V(\F)=V(\F\crt^k) + \frac{\d V}{2\d \Ft aa \d\Ft bb}\,\gd \vf,a,\td a,\gd
\vf,b,\td b,+\ldots
\e
The matrix
\beq3.121
{\gdg m^2(\F\crt^k),\td a,a,\td b,}_b = -\frac{\d^2V}{\d \Ft aa\d\Ft bb}
\Bigr|_{\F=\F\crt^k}
\e
can be calculated for $k=0$. One gets
\bml3.122
{\gdg m,2\td h,f,\td e,}_a = \frac{-1}{8\pi r^2V_2} \int_M \biggl\{
\frac{8\a_s^2}{\hbar c}\,\dg Q,sk,[\td e\td a][\td n\td q],\gd C,s,ac,
\gd C,k,ef,\gd (\F\crt^0),c,\td a,\gd (\F\crt^0),e,\td q,
-2\,\frac{\a_s}{\sqrt{\hbar c}}\,\dg Q,as,[\td p\td q][\td n\td a],
\gd f,\td e,\td p\td q,\gd C,s,ef,\gd (\F\crt^0),e,\td a,\\
{}+\frac{4\a_s}{\sqrt{\hbar c}}\,\dg Q,sf,[\td e\td a][\td p\td q],
\gd f,\td n,\td p\td q,\gd C,s,ea,\gd (\F\crt^0),a,\td a,
+\dg Q,af,[\td c\td d][\td p\td q],\gd f,e,\td c\td d,\gd f,\td n,\td p\td q,
\biggr\}\sqrt{|\wt g|}\,d^{n_1}x.
\e
For $k=1$, $\gd H,a,\td p\td q,(\F\crt^1)\ne0$ and $\F\crt^1$ (if exists)
\sf ies the following \e
\beq3.123
\frac{2\a_s}{\sqrt{\hbar c}}\,\dg Q,sk,[\td e\td a][\td p\td q],\gd C,s,ac,
\gd(\F\crt^1),c,\td a,=\dg Q,ak,[\td c\td d][\td p\td q],\gd f,\td e,\td c\td
d,
\e
and the supplementary condition \er{3.97}.

A mass matrix for Higgs' bosons looks like
\beq3.124
{\gdg m,2\td h,f,\td e,}_a = \frac{-1}{8\pi r^2V_2} \int _M
\biggl( \frac{4\a_s}{\sqrt{\hbar c}}\,\dg Q,sk,[\td e\td h][\td p\td q],
\gd H,k,\td p\td q,(\F\crt^1)\gd C,s,af,\biggr)\sqrt{|\wt g|}\,d^{n_1}x.
\e
We can diagonalize the mass matrix and we get
\beq3.125
{\gdg m^2(\F\crt^k),a,\td a,b,}_{\td b}\gd \vf,\td a,a,\gd\vf,\td b,b,
=\sum_{j=1}^{l_k}\gd m_j^2(\F\crt^k),a,\td a, \gd \psi,a,\td a,
\gd \psi,a,\td a,
\e
where
\beq3.126
\gd \psi,a,\td a,= \sum_{b,\td b}{\gdg A,\td b,b,\td a,}_a \dg \vf,\td b,b,.
\e
For the mass matrix one has
\beq3.127
{\gdg (A^{-1}),\td c,c,a,}_{\td a}{\gdg m^2(\F\crt^k),\td a,a,\td b,}_b
{\gdg A,\td d,\vphantom{\td b}d,b,}_{\td b} = (\gd m^2(\F\crt^k),\td c,c,)\gd \d,\td d,\td c,
\gd \d,c,d,.
\e
The eigenvalue problem for $m^2(\F\crt^k)$ can be posed as follows
\beq3.128
{\gdg m^2(\F\crt^k),\td a,a,\td b,}_b \gd X,a,\td a,
=m^2(\F\crt^k) \gd X,\td b,b,.
\e
One gets the mass spectrum of Higgs' \pc s from the secular \e
\beq3.129
\det\bigl({\gdg [m^2(\F\crt^k)],\td a,a,\td b,}_b - m^2(\F\crt^k)
{\gdg I,\td a,a,\td b,}_b\bigr)=0
\e
where
\beq3.130
{\gdg I,\td a,a,\td b,}_b = \d_{ab}\d^{\td a\td b}.
\e

The diagonalization procedure of the matrix ${\gdg m^2(\F\crt^k),\td a,a,\td
b,}_b$ can be achieved in the two following ways. The matrix defines a quadratic
form on the \rp ation space $N$ for Higgs' field. Moreover, the space~$N$ can
be decomposed into Higgs' multiplets $\ul m{}_j$ and according to this de\cm\
the matrix can be written in a block diagonal form
\beq3.131
[m^2(\F\crt^k)]=\sum_j \oplus m_j^2(\F\crt^k).
\e
We can diagonalize every matrix $m_j^2(\F\crt^k)$ corresponding to the
multiplet~$\ul m{}_j$.

Let us consider a problem of the Higgs' multiplet $\gd \F,c,\td a,$ on~$E$.
One can find a \rp ation space $N$ of $\gd\F,c,\td a,$ in the following way
(see Ref.~\cite{23}). Let
\beq3.132
\Ad_G  \to \sum_i \oplus\ul n{}_i \oplus \Ad_{G_0}
\e
be the de\cm\ of the adjoint \rp ation of~$G$, where $\ul n{}_i$ are
irreducible \rp ations of~$G_0$ and let us consider the branching rule
of~$\Ad_H$
\beq3.133
\Ad_H \to \sum_j \oplus(\ul n{}'_j \otimes \ul m{}_j)
\e
where $\ul n{}'_j$ are irreducible \rp ations of $G_0$ and $\ul m{}'_j$ are
irreducible \rp ations of $G$. The latest formula comes from the known fact
$G_0 \ot G \subset H$. Thus for every pair $(\ul n{}_i,\ul n{}_j')$ where
$\ul n{}_i$ and~$\ul n{}_j'$ are identical irreducible \rp ations of $G_0$
there is an $\ul m{}_j$ multiplet of Higgs' field on~$E$. In this way we can
decompose $\F$ into a sum
\beq3.134
\F=\sum_{(\ul n{}_i,\ul n{}_j')} \op \F_{\ul m{}_j}^{(\ul n{}_i,\ul n{}_j)}
\e
or
\beq3.135
N=\sum_{(\ul n{}_i,\ul n{}_j')} \op m{}_j.
\e
Thus the multiplet of Higgs' field is quite complicated in contradiction to
the usual case where the Higgs' field belongs to the adjoint \rp ation of
chosen group. Moreover, in our case we have to do with smaller number of
parameters in the theory. The theory is established by a coupling \ct~$\a_s$,
a~radius~$r$, parameters coming from the \nos ity of the theory $\xi,\z$, a
homo\mo sm~$\mu$, an embedding of~$G$ in $H$, $\a_i^c$ ($\fg$ in $\fh$) and an embedding
of~$G_0$ in~$G$ (i.e.\ the manifold~$M$).

The second way of diagonalization of the matrix $[m^2(\F\crt^k)]$ is based on
the following observation.

The matrix ${\gdg m^2(\F\crt^k),\td a,a,\td b,}_b$ can be transformed into a
different matrix $(n_1 n)\times (n_1 n)$ forming an index from two
indices $\wt a$ and~$a$
\beq3.136
\ov a=\a a+\b \wt a+\g
\e
where $\a,\b,\g$ are integers. The new index $\ov a$ should be unambigous.
Thus we must choose $\a,\b,\g$ in such a way that for every $\ov a\in
N_1^{n_1n}$ the \e\ \er{3.136} has only one \so\ for $a\in N_1^n$, $\wt a
\in N_1^{n_1}$.

After this we diagonalize $[m^2(\F\crt^k)]$ as an ordinary matrix. What is
a scale of masses in our theory? It is easy to see that
\beq3.137
m_{\td A}=\frac{\a_s}r\,\Bigl(\frac \hbar c\Bigr)
\e
is this scale where $m_{\tA}$ is a typical vector boson mass obtained due to
\Hm.

Let us consider the following de\cm\ of the \cn\ $\o_E$ defined on the
principal fiber bundle~$Q$:
\beq3.138
\o_E=\gd \o,0,E, + \si_E, \q \gd \o,0,E,\in \fg_0, \q \si_E\in \mathfrak m,
\e
corresponding to the de\cm\ of the \Li~$\fg$,
$$
\fg=\fg_0\dot+\mathfrak m.
$$
In this way we consider a reduction of a bundle $Q$ to~$Q_0$ induced by an
embedding of~$G_0$ into~$G$. The form $\gd \o,0,E,$ is a \cn\ defined
on~$Q_0$ and $\si_E$ is a tensorial form defined on $Q(E,G)$. We suppose that
the reduction of the bundle $Q$ to~$Q_0$ is possible.

The form $\gd \o,0,E,$ corresponds to the \YM' field (massless vector bosons)
which remains after \s y breaking. The tensorial form $\si_E$ corresponds to
massive vector bosons.

One gets for the \cvt\ form
\beq3.139
\O_E=\gd\O,0,E, + D^0\si_E + [\si_E,\si_E],
\e
where $\gd\O,0,E,$ is a \cvt\ form for $\gd\o,0,E,$ and $D^0$ means a \ci\
exterior \dv\ \wrt $\gd\o,0,E,$. Thus
\bea3.140
\gd\O,0,E,&=\frac12\,\gd\wt H,\hi,\m, \ov\t{}^\mu \land \ov\t{}^\nu\gd\b,i,\hi,
Y_i\\
\si_E&=\si^{\td a}Y_{\td a} \lb3.141 \\
e^* D^0\si_E&= \gv{\n^0_{[\mu}}\gd\ov \si{},a,\nu], \ov\t{}^\mu \land \ov\t{}^\nu
Y_{\td a} \lb 3.142 \\
\gv{\n^0_{[\mu}}\gd\ov \si{},a,\nu],&=\pa_{[\mu}\gd\ov\si,a,\nu], +
\frac g{\hbar c}\,\gd f, \td a,\td bi, \gd\b,i,\hi,
\gd \ov\si,\td b,[\nu,\gd \wt A,\hi,\mu], \lb3.143 \\
\si^*\si_E&=\gd\ov\si,\td a,\nu, \ov\t{}^\nu Y_{\td a} = \gd\ov\si,i,\nu,
\gd\b,\td a,i, Y_{\td a}, \lb3.144 \\
e^*\gd\o,0,E,&=\gd\wt A,i,\mu, \gd\b,i,\hi, \ov\t{}^\mu Y_i \lb3.145
\e
where $e$ is a local section of the principal bundle~$Q$, matrices $\gd\b,
\td a,i,$, $\gd \b,i,\hi,$ define an embedding of $\fg_0$ into~$\fg$, $\gv
{\n^0_\mu}$ means a gauge \dv\ \wrt a \cn\ $\gd\o,0,E,$.

If the \s y is broken from $G$ to $G_0'$ we have a different de\cm
\beq3.146
\o_E=\gd\o,\prime0,E,+\si'_E.
\e
One can easily connect $\si_E$ or $\si'_E$ with $\wt B$ (i.e.\ fields with
defined non-zero rest mass, because $\wt A$ have not defined masses) fields.
One gets
\beq3.146a
e^*\si_E = \frac g{\hbar c}\,\gd\wt B,i',\mu, \ov\t{}^\mu Y_{i'}
\e
where
\beq3.147
Y_{i'}=\gd(A^{-1}),i,i', Y_i =\gd(A^{-1}),i,i',\gd\b,\td a,i,Y_{\td a}.
\e
The matrix $A$ is defined by \er{3.117}. The same holds for $\si'_E$.

Let us consider the \fw\ gauge \tf, i.e.\ a change of a local section of~$Q$
from $e$ to~$f$,
\beq3.147a
e(x)=\nadd{(k)}U{}^{-1}(x)f(x),
\e
where
\beq3.148
\nadd{(k)}U(x)=\exp\Bigl(\sum_{\check a}\nadd{(k)}\eta_{\check a}(x)
Y_{\check a}\Bigr)
\e
for $k=0$, $Y_{\check a}\in \fm$, i.e.\ $\check a=\wt a$; for $k=1$,
$Y_{\check a}\in\fm$, and $\nadd{(k)}\eta_{\check a}(x)$ is a multiplet of
scalar fields on~$E$ transforming according to the $\Ad G_0$ ($\Ad G_0'$).
$Y_{\check a}$~span $\fm$ or~$\fm'$ ($k=0$ or $k=1$). Such a gauge \tf\ (a
condition) is a ``unitary gauge''.

Let us consider the \fw\ parametrization of the Higgs' field
\beq3.149
\Ft ca=\Ad_G (\nadd{(k)}U(x))^{\td a'}_a \Ad_H(\nadd{(k)}U(x))^{\td c}_{c'}
\bigl(\gd(\F\crt^k),c',\td a', + \gd\nadd{(k)}\vf,c',\td a',(x)\bigr).
\e
We transform Higgs' and gauge fields
\bg3.150
\Bigl(\F^{\nadd{(k)}U}\Bigr)^c_{\td a}=\Ad_H(\nadd{(k)}U{}^{-1})^c_{c'}
\Ad_G(\nadd{(k)}U{}^{-1})^{\td a'}_{\td a}\F^{c'}_{\td a'}
=(\F\crt^k)^a_{\td a}+\gd\nadd{(k)}\vf,c,\td a,(x)\\
\gd \wt B,\prime i,\mu,(x)Y_i = \gd\Ad_H(\nadd{(k)}U(x)),i,j,
\gd\wt B,j,\mu, Y_i - \frac{\hbar c}g\,\pa_\mu \nadd{(k)}U(x)
\nadd{(k)}U{}^{-1}(x).
\e
One easily gets
\beq3.152
\gv{\n^0_\mu}\Ft ca=\gd\Ad_H(\nadd{(k)}U(x)),c,c',
\gd\Ad_G(\nadd{(k)}U{}^{-1}(x)),\td a',\td a,\gd\gv{\n^0_\mu}\bigl(
\F^{\nadd{(k)}U}\bigr),c',\td a',.
\e
On the level of a tensorial form one gets
\beq3.153
f^*\nadd{(k)}\si_E = \Ad_G(\nadd{(k)}U{}^{-1}(x))
\bigl(e^*\nadd{(k)}\si_E-d\nadd{(k)}U(x)\nadd{(k)}U{}^{-1}(x)\bigr)
\e
and
\beq3.154
\cLY(\wt A)=\cLY(\wt B)=\cLY(\wt B{}').
\e

It is important to notice that we should consider a local section for
$\ell_{ij}$, i.e.\ $a_{ij}=e^*\ell_{ij}$ in the lagrangian for the \YM'
field. The fields $\wt B{}'$ are massive with the same masses as~$\wt B$. The
important point to notice is that the full lagrangian is still $G$-gauge \iv
t. Moreover, a choice of a particular value of $\F\crt^k$ (which is $G_0$ \iv
t) reduces \s y from~$G$ to~$G_0$ (\sn ly). The fields $\eta_{\check a}(x)$
disappear. They are eaten by the gauge \tf\ and due to this the massive
vector fields have three polarization degrees of freedom. Sometimes
$\eta_{\check a}(x)$ are called ``would-be Goldstone bosons''. In the matrix
of masses they correspond to zero modes.

Let us come back to field \e s in our theory. From the Palatini \v al \pp\
for the action~$S$ (see Eqs \er{3.76}--\er{3.77}) one gets (\v\ \wrt $\gd\ov
W,\la,\mu,,g_\m,\o_E$ and~$\F$)
\bg3.155
\ov R_\m(\ov W)-\frac12 \,g_\m \ov R(\ov W)=\frac{8\pi G_N}{c^4}
\bigl(\gv {T_\m}+T_\m(\F)+\nad{int}{T_\m}+g_\m \La\bigr) \\
\gd\falg,[\m],{,\nu},=0 \lb3.156 \\
\ov\n_\nu g^\[\m]=0 \lb3.156a \\
g_{\m,\si}-g_{\xi\nu}\gd\ov\G,\xi,\mu\si, -g_{\mu\xi}\gd\ov\G,\xi,\si\nu,=0
\lb3.157
\e
\bml3.158
\gv\n(\wt\ell_{ij}\fal{\wt L}{}^{i\a\mu})
=2\falg{}^\[\a\b]\gv{\n_\b}(\wt h_{ij}g^\[\m]\gd\wt H,i,\m,)\\
{}+\frac2{r^2}\sqrt{-g}\,\frac{\a_s}{\sqrt{\hbar c}}\biggl[
\ell_{ab}g^{\td b\td n}g^{\mu\a}\gd L,a,\mu \td b,
\bigl(\Ft dc\gd C,b,dc,\gd \a,c,j, +\Ft ba\gd f,\td a,\td nj,\bigr)\\
{}+\biggl(\frac{\d \gd L,a,\b\td b,}{\d\gv{\n_\a}\Ft wv}\biggr)\ell_{ab}
g^{\td b\td n}g^{\b\mu}(\gv{\n_\mu}\Ft bn)\bigl(\Ft dw\gd C,w,dc,\gd \a,c,j,
+\Ft wa\gd f,\td a,nj,\bigr)\biggr]_{av}\\
{}+\frac4{r^2} \sqrt{-g}\,h_{ab}\gd \mu,a,k,\wt\ell_{ij}\wt\ell{}^{ki}
\wt g{}^\[\td a\td b]\gv{\n_\mu}
\biggl\{g^\[\mu\a]\biggl[\frac{\sqrt{\hbar c}}{\a_s}\,\gd C,b,cd,\Ft ca
\Ft db - \a_s\Bigl(\frac\hbar c\,\gd \mu,b,\hi,\gd f,\hi,\td a\td b,
-\frac{\a_s}{\hbar c}\,\Ft bd\gd f,d,\td a\td b,\Bigr)\biggr]\biggr\}.
\e
\bml3.159
\gv{\n_\mu}(\ell_{ab}\gd\fal L,a\mu,\td b,)_{av}
=-\frac{\sqrt{-g}}{2r^2}\Bigl\{\Bigl(\frac{\d V'}{\d \Ft bn}\Bigr)g_{\td b\td
n}\\
{}+2\sqrt{-g}\,\gd \mu,e,i,(\gd \wt H,i,\m,g^\[\m])h_{ed}\Bigl(
\frac{2\sqrt{\hbar c}}{\a_s}\,g^\[\td a\td n]\gd C,d,cb,\Ft ca g_{\td b\td n}
-\frac{\a_s}{\sqrt{\hbar c}}\,g^\[\td c\td d]\gd f,\td n,\td c\td d,
g_{\td b\td n}\Bigr)\Bigr\}
\e
where
\bml3.160
\gv {T_{\a\b}}=-\frac{\wt \ell_{ij}}{4\pi} \Bigl\{g_{\g\b}g^{\tau\rho}
g^{\ve\g}\gd \wt L,i,\rho\a,\gd \wt L,j,\tau\ve,
-2g^\[\m]\gd \wt H,(i,\m,\gd \wt H,j),\a\b,\\
{}-\frac14\,g_{\a\b}\Bigl(\wt L{}^{i\m}\gd \wt H,j,\m,
-2(g^\[\m]\gd \wt H,i,\m,)(g^\[\g\si]\gd \wt H,j,\g\si,)\Bigr)\Bigr\}
\e
is the energy momentum tensor for the gauge (\YM') field with a zero trace
\bg3.161
\gv {T_{\a\b}}g^{\a\b}=0\\
T_\m(\F)=\frac1{4\pi r^2}\,\bigl(\ell_{ab}g^{\td b\td n}\gd L,a,\mu\td b,
\gv{\n_v}\Ft bn\bigr)_{av} \lb3.162 \\
{}-\frac12\,g_\m\Bigl(-\frac1{8\pi r^4}\,V(\F)+\frac1{4\pi r^2}\,\ell_{ab}
(g^{b\td n}g^{\a\b}\gd L,a,\a\td b, \gv{\n_\b}\Ft bn)_{av}\Bigr).\lb3.162a
\e

It is an energy-momentum tensor for a Higgs' field
\bml3.163
\nad{int}{T_\m} = -\frac1{2\pi r^2}\,h_{ab}\gd \mu,a,\hi,\gd \wt H,i,\m,
\biggl(\wt g{}^\[\td a\td b]\Bigl(\frac{\sqrt{\hbar c}}{\a_s}\,
\gd C,b,cd,\Ft ca\Ft db - \frac{\a_s}{\sqrt{\hbar c}}\,\gd\mu,b,\hi,
\gd f,\hi,\td a\td b, - \frac{\a_s}{\sqrt{\hbar c}}\,\Ft bd
\gd f,\td d,\td a\td b,\Bigr)\biggr)_{av}\\
{}+\frac{g_\m}{4\pi r^2}\Bigl[h_{ab}\gd \mu,a,i,\gd \wt H,i,\a\b,
g^\[\a\b]\wt g{}^\[\td a\td b]\Bigl(
\frac{\sqrt{\hbar c}}{\a_s}\,\gd C,b,cd,\Ft ca\Ft db
-\frac{\a_s\hbar}{c} \,\gd \mu,b,\hi,\gd f,\hi,\td a\td b,
-\frac{\a_s}{\sqrt{\hbar c}}\,\Ft bd\gd f,\td d,\td a\td b,\Bigr)\Bigr]_{av}.
\e

It is an energy-momentum tensor corresponding to the non-minimal \ia\ term
$\cL_{\rm int}(\wt A,\F)$.
\beq3.164
\La=\frac{c^4}{16\pi G_N} \Bigl(\frac{\a_s^2 \wt R(\wt\G)}{\ell\pl^2}
+\frac{\ul{\wt P}}{r^2}\Bigr)=\frac{16\pi G_N}{c^4}\,\ov\la_c.
\e
It plays a role of the ``\co ical \ct''
\bg3.165
\fal{\wt L}{}^{i\m} = \sqrt{-g}\,g^{\b\nu}g^{\g\nu}\gd \wt L,i,\b\g,\\
\falg^\[\m] = \sqrt{-g}\, g^\[\m] \lb3.166 \\
(\dots\dots)_{av}=\frac1{V_2} \int_M \sqrt{|\wt g|}\,dx^{n_1}\,(\dots\dots).
\lb3.167
\e
We can write
\beq3.168
\gv{\n_\mu}(\wt\ell_{ij} \fal{\wt L}{}^{i\a\mu})
=\sqrt{-g}\gv{\ov\n_\mu}(\wt \ell_{ij}\wt L{}^{i\a\mu})
\e
where $\gv{\ov\n_\mu}$ means a \ci\ \dv\ \wrt a \cn\ $\gd\ov\o,\a,\b,$ on~$E$
and $\o_E$ at once.

Let us come back to the \e\ of motion for test \pc s in our theory. According
to the usual interpretation we write down a geodetic \e\ on~$P$ \wrt a \LC
\cn\ induced by a \s ic part of $\k_\(\tA\tB)$.

One writes
\beq3.169
u^{\tA}\wt\n_{\tA} u^\tB=0
\e
where $\wt\n_\tA$ means a \ci\ \dv\ \wrt a \LC \cn\ induced by $\k_\(\tA\tB)$
on~$P$.

One finds
\bg3.170
\frac{\wt{\ov D}u^\a}{d\tau} + \Bigl(\frac{q^c}{m_0}\Bigr)u^\b h_{cd}\wt
g{}^\(\a\d) \gd H,d,\b\d, + \Bigl(\frac{q^c}{m_0}\Bigr)u^{\td b}h_{cd}
\wt g{}^\(\a\d)\gv{\n_\d}\Ft db=0 \\
\frac{\wt Du^{\td a}}{d\tau} + \frac1{r^2}\Bigl(\frac{q^c}{m_0}\Bigr)
u^\b h_{cd}h^{0\td a\td d}\gv{\n_\b}\Ft dd + \frac1{r^2}
\Bigl(\frac{q^c}{m_0}\Bigr)u^{\td b}h_{cd}h^{0\td a\td d}\gd H,d,\td d\td b,
=0 \lb3.171 \\
\frac d{d\tau}\Bigl(\frac{q^b}{m_0}\Bigr)=0 \lb3.172
\e
where $\wt{\ov D}$ means a \ci\ \dv\ along a line \wrt the \cn\
$\gd\wt{\ov\o},\a,\b,$ on~$E$. $\wt D$~means a \ci\ \dv\ along a line \wrt the
\cn\ $\gd\wt\o,a,b,$ on~$G/G_0$ ($r=\rm const$),
\bg3.173
u^\tA = (u^\a,u^{\td a},u^a)\\
2u^a=\frac{q^a}{m_0}, \lb3.174
\e
$q^a$ is a \YM' charge known from the \E\nA\ Kaluza--Klein Theory (color
(isotopic) charge), $u^\a$ is a four-velocity of a test \pc.

$u^{\td a}$ is a charge associated with a Higgs' field. This charge
transforms according to the properties of a complement $\fm$ \wrt $G_0$
and~$G$. Eq.~\er{3.171} describes a movement of a test \pc\ in a \gr al,
gauge and Higgs' field. Eq.~\er{3.172} is an \e\ for a charge associated with
Higgs' field. This charge describes a coupling between a test \pc\ and a
Higgs' field.
Eq.~\er{3.173} has a usual meaning (a constancy of a color (isotopic) charge).
In this way we get a \gn\ of \KWK \e\ to the presence of a Higgs' field.
We have a normalization of a four-velocity~$u^\a$, $g_\(\a\b)u^\a u^\b=1$.

Let us project the \e\ on a \spt~$E$, i.e.\ we take a section $e:E\to P$. One
gets
\bg3.175
\frac{\wt{\ov D}u^\a}{d\tau} + \Bigl(\frac{Q^c}{m_0}\Bigr)u^\b \wt g{}^\(\a\d)
\gd F,d,\b\d,+ \Bigl(\frac{Q^c}{m_0}\Bigr)u^{\td b}h_{cd}\wt g{}^\(\a\d)
e^*(\gv{\n_\d}\Ft db)=0 \\
\frac{\wt Du^{\td a}}{d\tau} + \frac1{r^2}\Bigl(\frac{Q^c}{m_0}\Bigr)u^\b
h_{cd}h^{0\td a\td d}e^* (\gv{\n_\b}\Ft dd)
+\frac1{r^2}\Bigl(\frac{Q^c}{m_0}\Bigr)u^{\td b}h_{cd}h^{0\td a\td d}
e^*(\gd H,d,\td d\td b,)=0 \lb3.176 \\
e^*\o= \gd A,a,\mu,\ov\t{}^\mu X_a + \Ft ab\wt\t{}^b X_a \lb3.177 \\
e^* (q^cX_c) = Q^cX_c. \lb3.178
\e
Equation \er{3.172} takes the form
\beq3.185
\bga
\frac{dQ^a}{d\tau} - \gd C,a,cb,Q^c \gd A,b,N, u^N=0,\\
\hbox{or}\q \frac{dQ^a}{d\tau} - \gd C,a,cb,Q^c\gd A,b,\nu, u^\nu
-\gd C,a,cb,Q^c\gd\F,b,\td n, u^{\td n}=0,
\ega
\e
$\wt g^\(\a\b)$ is defined by Eq.~\er{2.24}.

Let us consider a \E\nos\ Hermitian Kaluza--Klein Theory with \sn\ \s y
breaking. We should introduce an Hermitian tensor on the manifold $V=E\tm
M=E\tm G/G_0$. It is
\beq3.186
\g_{AB}=\left(\begin{array}{c|c}
g_\m&0 \\\hline 0&r^2g_{\td a\td b} \end{array}\right)
\e
but now
\bea3.187
g_{\td a\td b}&=\gd h,0,\td a\td b, + i\z \gd k,0,\td a\td b,\\
g^*_{\td a\td b}&=g_{\td b\td a} \lb3.188 \\
\noalign{\noindent and}
\g^*_{AB}&=\g_{BA}. \lb3.189
\e

The tensor (in a nonholonomic frame)
\beq3.190
\k_{\tA\tB}=\left(\begin{array}{c|c}
\g_{AB} & 0 \\\hline 0 & \ell_{ab}  \end{array}\right)
\e
is \st
\bg3.191
\ell_{ab}=h_{ab}+i\xi k_{ab}\\
\ell^*_{ab}=\ell_{ba} \lb3.192 \\
\noalign{\noindent and}
\k^*_{\tA\tB}=\k_{\tB\tA}. \lb3.193
\e
The \cn\ $\gd\G,\tA,\tB\tC,$ is compatible with $\k_{\tA\tB}$ and we have
\beq3.194
\gd\G,*\td N,\td M\td W,=\gd\G,\td N,\td W\td M,+(\gd\wt\G,\td N,\td M\td W,
-\gd\wt\G,\td N,\td W\td M,)
\e
where $\wt N,\wt M,\wt W=1,2,3,\dots,(m+4)$.

All the formulae derived here (in this section) are the same but we should
consider $g_\[\m]$ as a pure imaginary tensor and put $i\z$ in place of~$\z$
and $i\xi$ in place of~$\xi$.

The Ricci (\MR) tensor and all energy-momentum tensors are Hermitian.

$\gd\wt\G,\td N,\td M\td W,$ is a \cn\ generated by $\k_\(\tA\tB)$. In this
theory we can consider K\"ahler structures on $M=G/G_0$.

In the case of the \NK\ with a \sn\ \s y breaking and \Hm\ we have more
possibilities. We can have a complex Hermitian \sc e as we describe above
or a hypercomplex Hermitian \sc e on a $P$ manifold. Moreover, we can define
on $M=G/G_0$ a hypercomplex Hermitian metric tensor or a complex Hermitian
metric. This means that we have $\xi\mapsto I\xi$ and $\z\mapsto i\z$ (a~pure
imaginary). The last possibility seems to be very interesting for we get
Hermitian Theory with a mixture of hypercomplex and ordinary complex. In this
way we get two disconnected real \sc es on~$E$ (a~\spt) coupled to
\YM' fields and to a Higgs' field. For a base manifold $V=E\tm M$ is a
Cartesian product of $E$ and~$M$ we have to do effectively with a real version
and only on~$M$ a tensor is complex (Hermitian). In some cases the geometry
of a whole space is effectively real and only on~$M$ we have even K\"ahlerian
geometry.

\section{GSW (Glashow--Salam--Weinberg) model in the \NK}
Let $\ul P$ be a principal fiber bundle
\beq4.1
\ul P=(P,V,\pi,H,H)
\e
over the base space $V=E\tm S^2$ (where $E$ is a \spt, $S^2$---a two-\di al
sphere) with a projection~$\pi$, a \sc al group~$H$, a typical fiber~$H$
and a bundle manifold~$P$. We suppose that $H$ is semisimple. Let us define
on~$P$ a \cn~$\o$ which has values in a \Li\ of~$H,\fh$. Let us suppose that
a group $\SO(3)$ is acting on~$S^2$ in a natural way. We suppose that $\o$ is
\iv t \wrt an action of the group $\SO(3)$ on~$V$ in such a way that this
action is equivalent to $\SO(3)$ action on~$S^2$. This is equivalent to the
condition \er{3.23}. If we take a section $e:E\to P$ we get
\beq4.2
e^*\o=\gd A,a,A,\ov\t{}^A X_a=A_A\ov\t{}^A
\e
where $\ov\t{}^A$ is a frame on $V$ and $X_a$ are generators of the \Li~$\fh$.
\beq4.3
[X_a,X_b]=\gd C,c,ab,X_c.
\e
We define a \cvt\ of the \cn\ $\o$
\beq4.3a
\O=d\o+\frac12[\o,\o].
\e
Taking a section $e$
\bg4.4
e^*\O = \frac12 \,\gd F,a,AB,\ov\t{}^A \land \ov\t{}^B X_a =
\frac12 \,F_{AB}\ov\t{}^A \land \ov\t{}^B \\
\gd F,a,AB, = \pa_A\gd A,a,B, - \pa_B\gd A,a,A, - \gd C,a,cb, \gd A,c,A,
\gd A,c,B,. \lb4.5
\e

Let us consider a local \cd\ systems on~$V$. One has $x^A=(x^\mu,\psi,\vf)$
where $x^\mu$ are \cd\ system on~$E$, $\ov\t{}^\mu=dx^\mu$, and $\psi$
and~$\vf$ are polar and azimuthal angles on~$S^2$, $\t^5=d\psi$, $\t^6=d\vf$.
We have $A,B,C=1,2,\dots,6$, $\mu=1,2,3,4$. Let us introduce vector fields
on~$V$ corresponding to the infinitesimal action of $\SO(3)$ on~$V$
(see Ref.~\cite{21}). These
vector fields are called $\d_m=(\d^A_m)$, $\ov m=1,2,3$, $A=1,2,\dots,6$.
Moreover, they are acting only on the last two \di s ($A,B=5,6$, $\wt a,\wt
b=5,6$). We get:
\beq4.6
\bal
\d^\mu_{\bar m }&=0 &\hbox{and}&\\
\d^\psi_1&=\cos\vf, &&\d^\vf_1=-\cot\psi \sin\vf,\\
\d^\psi_2&=-\sin\vf, &&\d^\vf_2=-\cot\psi \cos\vf,\\
\d^\psi_3&=0, &&\d^\vf_3=1.
\eal
\e
They \sf y commutation relation of the \Li\ $A_1$ of a group $\SO(3)$,
\beq4.7
\d^A_{\bar m}\pa_A\d^B_{\bar n} - \d^A_{\bar n}\pa_A\d^B_{\bar m} =
\ve_{\bar m\bar n\bar p}\d^B_{\bar p}.
\e
The gauge field $A_A$ is spherically \s ic (\iv t \wrt an action of a group
$\SO(3)$) iff for some $V_{\bar m}$---a field on~$V$ with values in the
\Li~$\fh$---
\beq4.8
\pa_B\d^A_{\bar m}A_A + \d^A_{\bar m}\pa_A A_B = \pa_BV_{\bar m}
-[A_B,V_{\bar m}].
\e
It means that
\beq4.9
\mathop{\cL}_{\d_{\bar m}}A_A = \pa_BV_{\bar m} - [A_A,V_{\bar m}],
\e
a Lie \dv\ of $A_A$ \wrt $\d_{\bar m}$ results in a gauge \tf\ (see also
Eq.~\er{3.23}).

Eq.~\er{4.9} is \sf ied if
\beq4.10
V_1=\F_3\,\frac{\sin\vf}{\sin\psi}, \q
V_2=\F_3\,\frac{\cos\vf}{\sin\psi}, \q V_3=0
\e
and
\beq4.11
A_\mu=A_\mu(x), \q A_\psi=-\F_1(x)=A_5=\F_5, \q A_\vf = \F_2(x)\sin\psi
-\F_3\cos\psi=A_6=\F_6
\e
with the \fw\ constraints
\beq4.12
\bal
{}[\F_3,\F_1]&=-\F_2, \\
[\F_3,\F_2]&=\F_1, \\
[\F_3,A_\mu]&=0.
\eal
\e
$A_\mu,\F_1,\F_2$ are fields on $E$ with values in the \Li\ of~$H(\fh)$,
$\F_3$ is a \ct\ \el\ of Cartan subalgebra of~$\fh$. Let us introduce some
additional \el s according to the \E\nos\ Hermitian Kaluza--Klein Theory.
According to Section~3 we have on~$E$ a \nos\ Hermitian tensor $g_\m$, \cn s
$\gd \ov\o,\a,\b,$ and $\gd\ov W,\a,\b,$. On~$S^2$ we have a \nos\ metric
tensor
\beq4.13
\g_{\td a\td b}=r^2g_{\td a\td b}=r^2\bigl(\gd h,0,\td a\td b,
+\z \gd k,0,\td a\td b,\bigr)
\e
\setbox9=\hbox{$-\sin^2\psi$}%
\noindent where $r$ is the radius of a sphere $S^2$ and $\z$ is considered to
be pure imaginary,
\bea4.14
\gd h,0,\td a\td b,&=\left(\begin{array}{c|c}
\hbox to\wd9{\hfil $-1$\hfil} & 0 \\ \hline 0 & -\sin^2\psi \end{array}\right)\\
\gd k,0,\td a\td b,&=\left(\begin{array}{c|c}
0 & \hbox to\wd9{\hfil$\sin\psi$\hfil} \\ \hline -\sin\psi & 0 \end{array}\right) \lb4.15
\e
\setbox0=\hbox{$-\z\sin\psi$}%
\def\vp{\vrule height\ht0 depth\dp0 width0pt}%
\noindent and a \cn\ compatible with this \nos\ metric
\bg4.16
g_{\td a\td b}=
\begin{array}{c}
\noalign{\vskip-10pt}
\begin{array}{ccc}
\hbox to\wd0{\hfil$\scriptstyle 5$\hfil}&\hbox to\wd0{\hfil$\scriptstyle 6$\hfil}\end{array}\\
\begin{array}{cc}
\left(\begin{array}{c|c}
-1 & \z\sin\psi \\ \hline -\z\sin\psi & -\sin^2\psi
\end{array}\right)&
\begin{array}{c}
\scriptstyle 5\vp\\\scriptstyle 6\vp
\end{array}
\end{array}
\end{array}\\
\wt g=\det(g_{\td a\td b}) = \sin^2\psi(1+\z^2) \lb4.17 \\
g^{\td a\td b} = \frac1{\sin^2\psi(1+\z^2)}
\begin{array}{c}
\noalign{\vskip-4pt}
\begin{array}{ccc}
\hbox to\wd0{\hfil$\scriptstyle 5$\hfil}&\hbox to\wd0{\hfil$\scriptstyle 6$\hfil}\end{array}\\
\begin{array}{cc}
\left(\begin{array}{c|c}
-\sin^2\psi & -\z\sin\psi \\ \hline \z\sin\psi & -1
\end{array}\right)&
\begin{array}{c}
\scriptstyle 5\vp\\\scriptstyle 6\vp
\end{array}
\end{array}
\end{array}, \lb4.18
\e
$\wt a,\wt b=5,6$. In this way we have to do with K\"ahlerian structure
on~$S^2$ (Riemannian, symplectic and complex which are compatible). This seems to
be very interesting in further research connecting \un\ of all fundamental
\ia s. On $H$ we define a \nos\ metric
\beq4.19
\ell_{ab}=h_{ab}+\xi k_{ab}
\e
where $k_{ab}$ is a right-\iv t skew-\s ic 2-form on~$H$.

One can rewrite the constraints \er{4.12} in the form
\beq4.20
\bal
{}[\F_3,\F]&=i\F\\
[\F_3,\wt\F]&=-i\wt\F\\
[\F_3,A_\mu]&=0
\eal
\e
where $\F=\F_1+i\F_2$, $\wt\F=\F_1-i\F_2$ (see Ref.~\cite{21}).

In this way our 6-\di al gauge field (a \cn\ on a fiber bundle) has been
reduced to a 4-\di al gauge one (a~\cn\ on a fiber bundle over a \spt~$E$)
and a collection of scalar fields defined on~$E$ \sf ying some constraints.
According to our approach there is defined on~$S^2$ a \nos\ \cn\ compatible
with a \nos\ tensor $g_{\td a\td b}$, $\wt a,\wt b=5,6$,
\beq4.21
\bga
\wh Dg_{\td a\td b}=g_{\td a\td d}\gd Q,\td d,\td b\td c,(\wh\G)\ov\t{}
^{\td c}\\
\gd Q,\td d,\td b\td d,(\wt\G)=0
\ega
\e
where $\wh D$ is an exterior \ci\ \dv\ \wrt a \cn\ $\gd\wh\o,\td a,\td b,
=\gd\wh\G,\td a,\td b\td c,\ov\t{}^{\td c}$ and $\gd Q,\td d,\td b\td c,
(\wh \G)$ its torsion.

Let us metrize a bundle $P$ in a \nos\ way. On~$V$ we have \nos\ tensor (see
Ref.~\cite1)
\beq4.22
\g_{AB}=\bma g_\m & 0 \\\hline 0 & r^2g_{\td a\td b} \ema
\e
and a \nos\ \cn\ $\gd\ov\o,A,B,=\gd \G,A,BC,\t^C$ compatible with this tensor
\beq4.23
\bga
\ov D\g_{AB}=\g_{AD}\gd Q,D,BC,(\ov\G)\t^C\\
\gd Q,D,BD,(\ov\G)=0.
\ega
\e
The form of this \cn\ is as follows
\beq4.24
\gd \ov\o,A,B,=\bma \gd \ov\o,\a,\b, & 0 \\\hline 0 & \gd \wh\o,\td a,\td b,
\ema
\e
where $\ov D$ is an exterior \ci\ \dv\ \wrt $\gd \ov\o,A,B,$ and $\gd Q,D,BC,
(\ov\G)$ its torsion.

Afterwards we define on $P$ a \nos\ tensor
\beq4.25
\k_{\tA\tB}\t^\tA \ot \t^\tB = \pi^* (\g_{AB}\ov\t{}^A \ot \t^B)
+\ell_{ab}\t^a \ot \t^b
\e
where
\beq4.26
\t^\tA = (\pi^*(\ov \t{}^A),\la\o^a),
\e
$\o=\o^0X_a$ is a \cn\ defined on $P$ ($\tA,\tB,\tC=1,2,\dots,n+6$).

We define on $P$ two \cn s $\gd \o,A,B,$ and $\gd W,A,B,$ \st $\gd \o,A,B,$
is compatible with a \nos\ tensor $\k_{\tA\tB}$,
\beq4.27
\bga
D\k_{\tA\tB}=\k_{\tA\td D}\gd Q,\td D,\tB\tC,(\G)\t^\tC\\
\gd Q,\td D,\tB\td D,(\G)=0,
\ega
\e
where $D$ is an exterior \ci\ \dv\ \wrt a \cn\ $\gd \o,\tA,\tB,$ and
$\gd Q,\td D,\tB\tC,(\G)$ its torsion.

The second \cn
\beq4.28
\gd W,\tA,\tB,=\gd \o,\tA,\tB, - \frac4{3(n+4)}\,\gd \d,\tA,\tB,\ov W
\q (n=\dim H).
\e

In this way we have all quantities known from Section~3. We calculate a
scalar of \cvt\ (\MR) for a \cn\ $\gd W,\tA,\tB,$ and afterwards an action
\bml4.29
S=-\frac1{V_1V_2} \int_U \sqrt{-g}\,d^4x \int_H\sqrt{|\ell|}\,d^nx
\int_{S^2}\sqrt{|\wt g|}\,d\O\,R(W) \\
{}=-\frac1{r^2V_1V_2} \int_U \sqrt{-g}\,d^4x \int_{S^2}\sqrt{|\wt g|}\,d\O\,
\Bigl(\ov R(\ov W)\\
{}+\frac{8\pi G_N}{c^4}\Bigl(\cLY+\frac1{4\pi r^2}\,
\cL_{\rm kin}(\n\F)-\frac1{8\pi r^2}\,V(\F)-\frac1{2\pi r^2}\,
\cL_{\rm int}(\F,\wt A)\Bigr)+\la_c\Bigr)
\e
where $V_1=\int_U\sqrt{|\ell|}\,d^nx$, $V_2=\int_{S^2}\sqrt{|\wt g|}\,d\O$,
$U\subset E$,
\beq4.30
\la_c=\Bigl(\frac{\a_s^2}{\ell\pl^2}\,\wt R(\wt \G)+\frac1{r^2}\,\ul{\wt P}
\Bigr) 
\e
where $\wt R(\wt\G)$ is a \MR\ \cvt\ scalar on a group~$H$ (see Section~3 for
details).
\beq4.32
\wt {\ul P}=\frac1{V_2}\int_{S^2}\sqrt{|\wt g|}\,d\O\,\wh R(\wh \G)
\e
where $\wh R(\wh \G)$ is a \MR\ \cvt\ scalar on $S^2$ for a \cn\
$\gd\wh\o,\td a,\td b,$.
\beq4.33
\cLY=-\frac1{8\pi}\,\ell_{ij}\bigl(\wt H{}^{(i}\wt H{}^{j)}- \wt L^{i\m}
\gd\wt H,j,\m,\bigr)
\e
where
\beq4.34
\ell_{ij}g_{\mu\b}g^{\g\mu}\gd\wt L,i,\g\a, + \ell_{ji}g_{\a\mu}g^{\mu\g}
\gd \wt L,i,\b\g, = 2\ell_{ji}g_{\a\mu}g^{\mu\g}\gd\wt H,i,\b\g,
\e
One gets from \er{3.44}
\bg4.39
\gd L,b,\td a\td b,=h^{bc}\ell_{cd}\gd H,d,\td b\td a,,\\
V(\F)=-\frac1{V_2}\int\sqrt{|\wt g|}\,d\O\,\bigl(2h_{cd}(\gd H,c,\td a\td b,
g^{\td a\td b})(\gd H,d,\td c\td d,g^{\td c\td d}) - \ell_{cd}g^{\td a\td m}
g^{\td b\td n}\gd L,c,\td a\td b,\gd H,d,\td m\td n,\bigr)\hskip90pt \nn\\
\hskip90pt{}=\frac1{V_2}\,\frac{2\pi^2}{\sqrt{1+\z^2}}\,\k\bigl(
(\ve_{\bar r\bar s\bar t}\F_{\bar t}+[\F_{\bar r},\F_{\bar s}]),
(\ve_{\bar r\bar s\bar t}\F_{\bar t}+[\F_{\bar r},\F_{\bar s}])\bigr)
\lb4.35 \\
\k_{de}=(1-2\z^2)h_{de}+\xi^2\gd k,c,d,k_{ce} \lb4.36
\e
where
\bg4.37
\gd k,c,d,=h^{cf}k_{fd} \\
V_2=\int_{S^2}\sqrt{|\wt g|}\,d\O=4\pi \sqrt{1+\z^2}, \lb4.38
\e
$\ov r,\ov s,\ov t=1,2,3$, $\ve_{\bar r\bar s\bar t}$ is a usual anti\s ic
symbol $\ve_{123}=1$.

We get also from \er{3.44}
\beq4.40
\ell_{dc}g_{\mu\b}g^{\g\mu}\gd L,d,\g\td a, + \ell_{cd}\gd L,d,\b\td a,
=2\ell_{cd}\gd F,c,\b\td a,.
\e
Using Eq.~\er{3.84} one gets
\bml4.41
\gd L,n,\o\td m,=\gv{\n_\o}\Ft nm + \xi\gd k,n,d, \gv{\n_\o}\Ft dm
-\wt g{}^\(\a\mu)\gv{\n_\a}\Ft nm g_\[\mu\o]\\
{}+\xi \gd k,n,d,\gv{\n_\b}\Ft dm\wt g{}^\(\d\b)g_\[\d\a]g_\[\o\mu]
\wt g{}^\(\a\mu) - \xi^2 k^{nb}k_{bd}\wt g{}^\(\a\b)\gv\n \Ft dmg_\[\o \b].
\e
Moreover, now we have to do with Minkowski space $g_\m=\eta_\m$ and
\beq4.42
\gd L,n,\o\td m,=\gd H,n,\o\td m,+\xi \gd k,n,d,\gd H,d,\o\td m,.
\e
We remember that $\wt m=5,6$ or $\vf,\psi$ and that
\beq4.43
\gd H,n,\mu\td m,=\gv{\n_\mu}\Ft nm.
\e
We have
\beq4.44
\cL_{\rm kin}(\gd H,n,\mu\td m,)=\frac1{V_2} \int \sqrt{|\wt g|}\,d\O\,
(\ell_{ab}\eta^{\b\mu}\gd L,a,\b\td b,\gd H,b,\mu\td a,g^{\td b\td a}).
\e
Finally we get
\bg4.45
\cL_{\rm kin}(\n_\mu\F_{\bar m})=\frac{2\pi^2}{V_2}\,\frac{\eta^\m}
{\sqrt{1+\z^2}}\,\bar\k\bigl(\gv{\n_\mu}\F_{\bar m},\gv{\n_\nu}\F_{\bar m}
\bigr)\\
\ov\k_{ad}=(h_{ad}+\xi^2k_{ab}\gd k,b,d,) \lb4.46
\e
where
\beq4.47
\gv{\n_\mu}\F_{\bar m}=\pa_\mu\gd \F,a,\bar m, - [A_\mu,\F_{\bar m}].
\e

Now we follow Ref.~\cite{21} and suppose $\mathop{\rm rank}H=2$ and
afterwards $H=G2$. In this way our lagrangian can go to the GSW model where
$\SU(2) \tm \U(1)$ is a little group of~$\F_3$ (see Appendix~B).
We get also a Higgs' field
complex doublet and \sn\ \s y breaking and mass generation for intermediate
bosons. For simplicity we take $\xi=0$ and also we do not consider an
influence of the \nos\ gravity on a Higgs' field.
We get also a mixing angle $\t_W$ (Weinberg angle). If we choose
$H=G2$ we get $\t_W=30^\circ$. We get also some predictions of masses
\beq4.49
\frac{M_H}{M_W} = \frac1{\cos\t_W}\cdot \sqrt{1-2\z^2}
\e
where $\z$ is an arbitrary \ct
\beq4.50
\frac{M_H}{M_W} = \frac{2\sqrt{1-2\z^2}}{\sqrt3}.
\e
We take $M_H\simeq125$\,GeV and $M_W\simeq80$\,GeV (see Refs
\cite{26,27,28,29,30}).

One gets
\beq4.51
\z=\pm 0.911622i.
\e
Thus $\z$ is pure imaginary.
This means we can explain mass pattern in GSW model. $r$ gives us a scale of
mass and is an arbitrary parameter.

Moreover, a scale of energy is equal to
$M=\frac{\hbar c}{r\sqrt{2\pi}\,\sqrt{1+\z^2}}$
which we equal to MEW (electro-weak) energy scale, i.e.\ to~$M_W$. One gets
$r\simeq 2.39\tm 10^{-18}$\,m. In the original Manton model Higgs' boson is too
light. We predict here masses for $W,Z^0$ and Higgs bosons in the theory taking
two parameters, $\z$ (Eq.~\er{4.51}) and $r\simeq 2.39\tm 10^{-18}$\,m in order
to get desired pattern of masses. The value of the Weinberg angle derived here
for $H=G2$ has nothing to do with ``GUT driven'' value $\frac14$ for $\frac14$
is a value of our $\sin^2\t_W$, not $\sin\t_W$.
According to Ref.~\cite{21} a Lie group $H$ should have a Lie algebra $\fh$
with rank~2. We have only three possibilities: G2, $\SU(3)$ and $\SO(5)$.
The angle between two roots plays a role of a Weinberg angle. For $\SO(5)$
$\t=45^\circ$ and for $\SU(3)$ $\t=60^\circ$. Only for $G2$, $\t=\t_W
=30^\circ$, which is close to the experimental value. In this way a \un\
chooses $H=G2$.

Let us notice that $\dim G2=14$ and for this $\dim P=20$.

Moreover, we have
\beq4.52
M_Z = \frac{M_W}{\cos\t} = \frac{M_W}{\cos\t_W} = \frac2{\sqrt3}\,M_W
\simeq 92.4
\e
and we get from the theory
\beq4.53
\sin^2\t_W=0.25 \q (\t_W=30^\circ).
\e
However from the experiment we  get
\beq4.54
\sin^2\t_W=0.2397\pm0.0013
\e
which is not $0.25$.

Moreover, from theoretical point of view the value $0.25$ is a value without
radiation corrections and it is possible to tune it at $Q=91.2$\,GeV/c in the
$\ov{\rm MS}$ scheme to get the desired value.

Let us notice the following fact. In the electroweak theory we have a
Lagrangian for neutral current \ia
\bml4.55
\cL_N = q J^{\rm em}_\mu A^\mu + \frac g{\cos\t_W}(J^3_\mu - \sin^2\t_W
J^{\rm em}_\mu)Z^{0\mu}
=qJ^{\rm em}_\mu A^\mu +\sum_f \ov\psi_f \g_\mu(g^f_V-g^f_A\g^5)\psi_f
Z^{0\mu}
\e
where $g^f_V$ and $g^f_A$ are coupling \ct s for vector and axial \ia s for a
fermion~$f$. One gets
\beq4.56
\bal
g^f_V&=\frac{2q}{\sin2\t_W}(T^3_f - 2q_f\sin^2\t_W)\\
g^f_A&=\frac{2q}{\sin2\t_W}
\eal
\e
where $T^3_f$ is the third component of a weak isospin of a fermion~$f$ and
$q_f$ is its electric charge measured in \el ary charge~$q$,
\beq4.57
q_f=T^3_f+\frac{Y_f}2
\e
where $Y_f$ is a weak hypercharge for~$f$. It is easy to see that for an
electron we get $g^f_V=0$ if $\t_W=30^\circ$.

Moreover, we know from the experiment that
\beq4.58
g^f_V\ne0
\e
(see Ref.\ \cite{26}).

In the original GSW model a Weinberg angle $\t_W$ is a
phenomenological parameter which has no geometrical interpretation in terms
of Lie algebraic theory. Here this parameter has this interpretation.
Moreover, this theory is still classical. How we can quantize it in a general
case including \nos\ gravity we describe in \ti{Conclusions and prospects
of further research}. Some quantum corrections can change many things going
effectively to Eq.~\er{4.58}. Moreover, in Minkowski space $g_\m=\eta_\m$
and with $\xi=0$ the
situation is much more simple and we can agree that radiative correction can
go to Eq.~\er{4.58} which is interpreted as a correction to $\sin\t_W$
for example in $\ov{\rm MS}$ scheme at $Q=91.2$\,GeV/c. This means that even
if $g^f_V(\t_W=30^\circ)=0$ the corrections change $g^f_V$ to be nonzero in
such a way that $\t_W$ is not exactly equal to $30^\circ$. Moreover, the \un\
scheme with $H=G2$ is still valid.

\def\Im{\mathop{\rm Im}}
Let us define a differential cross-section for $f^+f^- \to f^{\prime+}
f^{\prime-}$ scattering
\beq1
\frac{d\si}{dt}\bigl(f^-(P)f^+ \to f^{\prime-}(P')f^+\bigr)
= \frac{4\pi\a_{\rm em}}s \,\k^2_{PP'} \bigl|\cM_{PP'}(-s)\bigr|^2
\e
where $\k^2_{PP'}$ is a kinematic factor from the Dirac algebra equal to
$(\frac us)^2$ for $L{\rm(left)} \to L{\rm(left)}$ and $R{\rm(right)} \to
R{\rm(right)}$ and  to $(\frac ts)^2$ for $L{\rm(left)}\to R{\rm(right)}$ and
vice versa. At $Z^0$ mass energies we can ignore mass of fermion~$f$
and~$f'$ ($m_{Z^0}> 2m_f$ and $m_{Z^0}>2m_{f'}$).

In this way the helicity is conserved. $\cM_{PP'}$~is an invariant
amplitude which contains all nontrivial information about a coupling. It is
defined in such a way that $\cM_{PP'}$~is equal to~1 independently of $P,P'$
for a simple $s$-channel photon exchange diagram of lowest order QED for
electrons. In GSW theory one gets
\bml2
\cM_{PP'}(Q^2) = q_f\biggl(\frac{-s}{Q^2}\biggr) q_{f'}\\ {}+
\biggl(\frac{T_f^3 - q_f\sin^2\t_W}{\cos\t_W \sin\t_W}\biggr)
\biggl(\frac{-s}{Q^2+M_{Z^0}^2 - \Im(\Pi^{\text{1-loop}}_{Z^0Z^0}(Q^2))}\biggr)
\biggl(\frac{T^3_{f'}-q_{f'}\sin^2\t_W}{\cos\t_W\sin\t_W}\biggr)
\e
where
\bml3
\Im(\Pi^{\text{1-loop}}_{Z^0Z^0})=\G^0_{Z^0}M_{Z^0}=
\frac{\a_{\rm em}}{3\sin^2\t_W\cos^2\t_W } \sum_f\biggl[\biggl[
\biggl(\frac{T_{fL}^3}2 -q_f\sin^2\t_W\biggr)\biggl(1+\frac{2m_f^2}{M^2_{Z^0}}\biggr)\\
+\biggl(\frac{T^3_{fL}}2\biggr)^2\biggl(1-4\,\frac{m_f^2}{M^2_{Z^0}}\biggr)\biggr]
\biggl(1-\frac{4m_f^2}{M^2_{Z^0}}\biggr)C_{QCD}(f)\biggr],
\e
where $T_{fL}^3$ is a left-handed isospin component for a fermion~$f$. The
factor $C_{QCD}$ is $C_{QCD}=3\bigl(1+\a_s(-M^2_{Z^0})/\pi\bigr)$ for quarks
and $1$~for leptons. These formulas are very well known in all textbooks
and as we mention above $\t_W$~is an arbitrary parameter. If we evaluate
the formulas for $\t_W=\frac\pi 6$ one gets
\beq4
\cM_{PP'}(Q^2) = q_f\biggl(\frac{-s}{Q^2}\biggr)q_{f'}
\e
which simply means that $g^f_V=0$ ($\sin^2\t_W=0.25$).

Moreover, we can introduce an effective Weinberg angle $\t_W=\frac\pi6+\delta$
in such a way that all the formulas are satisfied. In this way radiative
corrections can be considered as corrections to $30^\circ$ Weinberg angle.
The formula \eqref2 can be evaluated in the following way:
\beq5
\cM_{PP'}(Q^2) = q_f\biggl(-\frac{s}{Q^2} + 4\delta^2\biggl(-\frac s
{(Q^2+M^2_{Z^0Z^0}-\Im(\Pi^{\text{1-loop}}_{Z^0Z^0}(Q^2)))}\biggr)
\biggr)q_{f'}
\e
($\delta$ is a small correction to $\t_W=\frac\pi6$).

One can use also some achievements from GSW model. Let us notice that
\beq4.63
M_W^2=\frac{\pi \a_{\rm em}(0)G_F}{\sqrt2 \sin^2\t_W (1-\D r)}
\e
where $\D r$ is the 1-loop correction and its dominant contributions are
\bea4.64
\D r&= \D r_0 - \frac{1-\sin^2\t_W}{\sin^2\t_W}\,\D\rho +\D r_{\rm rem}\\
\D{r_0} &=1- \frac{\a_{\rm em}(0)}{\a_{\rm em}(M_{Z^0}^2)} \lb4.65 \\
\D\rho &= \frac{3G_F \sum_f(m_{f_1}^2-m_{f_2}^2)}{8\pi^2\sqrt2} \simeq
\frac{3G_F(m_t^2-m_b)^2}{8\pi^2\sqrt2} \lb4.66 \\
\D_{\rm rem}&= \frac{\sqrt2\,G_FM_W^2}{16\pi^2}\cdot\frac{11}3 \Bigl(
\ln\Bigl(\frac{M_H^2}{M_W^2}\Bigr)-\frac56\Bigr), \lb4.67
\e
$G_F$ is a Fermi \ct, $m_t$ and $m_b$ are top and bottom quark masses.

The term $\D r_0$ corresponds to the running of $\a_{\rm em}$ from zero (it
means, from $Q^2=m_e^2\simeq0$) to the electroweak scale $Q^2=M^2_{Z^0}$.
$\D\rho$ depends quadratically on the mass difference between the members
of the same fermion doublet. $\D r_{\rm rem}$ (the~remainder) is dominated
by Higgs' boson effects and depends logarithmically on~$M_H$.

We evaluate these formulas for $\t_W=\frac\pi6$ getting
\beq4.68
M_W^2 = \frac{4\pi \a_{\rm em}G_F}{\sqrt2(1-\D r)}\,.
\e
Now we proceed as before writing
\beq4.69
M_W^2 = \frac{4\pi \a_{\rm em}G_F}{\sqrt2\sin^2\t_W}
\e
where
$$
\t_W = \frac\pi 6+\d \qh{(as above).}
$$
$\d$ is not a new phenomenological parameter. It is an effect of 1-loop
corrections and a running of~$\a_{\rm em}$. In Eq.~\er{4.64} we can write
$\t_W=\frac\pi6$ getting
\beq4.70
\D r=\D r_0-3\D\rho+\D r_{\rm rem}.
\e
In the formula \er{4.67} we can put the value of Higgs' mass and bare value
of~$M_W$ obtained by us. In this way we get the desired value of $\sin^2\t_W$
\beq4.71
\sin^2\t_W=4(1-\D r).
\e

In terms of $\d$ one gets
\beq4.72
\d=-\frac{\sqrt3 \,\D r}6
\e
or
\beq4.73
\d=-\frac{\sqrt3}6 \Bigl(1-\frac{\a_{\rm em}(0)}{\a_{\rm em}(M^2_{Z^0})}
-3\D \rho+\D r_{\rm rem}\Bigr).
\e
We get exactly the same results if we use the $\ov{\rm MS}$ definition
of $\sin^2\t_W=1-\frac{M_W^2}{M_{Z^0}^2}$ which is also an effective value
of $\sin^2\t_W$.

Using results from Ref.~\cite{26} we can evaluate $\d$ from the formula
\er{4.73} getting
\beq4.74
\d=-0.00748550.
\e
We have
$$
\d=-25'44'' \qh{and } \t_W=29^\circ 34'16''.
$$
This gives
\beq4.75
\sin^2\t_W = 0.243546
\e
which is not bad as compared with the experimental value for an effective
Weinberg angle. The formula \er{4.73} can be improved getting
\beq4.76
\d=-\frac{\sqrt3}6 \Bigl(1-\frac{\a_{\rm em}(0)}{\a_{\rm em}(M^2_{Z^0})}
-3\D\rho +\D r_{\rm rem}\Bigr)/(1-4\D\rho).
\e
We get
\bg4.77
\d=-0.0074855\\
\sin^2\t_W = 0.249162 \lb{4.78}
\e

More precise quantum field calculations can improve the result. The
conclusion is as follows. The Weinberg angle is coming from the \un\
theory with $G2$ group. The value of this parameter is equal to~$\frac\pi6$.
The $\d$ correction is coming from radiative corrections.

Now we have a physical relevance and correct description of the Nature. The
results can be improved starting from the formula
\beq4.79
M_W^2\Bigl(1-\frac{M_W^2}{M_{Z^0}^2}\Bigr) = \frac{\pi \a_{\rm em}(0)}{\sqrt2\,
G_F}\,(1+\D r)
\e
using results from Refs \cite{x2, y2, z2} and references cited therein.
The numerical results obtained here do not change significantly the full
quantization scheme. Eventually we get some remarks.

We have here to do with a finite renormalization of a parameter in the theory,
i.e.\ with a finite renormalization of a Weinberg angle. According to the
idea of a renormalization of any parameter due to quantum \ia s this is
correct. We should renormalize not only masses or charges (as in QED, an
electron charge and its mass, which is an infinite  renormalization), but
really any physical quantity as in solid state physics (an effective mass of
an electron). An infinite renormalization in QED showed us an impossibility
to avoid a renormalization in general.

The second remark is as follows. In classical field theory as our model for
$g_\m=\eta_\m$ we have to do with parameters which have an interpretation as
tree values. They should be renormalized. Only in a superrenormalizable
theory they can remain the same in any order of perturbation calculus. Our
theory is not superrenormalizable.

In our approach on a classical level we have the following parameters:
$r_0,\z$ ($\sin\t_W$ is known from the theory). In order to get a precise
prediction we should translate them into~$G_F$ and $\a_{\rm em}(0)$, in
particular $G_F=G_{\mu}$. We take from Particle Data (see Ref.~\cite{26})
all the \ia\ \ct s (coupling parameters) which can change running parameters.

Let us use the above results to recalculate $\z$ and $r_0$ in terms of $M_H$
and $M_{Z^0}$. One gets
\bea4.80
M_W &= M_{Z^0}\cos\t_W \\
\frac{M_H}{M_{Z^0}}&= \sqrt{1-\z^2} \lb4.81
\e
and
\bg4.82
\z=\pm 0.948735i \\
r_0=\frac{\hbar c}{M_{Z^0} \sqrt{2\pi} \sqrt{1+\z^2}} =
2.73126 \tm 10^{-18}\,{\rm m}. \lb4.83
\e
In this way
\beq4.81a
M_W=M_{Z^0}\cos(\tfrac\pi6+\d)
\e
gives us for the value \er{4.75}
\beq4.84
M_W=79.3119\,{\rm GeV},
\e
for the value \er{4.78}
\beq4.85
M_W=79.3321\,{\rm GeV}.
\e
In the above formulas we take for $M_H=125.7\,$GeV and for $M_{Z^0}=91.19
\,$GeV. This means that for $\z$ given by \er{4.82} and $r_0$ given by
\er{4.83} we get desired values of $M_H$ and $M_{Z^0}$. The predicted value
for $M_W$ is a little smaller than the experimental value 80.385\,GeV.
Moreover, it seems that consideration of higher order corrections of
perturbation calculus ($2$-loop corrections) can improve the result to tune
it to the experimental value. This has been done in Appendix~E. It seems
that everything is self-consistent. The value of $M_W$ and $\sin^2\t_W$
obtained in Appendix~E indicates that higher order corrections improve an
agreement with an experiment. This means that our 20-\di al model works
pretty well. The Weinberg angle is not here a phenomenological parameter and
we have a confiance that a \un\ group~$H$ is~$G2$. This our future
development is justified by an experiment.

\section*{Appendix A}
\def\theequation{A.\arabic{equation}}
\setcounter{equation}0

In this appendix we find formulae for $\gd L,n,\o\nu,$, $\gd L,n,\td w\td
n,$, $\gd L,n,\o\td n,$. We get these formulae using a general formula from
$n$-\di al \gn\ of Einstein Unified Field Theory obtained by Hlavat\'y and
Wrede (see Refs \cite{16,32}). One gets
\bg aaa
\gd\G,N,WM,=\gd\wt\G{},N,WM,+\tfrac12\bigl(\dg K,WM,N, -2\dg k,[M\cdot,A,
K_{W]AB}k^{NB}\bigr)\hskip80pt \nn\\
\hskip20pt {} +h^{NE}\bigl\{\dg K,E(W\cdot,A, k_{M)A}+
\dg k,C\cdot,B, \bigl[\dg k,(M\cdot,C, K_{W)AB}\dg k,E\cdot,A,
+K_{EAB}\dg k,(W\cdot,A, \dg k,M)\cdot,C,\bigr]\bigr\} \lb A.1 \\
K_{ABC}=-\wt\nabla_A k_{BC}-\wt\nabla_Bk_{CA}+\wt\nabla_C k_{AB},\lb A.2
\e
where
\bea A.3
\g_{AB}&=h_{AB}+k_{AB}, \\
h_{AB}&=h_{BA}, \q k_{AB}=-k_{BA}, \lb A.4
\e
$\gd \wt\G,N,WM,$ is the \LC \cn\ generated by $h_{AB}=\g_\(AB)$ ($\g_\[AB]
=k_{AB}$), $\wt\n_A$ is a \ci\ \dv\ \wrt the \cn~$\gd \wt\G,N,WM,$.

The \cn\ $\gd \G,N,WM,$ is the \so\ of the \e
\beq A.5
\bga
D\g_{A+B-} = D\g_{AB} - \g_{AD}\gd Q,D,BC,(\G)\t^C=0, \q
A,B,C,D,N,M=1,2,\dots,\ov N,\\
\gd Q,D,BD,(\G)=0,
\ega
\e
where $D$ is an exterior \ci\ \dv\ \wrt the \cn\ $\G$.
\beq A.6
h^{AB}h_{BC}=\gd \d,A,C,
\e
and all indices are raised by $h^{AB}$. (E.~Schr\"odinger was surprized that
it was possible to find a \so\ to~\er{A.5} in a \ci\ form.)

\E\e\ \er{A.1} is more general than that form Refs \cite{16,32} for in
Eq.~\er{A.1} $\gd \wt\G,N,WM,$ are \cf s of the \LC \cn. This \cn\ can be
\nos\ in indices $W,M$ for it can be considered in nonholonomic frame. In
Refs \cite{16,32} $\gd \wt\G,N,WM,$ mean Christoffel symbols.

Moreover, the proof is exactly the same as in Refs \cite{16,32}. The authors
of Refs \cite{16,32} are using the natural nonholonomic frame connected to
the \nos\ metric $\g_{AB}$ in order to find \er{A.1}. Moreover, this
nonholonomic frame has nothing to do with the frame we consider.

V. Hlavat\'y and C. R. Wrede were first to consider $n$-\di al \gn\ of the
geometry from Einstein Unified Field Theory with the \nos\ real tensor
$\g_{AB}$. Thus we can find $\gd L,a,\m,$ from the \nos\ \nA\ Kaluza--Klein
theory where $\ov N=n+4$ (see Section~2). We can also consider \nA\ theory
with a \sn\ \s y breaking where $\ov N=4+n+n_1=4+m$ (see Section~3).

In order to find $\gd L,a,\m,$ we should calculate $\gd \G,n,\o\mu,= \gd
L,n,\o\mu,$. We should know a \LC \cn\ generated by $\g_\(AB)$
(and~$\k_\(\tA\tB)$) which is easy to find from Eqs \er{2.19} and \er{3.43}
(e.g.\ $\gd \wt\G,n,\o\mu,=\gd H,n,\o\mu,$) in order to find \ci\ \dv\ of
anti\s ic part of the metric. Thus one eventually finds:
\begin{gather}
\gd L,n,\o\nu,=\gd H,n,\o\nu, + \xi\gd H,f,\nu\o, k_{fe}h^{ne}
+\bigl(\gd H,n,\a\o,\wt g{}^\(\a\d)g_\[\d\nu]
-\gd H,n,\a\nu,\wt g{}^\(\a\d)g_\[\d\o]\bigr)\nn\\
{}-2\xi h^{na}k_{ad}\wt g{}^\(\d\tau)\wt g{}^\(\a\b)\gd H,d,\d\a,g_\[\tau\o]g_\[\b\nu]
-2\xi h^{na}k_{ad}\wt g{}^\(\b\d)\wt g{}^\(\a\tau)\gd H,d,\b[\o,g_{\nu]\tau}g_\[\d\a]\nn\\
{}+2\xi^2 h^{na}h^{bc}k_{ac}k_{bd}\wt g{}^\(\a\b)\gd H,d,\a[\o,g_{|\mu|\b]}\lb A.7
\end{gather}
(One can try to get formula \er{A.7} using a different approach. It means to use an
approximation formula from Ref.~\cite{X}. This is similar to our second approach
from Ref.~\cite8 in the case of an \elm c field (see Appendix~B of Ref.~\cite8).
Moreover, this approach in the
case of the \NK\ seems to be much more complex.)
\begin{gather}
\gd L,n,\td w\td n,=\gd H,n,\td w\td n, +\xi \gd H,f,\td n\td w,k_{fe}h^{ne}
+\xi^2\bigl(g_\[\td w\td b]\wt g{}^\(\td b\td a)\gd H,f,\td n\td a,
-g_\[\td n\td b]\td g{}^\(\td b\td a)\gd H,f,\td w\td a,\bigr)
k_{fb}k_{cd}h^{cn}h^{db}\hskip40pt \nn \\
\hskip50pt {}=\gd H,n,\td w\td n,+\xi \gd H,f,\td n\td w,k_{ef}h^{ne}
+\xi^2\z\bigl(\gd k,0,\td w\td b,\wt g{}^\(\td b\td a)\gd H,f,\td n\td a,
-\gd k,0,\td n\td b,\wt g{}^\(\td b\td a)\gd H,f,\td w\td a,\bigr)
k_{fb}k_{cd}h^{cn}h^{db} \lb A.8 \\
\gd L,n,\o\td n,=\gd H,n,\o\td n, - \xi\gd H,f,\o\td n,k_{fe}h^{ne}
-\xi^2\bigl(g_\[\o\b]\wt g{}^\(\b\a)\gd H,f,\a\td n,
+\z \gd k,0,\td n\td b,\wt g{}^\(\td b\td a)\gd H,f,\o\td a,\bigr)
k_{fb}k_{cd}h^{cn}h^{db} \hskip40pt \nn \\
\hskip20pt {}=\gv{\n_\o}\Ft nn - \xi\gv{\n_\o}\Ft fnk_{fe}h^{ne}
-\xi^2\bigl(g_\[\o\b]\wt g{}^\(\b\a)\gv{\n_a}\Ft fn
+ \z \gd k,0,\td n\td b,\wt g^\(\td b\td a)\gv{\n_\o}\Ft fa\bigr)
k_{fb}k_{cd}h^{cn}h^{db} \lb A.9
\end{gather}
where $\gv\n$ means a gauge \dv\ \wrt a \cn~$\o$ on~$E$,
\beq A.10
\gd H,c,\td a\td b,=\gd C,c,ab,\Ft aa\Ft bb - \gd \mu,c,\hi,\gd f,\hi,\td
a\td b, - \Ft cd\gd f,\td d,\td a\td b,.
\e
Working in the same way we get Einstein--Kaufmann \cn s on a group~$G$ and on
a homogeneous manifold $M=G/G_0$. This is important to find \co ical terms in
the theory.

We have \LC \cn s
\beq A.10a
\gd \wt\o,A,B, = \bma
\pi^*(\gd \td \o,\a,\b,) - h_{db}\wt g{}^\(\mu\a)\gd H,d,\mu\b,\t^b &
\gd H,a,\b\g,\t^\g \\ \hline
h_{bd}\wt g{}^\(\a\b)\gd H,d,\g\b,\t^\g & \gd \wt \o,a,b,(G) \ema
\e
where $\gd \wt\o,a,b,(G)$ is a \LC \cn\ on~$G$, $\gd\wt\o,\a,\b,$ is defined
on~$E$ and $\gd\wt\o,A,B,$ is defined on~$\ul P$, $A,B=1,2,\dots,n+4$;
\beq A.11
\gd \wt\o,\tA,\tB, = \bma
\pi^*(\gd \td \o,A,B,) - h_{db}\wt \g{}^\(MA)\gd H,d,MB,\t^b &
\gd H,a,BC,\t^C \\ \hline
h_{bd}\wt \g{}^\(AB)\gd H,d,CB,\t^C & \gd \wt \o,a,b,(H) \ema
\e
$\gd \wt\o,a,b,(H)$ is a \LC \cn\ on a group $H$,
\beq A.12
\wt\g{}^\(AB)\g_\(AC)=\gd\d,B,C,.
\e
Now $\gd \wt\o,A,B,$ is defined on $V=E\tm G/G_0$ and $\gd\wt\o,\tA,\tB,$
on~$\ul P$, $A,B=1,2,\dots,n_1+4$, $\wt A,\wt B=1,2,\dots,n_1+n+4$.

Using \er{A.10a} and \er{A.11} one can easily calculate \ci\ \dv s $\wt\n_A
k_{BC}$ or $\wt\n_\tA k_{\tB\tC}$ and afterwards $K_{ABC}$ or $K_{\tA \tB
\tC}$ in order to find desired \cn\ \cf s of $\gd \G,N,WM,$ and $\gd \G,\td
N,\td W\td M,$.

Let us do it for \er{A.10a}. One gets
\beq A.13
\bal
\gd \wt\G,d,\b\g,&=\gd H,d,\b\g,\\
\gd \wt\G,\b,\g b,&=-h_{db}h^{\a\b}\gd H,d,\a\g,\\
\gd \wt\G,\mu,a\g,&=h_{ad}h^{\a\b}\gd H,d,\g\b,\\
\gd \wt\G,a,\a c,&=\gd \wt\G,\d,ac,=\gd \wt\G,b,c\b,=0
\eal
\e
(see also Ref.~\cite{33}).

Using \er{A.1} one gets
\bml A.14
\gd \G,n,\o\mu, = \gd \wt\G,n,\o\mu, + \frac12\,\Bigl(\dg K,\o\mu\cdot,n,
-2\dg K,[\mu\cdot,\a,K_{\o]\a b}k^{nb} - 2\dg K,[\mu\cdot,\a,K_{\o]ab}
k^{nb} + h^{ne}\Bigl(\dg k,e(\o\cdot,a, k_{\mu)a}\\
{}+\dg k,\g\cdot,\b, \bigl[\dg k,(\mu\cdot,\g,K_{\o)a\b}\dg k,e\cdot,a,
-K_{e\a\b}\dg k,(\o\cdot,\a,\dg k,\mu)\cdot,\g,\bigr]\Bigr)\Bigr).
\e
Moreover, we have
\bg A.15
K_{\o\mu e}=-\wt \n_\o k_{\mu e}-\wt \n_\mu k_{e\o}-\wt \n_e k_{\o\mu}
=2\gd H,f,\mu\o,k_{fe}\\
K_{\o ab}=-\wt \n_\o k_{ab}-\wt \n_a k_{b\o}-\wt \n_b k_{\o a}=0. \lb A.16
\e

In all the formulae we keep original notation from Refs \cite{16,32} and
after all calculations we switch to our notation.
\beq A.16a
h^{\a\b} \to \wt g{}^\(\a\b), \q k_{ab} \to \xi k_{ab}.
\e
In this way
\beq A.17
K_{\o\mu e} = 2\xi \gd H,f,\mu\o,k_{fe}
\e
and we get Eq.~\er{A.7}.

Using \er{A.13} and switching according to \er{A.16a} one gets
$$
\bal
\wt\n_\o k_{\mu e}&=-h_{ed}\wt g{}^\(\nu\b)\gd H,d,\o\b, g_\[\m]
-\xi \gd H,m,\mu\o,k_{me}\\
\wt\n_\mu k_{e\o}&=-h_{ed}\wt g{}^\(\ve\g)\gd H,d,\g\mu, g_\[\ve\o]
-\xi \gd H,f,\mu\o,k_{me}\\
\wt\n_e k_{\o\mu}&=-h_{ed}\wt g{}^\(\nu\b)\gd H,d,\o\b,g_\[\m]
-h_{ed}\wt g^\(\nu\b)\gd H,d,\mu\b,g_\[\o\nu]\\
\wt\n_\o k_{ab}&=\wt\n_a k_{b\o}= \wt\n_b k_{\o a}=0.
\eal
$$
We quote these formulae for a convenience of a reader.

Working similarly we get \er{A.8} and \er{A.9}.

Let us come back to \co ical terms and calculate a \cn~\er{A.1} on~$G$ and
$G/G_0$. On a group~$G$ a right-\iv t Einstein--Kaufmann \cn\ reads
\bml A.18
\gd \G,n,wm, = -\frac12\,\gd C,n,wm, + \frac12\bigl(
\dg K,wm\cdot,n, - 2\mu^2 \dg k,[m\cdot,a, K_{w]ab}k^{nb}\bigr)\\
{}+h^{ne}\Bigl\{\mu \dg K,e(w\cdot,a,k_{m)a} + \mu^2\dg k,c\cdot,b,
\bigl[\dg k,(m\cdot,c, K_{w)ab}\dg k,e\cdot,a, - K_{eab}\dg k,(w\cdot,a,
\dg k,m)\cdot,c,\bigr]\Bigr\}.
\e
where
\beq A.19
K_{abc}=-\mu\bigl(\wt\n_a k_{bc}-\wt\n_b k_{ca} + \wt\n_c k_{ab}\bigr) \q
(\ell_{ab}=h_{ab}+\mu k_{ab}).
\e
$\wt\n_a$ means a Riemannian \ci\ \dv\ on a semisimple Lie group~$G$ \wrt a
bi\iv t Killing tensor $h_{ab}$.

One gets
\beq A.20
\wt\n k_{bc} = -\frac12(\gd C,f,bc,k_{fe}+\gd C,f,ec,k_{bf})
\e
and
\beq A.21
K_{abc}=\mu(\gd C,f,ba,k_{fc} + \gd C,f,ac,k_{fb} - \gd C,f,bc,k_{fa}).
\e
If we write a \cn\ on $\G$ in the form
\beq A.22
\gd \G,n,wm, = -\frac12\,\gd C,n,wm,+\gd u,n,wm,
\e
one gets for a \MR\ tensor on $G$
\beq A.23
R_{bd} = \wt R{}_{bd} + \wt\n_a \gd u,a,bd, - \wt\n_d \gd u,a,ba,
+\frac12\bigl(\wt\n_b \gd u,a,ad, - \wt\n_d \gd u,a,ab,\bigr)
\e
where
\bg A.24
\wt\n_a \gd u,c,ed, = -\frac12\bigl(\gd C,f,ea,\gd u,c,fd, + \gd C,f,da,
\gd u,c,ef, - \gd C,c,fa,\gd u,f,ed,\bigr)\\
\gd u,n,wm, = -\frac12\,\mu \bigl(\dg L,wm,n, - 2\mu \dg k,[m\cdot,a,
L_{w]ab} k^{nb}\bigr) \hskip100pt  \nn\\
\hskip80pt {}-\mu^2 h^{ne}\bigl(\dg L,e(w\cdot,a,k_{m)a} + \mu^2 \dg k,c\cdot,b,
\bigl[\dg k,(m\cdot,e, L_{w)ab}\dg k,e,a, - L_{eb}\dg k,(w\cdot,a,
\dg k,m)\cdot,c,\bigr]\bigr) \lb A.25
\e
where
\beq A.26
L_{abc} = \gd C,f,ba,k_{fc} + \gd C,f,ac,k_{fb} - \gd C,f,bc,k_{fa}.
\e

Eventually we find
\beq A.27
R_{bd} = \wt R_{bd} - \frac12\,\gd C,f,db,(\gd u,a,af,+\gd u,a,fa,)
-\frac14\,\gd C,a,fd,(2\gd u,f,ba,+\gd u,f,ab,)
+\frac14(\gd C,a,fb,\gd u,f,ad, - \gd C,f,bd,\gd u,a,fa,).
\e
$\wt R_{bd}$ is a \MR\ (equals to Ricci tensor) for a \LC \cn\ on~$G$
generated by $h_{ab}$
\beq A.28
\wt R_{bd} = -\frac14\,h_{bd}.
\e
Moreover, if
\beq A.29
k_{ab} = \gd C,f,ab,V_f
\e
where
\beq A.30
\wt\n_k V_f= -\frac12\,\gd C,e,fk,V_e
\e
we get
\bml A.31
\gd u,n,wm, = \frac\mu2 \,\gd C,sn,w,\gd C,p,ms,V_p
-\mu^3 \gd C,p,as,V_p C^{rnb} V_r \gdg C,q,[m\cdot,a, V_{|q|}\gd C,s,|b|w],\\
{}+\mu^4 \gdg C,f,c,b, V_f \bigl[ \gdg C,p,(m\cdot,c, V_{|p|} \gd C,s,|b|w),
\gd C,q,as,V_q C^{rna}V_r - \gdg C,s,b,n, \gd C,p,as,V_p
\gdg C,q,(w,a, V_q\gd C,r,m),V_r\bigr].
\e
$\gd u,n,wm,$ can be calculated explicitly in a general form. One gets
\bml A.32
\gd u,n,wm, = \frac12\,\mu \bigl(\gd C,f,mw,\dg k,f,n,
+\gdg C,f,\cdot w\cdot,n, k_{fm} - \gdg C,f,m,n, k_{fw}\bigr)\\
{}-\frac12\,\mu^2 \Bigl[\gd C,f,bw,k_{fa}(k^{na}\dg k,m,b, - k^{nb}\dg k,m,a,)
+\gd C,f,mb,k_{fa}(k^{na}\dg k,w,b,-k^{nb}\dg k,w,a,)
-2k^{nb}\dg k,m,a,\gd C,f,ab,k_{fw}\\
{}-\bigl(\dg k,f,a,k_{mc}\gdg C,f,w,n, + 2k_{fm}C^{anf}k_{wa}
-\gdg C,f,w,n, \dg k,f,a,k_{ma} + \gdg C,f,m,n, \dg k,f,a,k_{wa}\bigr)\Bigr]\\
{}+\frac12\,\mu^4\Bigl[3\dg k,c,b,\gd C,f,ab,\dg k,f,n, \dg k,w,a, \dg k,m,b,
+\gd C,f,bw,k_{fa}\dg k,m,c,(\dg k,c,a,k^{nb} - \dg k,c,b,k^{na})
+\gd C,f,bm,k_{fa}\dg k,w,c,(\dg k,c,a,k^{nb} - \dg k,c,b,k^{na})\\
{}+\gd C,f,ab,\dg k,c,b, \dg k,w,c, \gd k,a,m,\gd k,n,f,
+\gd C,fn,b,k_{fa}\dg k,m,c, (\dg k,c,a,\dg k,w,b, - \dg k,b,c,\dg k,w,a,)
+\gd C,an,f,\dg k,b,f,(\dg k,c,b,k_{ma}\dg k,w,c, - \dg k,w,b,\dg k,m,c,k_{ab})
\Bigr]
\e
and
\beq A.33
R_G=\ell^{ab}R_{ab}.
\e

In the case of the Einstein--Kaufmann \cn\ on $M=G/G_0$ manifold one gets
\beq A.34
\gd\wh{\ov\G},\td n,\td w\td m,= {\wt n\atopwithdelims\{\}\wt w\wt m}
+\gd u,\td n,\td w\td m,
\e
where ${\wt n\atopwithdelims\{\}\wt w\wt m}$ is the Christoffel symbol built
from $\gd h,0,\td a\td b,$. In this way a \co ical term reads
\bg A.35
\ul P=\frac1{V_1}\int_M \sqrt{|\wt g|}\,\wh{\ov R}(\wh{\ov\G})\,d^{n_1}x\\
\wh{\ov R}(\wh{\ov\G})=g^{\td a\td b}\wh{\ov R}{}_{\td b\td d}(\wh{\ov\G})
\lb A.36 \\
\wh{\ov R}_{\td b\td d} = \wt R{}_{\td b\td d} + \wt\n_{\td a}\gd u,\td a,
\td b\td d, - \wt\n_{\td d}\gd u,\td a,\td b\td a,
+\frac12\bigl(\wt\n_{\td a}\gd u,\td a,\td a\td d,
-\wt\n_{\td d}\gd u,\td a,\td a\td b,\bigr) \lb A.37
\e
where $\wh{\ov R}_{\td b\td d}$ is a \MR\ tensor for a \cn\ $\gd \wh{\ov \G},
\td a,\td b\td c,$ and $\wt R_{\td b\td d}$ is a Ricci tensor of a \LC \cn\
formed for a metric tensor $\gd h,0,\td a\td b,$, where
\bml A.37a
\gd u,\td n,\td w\td m, = \frac12 \bigl(\dg K,\td w\td m,\td n,
-2\dg\wt g,[m\cdot,\td a], K_{\td w]\td a\td b} \dgd \wt g,[,\td n\td b,],
\bigr)\\
{}+h^{0\td n\td e}\Bigl\{\dgd K,\td e,\td a,(\td w,\wt g_{|\td m|\td a)}
+\dgd \wt g,[\td c\cdot,\td b,],
\Bigl[\dgd \wt g,[(|\td m|\cdot,\td c,], K_{\td w)\td a\td b}
\dgd\wt g,[\td e\cdot,\td a,],  - K_{\td c\td a\td b}
\dgd\wt g,[(\td w\cdot,\td a,], \dgd\wt g,[\td m\cdot,\td c,], \Bigr]\Bigr\}.
\e

During the calculations in Section 4 we used the following identities:
\bea A.41
g_{\td a\td m}g^{\td m\td c} &= g_{\td m\td a}g^{\td m\td c} =
\d_{\td a}^{\td c}\\
g_{\td m\td a}g^{\td c\td m} &= g_{\td a\td m}g^{\td c\td m} =
\d_{\td a}^{\td c} \lb A.42
\e
where $\wt m,\wt a,\wt b=5,6,(\vf,\psi)$ and
\bea A.43
{}&F_{\mu\psi}=F_{\mu5}=-F_{5\mu}=-F_{\psi\mu}=-\gv{\n_\mu}\F_1(x)=
e^*(H_{\mu\psi}) \\
&F_{\mu\vf}=\sin\psi \gv{\n_\mu}\F_2(x)=-F_{\vf\mu}=F_{\mu6}=-F_{6\mu}=
e^*(H_{\mu\vf}) \lb A.44
\e
We have also
\bml A.45
F_{\psi\vf} = F_{65} = \cos\psi \bigl(\F_2(x)-[\F_3,\F_1(x)]\bigr)
+\sin\psi\bigl(\F_3+[\F_2(x),\F_1(x)]\bigr)\\
{}=e^*(H_{65})=-F_{\vf\psi}=-e^*(H_{56}).
\e

\section*{Appendix B}
\def\theequation{B.\arabic{equation}}
\setcounter{equation}0

Following Ref.~\cite{21} we use the following formulae
\bea B.1
\F_5&=\frac12(\vf^*_1x_{-\a}+\vf^*_2x_{-\b}-\vf_1x_\a-\vf_2x_\b)\\
\F_6&=\frac{\sin\psi}{2i} (\vf_1x_\a+\vf_2x_\b+\vf^*_1x_{-\a}+\vf^*_2x_{-\b})
-\F_3\cos\psi. \lb B.2
\e
$\F_3$ is \ct\ and commutes with a reduced \cn. $\SU(2)\tm U(1)$ is a little
group of~$\F_3$,
\beq B.3
\F_3=\frac12\,i(2-\langle \g,\a\rangle)^{-1}(h_\a+h_\b),
\e
$x_\a$, $x_{-\a}$, $x_\b$, $x_{-\b}$ are \el s of a \Li~$\fh$ of~$H$ (see Ref.~\cite{b})
corresponding to roots $\a,-\a,\b,-\b$, $h_\a$~and~$h_\b$ are \el s of Cartan
subalgebra of~$\fh$ \st
\beq B.4
h_\a = \frac{2\a_i}{\a\cdot\a}\,H_i = [x_\a,x_{-\a}],
\e
where $\a=(\a_1,\dots,\a_k)$, $k=\mathop{\rm rank}(\fh)$, $\g=\a-\b$,
$[H_i,x_\o]=\o_ix_\o$, $H_i$~form Cartan subalgebra of~$\fh$, $[x_\o,x_\tau]
=C_{\o,\tau}x_{\o+\tau}$ if $\o+\tau$ is a root, if $\o+\tau$ is not a root
$x_\o$ and $x_\tau$ commute. We take $k=2$.
\beq B.5
\langle \g,\a \rangle = \frac{2\g\cdot\a}{\a\cdot \a} =
2\,\frac{|\g|}{|\a|}\cos\t.
\e
In this way we get a Higgs' doublet $\binom{\vf_1}{\vf_2}=\wt\vf$.

\allowdisplaybreaks
The $\SU(2)\tm \U(1)$ generators are given by
\beq B.6
\bal
t_1&=\frac12\,i(x_\g+x_{-\g})\\
t_2&=\frac12(x_\g-x_{-\g})\\
t_3&=\frac12\,ih_\g\\
y&=\frac12\,ih.
\eal
\e
$h$ is an \el\ of Cartan subalgebra orthogonal to $h_\g$ with the same norm.
Now everything is exactly the same as in Ref.~\cite{21} except the fact that
\bea B.7
\bar k_{ad}&=h_{ad}-\xi^2 k_{ab}\gd k,b,d,\\
k_{ad}&=(1-2\z^2)h_{ad}-\xi^2k_{ab}\gd k,b,d,. \lb B.8
\e
In Ref.~\cite{21}
\beq B.9
\bar k_{ad}=k_{ad}=h_{ad}.
\e
A four-\pt\ of \YM' field (a \cn\ $\o_E$) can be written as
\bg B.10
A_\mu=\sum_{i=1}^3 A_\mu t_i + B_\mu y \\
\hbox{or}\q
A_\mu=\frac12 \, i(A_\mu^- x_\g + A_\mu^+ x_{-\g} + A_\mu^3 h_\g
+ B_\mu h) \lb B.11 \\
A_\mu^\pm = A_\mu^1 \pm i A_\mu^2. \lb B.12
\e
We have (see Ref.~\cite{21})
\begin{align}
{}&h(t_i,t_j)=-\frac1{\g\cdot\g}\,\d_{ij} \nn\\
&h(y,y)=-\frac1{\g\cdot\g} \nn\\
&h(t_i,y)=0\nn\\
&F_\m=\bigl(\pa_\mu A_\nu^a - \pa_\nu A^a_\mu + \gd\ve,a,bc,A_\mu^b
A_\nu^c\bigr)t_a + (\pa_\mu B_\nu-\pa_\nu B_\mu )y
= \gd F,a,\m,t_a + B_\m y \lb B.13 \\
&h(F_\m,F_\m)=-\frac{\d_{ab}}{\g\cdot\g}\,\gd F,a,\m,F^{b\m}-
\frac1{\g\cdot\g}B_\m B^\m \lb B.14 \\
&\gv{\n_\mu}\F=\Bigl(\pa_\mu \vf_1 - \frac12 \,iA_\mu^- \vf_2
-\frac12\,i A_\mu^3 \vf_1 
-\frac12\,i\tan \t B_\mu \vf_1\Bigr)x_\a \nn \\
&\hskip30pt {}+\Bigl(\pa_\mu \vf_2 - \frac12 \,iA_\mu^+ \vf_1
+\frac12\,iA_\mu^3\vf_2 -\frac12\,i\tan \t B_\mu
\vf_2\Bigr)x_\b \lb B.15 \\
&\gv{\n_\mu}\wt{\F} = -\Bigl(\pa_\mu \vf_1^* + \frac12 \,iA_\mu^+ \vf_2^*
+\frac12\,i A_\mu^3 \vf_1^* +\frac12\,i\tan \t B_\mu \vf_1^*\Bigr)x_{-\a}\nn\\
&\hskip30pt {}-\Bigl(\pa_\mu \vf_2^* + \frac12 \,iA_\mu^- \vf_1^*
-\frac12\,i A_\mu^3 \vf_2^* +\frac12\,i\tan \t B_\mu\vf_2^*\Bigr)x_{-\b}
\lb B.16
\end{align}
We redefine the fields $A_\mu^a$, $B_\mu$ and $\wt\vf$ with some rescaling
($g$ is a coupling \ct)
\beq B.17
A_\mu^{\prime a}  = L_1 A_\mu^a, \q B_\mu'  = L_1 B_\mu, \q
\wt\vf{}'  = L_2\wt\vf
\e
where
\bea B.18
L_1 &= \frac1g\,\frac1{(\g\cdot \g)^{1/2}}\\
L_2 & = \frac1g\Bigl(\frac{\g\cdot\g}{\a\cdot\a}\Bigr)^{1/2} \lb B.19
\e
We proceed the following \tf
\beq B.20
\left(\begin{matrix}
Z_\mu^0 \\ A_\mu \end{matrix}\right) =
\left(\begin{matrix}
\cos\t &\ & -\sin\t \\ \sin\t && \cos \t \end{matrix}\right)
\left(\begin{matrix}
A_\mu^3 \\ B_\mu \end{matrix}\right).
\e
\goodbreak

According to the classical results we also have $\frac {g'}g = \tan\t$,
assuming $q=g\sin\t$, where $q$ is an \el ary charge and $g$~and~$g'$ are
coupling \ct s of $A_\mu^a$ and~$B_\mu$ fields. The \sn\ \s y breaking and
\Hm\ in the Manton model works classical if we take for minimum of the \pt
\beq B.21
\wt\vf_0=\left(\begin{matrix} 0 \\ \frac v{\sqrt2} \end{matrix}\right)
e^{i\a}, \q \a \hbox{ arbitrary phase,}
\e
and we parametrize $\wt\vf=\binom{\vf_1}{\vf_2}$ in the following way
\beq B.22
\wt\vf(x)=\exp\Bigl(i\,\frac1{2v}\,\si^a t^a(x)\Bigr)
\left(\begin{matrix} 0 \\ \frac {v+H(x)}{\sqrt2} \end{matrix}\right).
\e
For a vacuum state we take
\beq B.23
\wt\vf_0=\left(\begin{matrix} 0 \\ \frac {v}{\sqrt2} \end{matrix}\right),
\e
$t^a(x)$ and $H(x)$ are real fields on~$E$. $t^a(x)$ has been ``eaten'' by
$A_\mu^a$, $a=1,2$, and~$Z_\mu^0$ fields making them massive. $H(x)$~is our
Higgs' field. $\si^a$~are Pauli matrices.

In the formulae \er{B.7}--\er{B.8} we take $\xi=0$. One gets in the
Lagrangian mass terms:
$$
M_W^2 W_\mu^+ W^{-\mu} + \frac12\, M_Z^2 Z_\mu^0Z^{0\mu} - \frac12\,M_H^2H^2,
$$
where $W_\mu^+=A_\mu^+$, $W_\mu^-=A_\mu^-$, getting masses for
$W^\pm$, $Z^0$ bosons and
a~Higgs boson (see Eqs \er{4.49}--\er{4.53}). For G2 $\langle \g,\a
\rangle=3$ and $\t=30^\circ$, $\t$~is identified with the Weinberg angle
$\t_W$.

In order to proceed a \Hm\ and \sn\ \s y breaking in this model we use the
following gauge \tf
\beq B.23a
\wt\vf (x) \mapsto U(x)\wt\vf(x)=\frac1{\sqrt2}
\left(\begin{matrix} 0 \\ v+H(x) \end{matrix}\right),
\e
where
\beq B.24
v=\frac{2\sqrt2}{rg}\cos\t
\e
a vacuum value of a Higgs field
\beq B.25
U(x)=\exp\Bigl(-\frac1{2v}\,t^a(x)\si^a\Bigr).
\e
$H(x)$ is the remaining scalar field after a \s y breaking and a \Hm. One gets
\bg B.26
A_\mu \mapsto A_\mu^u = {\rm ad}'_{U^{-1}(x)}A_\mu + U^{-1}(x)\pa_\mu U(x)\\
F_\m \mapsto F_\m^u = {\rm ad}'_{U^{-1}(x)}F_\m. \lb B.27
\e

\section*{Appendix C}
\def\theequation{C.\arabic{equation}}
\setcounter{equation}0
In this appendix we derive a \KWK \e\ in GWS-model. One gets (see Eqs
\er{3.170}--\er{3.172})
\bg C.1
\frac{\wt{\ov D}u^\a}{d\tau} + \frac{u^\b}{m_0} h(q,H_{\b\d})
+\frac1{m_0}\,\wt g{}^\(\a\d)\Bigl(u^5\cdot h(q,\gv{\n_\d}\F_5)
+u^6h(q,\gv{\n_\d}\F_6)\Bigr)=0 \\
\frac{\wt Du^5}{d\tau} - \frac1{r^2}\,\frac{u^\b}{m_0}\,h(q,\gv{\n_\b}\F_5)
-\frac1{r^2}\,u^6h(q,H_{56})=0 \lb C.2 \\
\frac{\wt Du^6}{d\tau}-\frac1{r^2}\,\frac{u^b}{m_0\sin^2\psi}\,h(q,\gv{\n_\b}\F_6)
-\frac1{r^2}\,\frac{u^5}{m_0\sin^2\psi}\,h(q,H_{65})=0 \lb C.3 \\
\frac d{d\tau}\Bigl(\frac q{m_0}\Bigr)=0. \lb C.4
\e
$q$ is an isotopic charge belonging to a \Li\ of~$H$ ($\fh$), $u^{\td a}=
(u^5,u^6)$ is a charge which couples a test \pc\ to Higgs' field,
$H_{\b\d}$~is a strength of $\SU(2)\tm U(1)$ \YM' field, $\F_5,\F_6$ are
scalar fields before a \sn\ \s y breaking (see Eq.~\er{4.11}). $\frac{\wt{\ov
D}}{d\tau}$ is a \ci\ \dv\ along a line \wrt a \cn~$\gd\wt{\ov\o},\a,\b,$
on~$E$, $\frac{\wt D}{d\tau}$ is a \ci\ \dv\ \wrt a \LC \cn\ on~$S^2$.
We have of course $g_\(\a\b)u^\a u^\b=1$.

Using some
additional fields $\F_1,\F_2,\F_3$ and also $\F$ and~$\wt\F$, we can write
$\gv{\n_\mu}\F_5$ and $\gv{\n_\mu}\F_6$ in terms of Higgs' fields $\vf_1$
and~$\vf_2$ (see Appendix~B), $m_0$ is the mass of a test \pc.
\begin{align}
\gv{\n_\mu}\F_5&=\frac12\gv{\n_\mu}(\F+\wt\F)=\frac12\Bigl[\Bigl(\pa_\mu\vf_1
-\frac12\,iA_\mu^- \vf_2 - \frac12\,iA_\mu^3 \vf_1 -\frac12\,i\tan\t B_\mu
\vf_1\Bigr)x_\a \nn \\
&+\Bigl(\pa_\mu \vf_2 - \frac12\,iA_\mu^+\vf_1 + \frac12\,iA_\mu^+\vf_2
-\frac12\,iB_\mu\vf_2\tan\t\Bigr)x_\b \nn \\
&-\Bigl(\pa_\mu\vf_1^* + \frac12\,iA_\mu^+ \vf_2^* + \frac12\,iA_\mu^3\vf_1^*
+\frac12\,iB_\mu\vf_1^*\tan\t\Bigr)x_{-\a} \nn \\
&-\Bigl(\pa_\mu\vf_2^* + \frac12 \,iA_\mu^- \vf_1^*
-\frac12\,i A_\mu^3 \vf_2^* +\frac12\,i\tan \t B_\mu\vf_2^*\Bigr)x_{-\b}\Bigr]
\lb C.4a
\end{align}
\begin{align}
\gv{\n_\mu}\F_6 &= \frac{\sin\psi}{2i} \gv{\n_\mu}(\F-\wt\F)=
\frac{\sin\psi}{2i}
\Bigl[\Bigl(\pa_\mu\vf_1
-\frac12\,iA_\mu^- \vf_2 - \frac12\,iA_\mu^3 \vf_1 -\frac12\,i\tan\t B_\mu
\vf_1\Bigr)x_\a \nn \\
&+\Bigl(\pa_\mu \vf_2 - \frac12\,iA_\mu^+\vf_1 + \frac12\,iA_\mu^+\vf_2
-\frac12\,iB_\mu\vf_2\tan\t\Bigr)x_\b \nn \\
&-\Bigl(\pa_\mu\vf_1^* + \frac12\,iA_\mu^+ \vf_2^* + \frac12\,iA_\mu^3\vf_1^*
+\frac12\,iB_\mu\vf_1^*\tan\t\Bigr)x_{-\a} \nn \\
&-\Bigl(\pa_\mu\vf_2^* + \frac12 \,iA_\mu^- \vf_1^*
-\frac12\,i A_\mu^3 \vf_2^* +\frac12\,i\tan \t B_\mu\vf_2^*\Bigr)x_{-\b}\Bigr]
\lb C.5
\end{align}

Let
\beq C.6
q=q_\g x_\g + q_{-\g}x_{-\g}+qh+\wt qh_\g+q_\a x_\a+q_{-\a}x_{-\a}+q_\b x_\b
+q_{-\b}x_{-\b}.
\e
It is easy to see that the first part of $q$,
\bg C.7
q=q_1+q_2, \\
q_1=q_\g x_\g +q_{-\g}x_{-\g}+qh + \wt qh_\g \lb C.8
\e
couples to \YM' field and the second part
\bg C.9
q_2=q_\a x_\a+q_{-\a}x_{-\a}+q_\b x_\b +q_{-\b}x_{-\b}
\e
to scalar fields $\F_5$ and $\F_6$.

In this way in a GSW model a test \pc\ has a weak isotopic charge, weak
hypercharge which are equivalent to weak charge and an electric charge. It
has also an additional weak charge which couples it to Higgs' field,
i.e.~$q_2$. Moreover, we have also $u^{\td a}=(u^5,u^6)$ charge. It would be
very interesting to observe this additional charges in an experiment.
\bea C.10
H_{56}&= \pa_5\F_6 - \pa_6\F_5 + [\F_5,\F_6] \\
H_{65}&= \pa_6\F_5 - \pa_5\F_6 + [\F_6,\F_5] \lb C.11
\e

We get
\begin{gather}
\frac{\wt{\ov D}u^\a}{d\tau} - \frac{u^\b}{m_0} \,\wt g{}^\(\a\d)
\wt Q{}^i \d_{ij}\gd F,i,\b\d, - \frac{u^b}{m_0}\,\wt g{}^\(\a\d)\cdot Q
\cdot B_{\b\d} \hskip120pt\nn \\
\hskip60pt {}+\frac1{m_0}\,\wt g{}^\(\a\d)\Bigl(u^5\cdot h\bigl(q,e^*(\gv{\n_\d}\F_5)\bigr)
+u^6 h\bigl(q,e^*(\gv{\n_\d}\F_6)\bigr)\Bigr)=0 \lb C.14  \\
\frac{\wt Du^5}{d\tau} - \frac1{r^2}\,\frac{u^\b}{m_0}\,h\bigl(Q,e^*(\gv{\n_\b})
\F_5\bigr) -\frac1{r^2}\,u^6h(Q,e^*(H_{56}))=0 \lb C.15 \\
\frac{\wt Du^6}{d\tau}-\frac1{r^2}\,\frac{u^b}{m_0\sin^2\psi}\,h\bigl(Q,
e^*(\gv{\n_\b}\F_6)\bigr)
-\frac1{r^2}\,\frac{u^5}{m_0\sin^2\psi}\,h(Q,e^*(H_{65}))=0 \lb C.16 \\
\frac{dQ^a}{d\tau} - \gd C,c,cb,Q^cA^b_M u^M=0 \lb C.17
\end{gather}
where
\bg C.18
e^*\o_E = A^i_\mu \ov \t{}^\mu t_i + B_\mu \ov \t{}^\mu y \\
e^*(q^cX_c)= Q^c X_c \lb C.19 \\
e^*\o = \a^c_i A^i_\mu \ov \t{}^\mu \wt t_i + \F^a_{\td a}\t^{\td a}X_a ,
\lb C.20
\e
$\wt Q{}^i=\frac{Q^i}{\g\cdot \g}$ is an isotopic charge, $\wt
Q=\frac{Q}{\g\cdot \g}$ is a weak hypercharge,
\bg C.21
\wt t_i=t_i,\ i=1,2,3, \q \wt t_4=y, \\
h(x,y)=h_{ab}x^a y^b. \lb C.22
\e

Let us consider the \tf\ \er{B.20} and the following \tf
\beq C.23
\left(\begin{matrix} Q^0 \\ q \end{matrix}\right) =
\left(\begin{matrix} \cos\t &\ & \sin\t \\ -\sin\t && \cos\t \end{matrix}\right)
\left(\begin{matrix} \wt Q{}^3 \\ Q \end{matrix}\right).
\e
$\t$ plays of course a role of the Weinberg angle $\t_W$. $Q^0$~is a neutral
weak charge and $q$~an electric charge. In this way we get in Eq.~\er{C.17} a
very familiar term
\beq C.23a
-\frac{u^\b}{m_0}\,\wt g{}^\(\a\d) qF_{\b\d}
\e
where
\beq C.24
F_{\b\d}=\pa_\b A_\d - \pa_\d A_\b
\e
is a strength of an \elm c field and $q$ an electric charge, i.e.\ a Lorentz
force term.

One gets for $H_{56}$
\bml C.25
H_{56}=-H_{65}=-i\sin\psi\Bigl(\vf_1\vf_1^* \,\frac{\a_i}{\a\cdot \a}
+\vf_2\vf_2^*\,\frac{\b_i}{\b\cdot\b} \Bigr) H_i
-i\cos\psi \vf_1x_\a - i\cos\psi\vf_2 x_b\\
{}-i\cos\psi\vf_1^* x_{-\a} - i\cos\psi\vf_2^* x_{-\b}
+\frac i2 \sin\psi \vf_1\vf_2^* C_{\a,-\b}x_\g
+\frac i2 \sin\psi \vf_2\vf_1^* C_{\b,-\a}x_{-\g}.
\e

Let us proceed a \sn\ \s y breaking and \Hm\ in our \KWK\ \e. In this way we
transform
\beq C.26
\gv{\n_\mu}\F_{\td a} \mapsto {\rm ad}'_{U^{-1}(x)}\gv{\n_\mu}\F_{\td a} = \gv{\n_\mu}\F_{\td a}^u\,,
\q \wt a=5,6,
\e
where
\begin{gather}
\gv{\n_\mu}\F_5^u = \frac1{2\sqrt2}\Bigl[\pa_\mu H(x)(x_\b-x_{-\b})\hskip200pt
\nn\\
\hskip40pt{}+\frac i2(v+H(x))\bigl(A_\mu^{3u}(x_\b+x_{-\b})+B_\mu
\tan\t(x_{-\b}-x_\b)
-A_\mu^{+u}x_{-\a}+A_\mu^{-u}x_\a\bigr)\Bigr] \lb C.27 \\
\gv{\n_\mu}\F_6^u = \frac{\sin\psi}{2i}\Bigl[\pa_\mu H(x)(x_\b+x_{-\b})
\hskip200pt \nn \\
\hskip40pt{}+\frac i2(v+H(x))\bigl(A_\mu^{+u}x_{-\a}-A_\mu^{-u}x_\a
+A_\mu^{3u}(x_\b-x_{-\b})+B_\mu \tan\t(x_{-\b}-x_\b)\bigr)\Bigr] \lb C.28 \\
H_{56}\mapsto {\rm ad}'_{U^{-1}(x)}H_{56}^u, \lb C.29
\end{gather}
where
\bg C.30
H_{56}^u =-\frac{\sin\psi(v+H(x))}2 \Bigl((v+H(x))\frac{\b_i}{\b\cdot \b}
\,H_i +\sqrt2 \cos\psi(x_\b+x_{-\b})\Bigr) \\
H_{56}^u = - H_{65}^u \lb C.31
\e
where
\bea C.32
A_\mu^+ \mapsto A_\mu^{+u} &= \bigl({\rm ad}'_{U^{-1}(x)}A_\mu\bigr)^+
+\frac i{2v}\,\pa_\mu t^+(x) \\
A_\mu^- \mapsto A_\mu^{-u} &= \bigl({\rm ad}'_{U^{-1}(x)}A_\mu\bigr)^-
+\frac i{2v}\,\pa_\mu t^-(x) \lb C.33 \\
A_\mu^3 \mapsto A_\mu^{3u} &= \bigl({\rm ad}'_{U^{-1}(x)}A_\mu\bigr)^3
+\frac i{2v}\,\pa_\mu t^3(x). \lb C.33b
\e
Simultaneously we proceed a \tf\ on charges
\beq C.33a
Q \mapsto Q^u ={\rm ad}'_{U^{-1}(x)}Q.
\e
In this way we have from Eqs \er{C.14}--\er{C.17}
\begin{gather}
\frac{\wt{\ov D}u^\a}{d\tau} - \frac{u^\b}{m_0}\,\wt g{}^\(\a\d) \wt Q{}^{iu}
\d_{ij}\gd W,i,\b\d, - \frac{u^\b}{m_0}\,\wt g{}^\(\a\d)Q^{0u}\gd Z ,0,\b\d,
-\frac{u^\b}{m_0}\,\wt g{}^\(\a\d)qF_{\b\d} \hskip100pt \nn \\
\hskip50pt{}+\frac{1}{m_0}\,\wt g{}^\(\a\d)\Bigl(u^5h(Q^u,\gv{\n_\d}\F_5^u)
+u^6 h(Q^u,\gv{\n_\d}\F_6^u)\Bigr)=0 \lb C.34 \\
\frac{\wt Du^5}{d\tau} - \frac1{r^2}\,\frac{u^\b}{m_0}\,
h(Q^u,\gv{\n_\b}\F_5^u) - \frac 1{r^2}\,u^6h(Q^u,H_{56}^u)=0 \lb C.35 \\
\frac{\wt Du^6}{d\tau} - \frac1{r^2}\,\frac{u^\b}{m_0\sin^2\psi}\,
h(Q^u,\gv{\n_\b}\F_6^u) - \frac 1{r^2}\,\frac{u^5}{m_0\sin^2\psi}\,
h(Q^u,H_{65}^u)=0. \lb C.36
\end{gather}
In Eqs \er{C.34}--\er{C.36} a test \pc\ is coupled to physical fields only,
i.e.\ $W_\m^i$, $F_\m$ and~$H$.

One derives a final form of $h(Q^u,e^*(\gv{\n_\mu}\F_5))$,
$h(Q^u,e^*(\gv{\n_\mu}\F_6))$, $h(Q^u,e^*(H_{56}))$, getting
\bml C.39
h(Q^u,e^*(\gv{\n_\mu}\F_5)) = \frac1{2\sqrt2}\Bigl(
\frac2{\a\cdot\a}\bigl(q_{-\a} W_\mu^{-u} - q_\a W_\mu^{+u}\bigr)
+ \frac2{\b\cdot\b}\Bigl({\pa_\mu H (q_{-\b} -q_\b)}\\
{}+\frac i2\,(v+H(x))(Z_\mu^{0u}\cos\t + A_\mu \sin\t)(q_{-\b}+q_\b)
+(-Z_\mu^{0u}\sin\t \tan\t + A_\mu \sin\t)(q_{-\b} - q_\b)\Bigr)\\
{}+ h(x_\a,x_\a)q_\a W_\mu^{-u}
+ h(x_\b,x_\b)q_\b \Bigl(\pa_\mu H +\frac i2 (v+H(x))
(Z_\mu^{0u}\cos\t + A_\mu\sin\t)\\
{}\hskip200pt +(Z_\mu^{0u}\sin\t \tan\t -A_\mu\sin\t)\Bigr)\\
{}+ h(x_{-\a},x_{-a}) q_{-\a} W_\mu^{+u}
+ h(x_{-\b,-\b})q_{-\b} \Bigl(-\pa_\mu H + \frac i2(v+H(x))
(Z_\mu^{0u}\cos\t + A_\mu \sin\t)\\
{} +(-Z_\mu^{0u}\sin\t \tan\t +A_\mu\sin\t)\Bigr)\Bigr)
\e
\bml C.40
h(Q^u,e^*(\gv{\n_\mu}\F_6)) = \frac{\sin\psi}{2\sqrt2 \,i}\biggl(
\frac{2i(v+H(x))}{\a\cdot\a}\,(q_\a W_\mu^{+u} - q_{-\a} W_\mu^{-u})\\
{}+ \frac2{\b\cdot\b}\Bigl(\pa_\mu H(q_\b+q_{-\b})
+(q_{-\b} - q_\b)\,\frac{Z_\mu^{0u}}{\cos\t}\Bigr)\\
{}-\frac i2\,h(x_\a,x_\a)q_\a W_\mu^{-u} + \frac i2\,h(x_{-\a},x_{-\a})
q_{-\a}(v+H(x))W_\mu^{+u} \\
{}+h(x_\b,x_\b)q_\b \Bigl(\pa_\mu H + \frac{Z_\mu^{0u}}{\cos\t}\Bigr)
+h(x_{-\b},x_{-\b})q_\b \Bigl(\pa_\mu H-\frac{Z_\mu^{0u}}{\cos\t}
\Bigr)\biggr)
\e
\bml C.41
h(Q^u,e^*(H_{56})) = -h(Q^u,e^*(H_{65}))\\ {}=
-\frac{\sin2\psi\sqrt2}4 \biggl((q_\b + q_{-\b})
\Bigl(\frac2{\b\cdot\b}+ h(x_\b,x_\b)+h(x_{-\b},x_{-\b})\Bigr)\biggr)
\e
Here the superscript $u$ means that all quantities are in a gauge $U$. In this
way all couplings of a test \pc\ are expressed by physical fields after a
\sn\ \s y breaking and \Hm.

Let us consider Eq.~\er{C.17} in more details using Eq.~\er{C.5} and let us
change a gauge using a gauge changing \f\ $U(x)$. One finds
\begin{gather}
\frac{dQ_\g^u}{d\tau} - i\bigl((Z_\mu^{0u} \cos\t - \sin\t A_\mu)Q_\g^u
- W_\mu^{-u}\wt q\bigr)u^\mu + \frac{v+H(x)}{\sqrt2}\langle \g,\a\rangle
\Bigl(u^5 + i\,\frac{\sin\psi}2\,u^6\Bigr)q_\a\nn\\
\hskip150pt {}-\frac i2\,u^6\cos\psi
\bigl(\langle\g,\b\rangle+\langle\g,\a\rangle\bigr)
Q_\g^u =0 \lb C.42 \\
\frac{dQ_{-\g}^u}{d\tau} - i\bigl(W_\mu^{+u}\wt Q{}^u -(Z_\mu^{0u}\cos\t
-\sin\t A_\mu)Q_{-\g}^u\bigr) u^\mu +
\frac1{\sqrt2}(v+H(x))\langle\g,\a\rangle q_{-\a}
\Bigl(u^5-i\,\frac{\sin\psi}2\,u^6\Bigr)\nn\\
\hskip150pt {}+
\frac i2\,u^6\cos\psi(\langle\g,\a\rangle+\langle\g,\b\rangle)Q_{\g}^u=0
\lb C.43 \\
\frac{dq}{d\tau}=0 \lb C.44 \\
\frac{d\wt Q{}^u}{d\tau}-\frac12(W_\mu^{+u}Q_\g^u - W_\mu^{-u}Q_{-\g}^u)=0
\lb C.45 \\
\frac{dq_\a}{d\tau}+\frac12\,i\Bigl((\cos\t A_\mu - Z_\mu^{0u}\sin\t)
\langle\a,\g\rangle q_\a - 2(Z_\mu^{0u}\cos\t - \sin\t A_\mu)q_\a
\,\frac{\g_1\a_2-\g_2\a_1}{\g\cdot\g}\Bigr)u^\mu \nn \\
\hskip80pt {}+\frac{u^5(v+H(x))}{\sqrt2}\,Q_\g^u \Bigl(u^5+\frac{i\sin\psi}
{2\sqrt2}\,u^6\Bigr) - \frac i2\,u^6\cos\psi q_\a(2+\langle\a,\b\rangle)=0
\lb C.46 \\
\frac{dq_{-\a}}{d\tau}+\frac12\,i\Bigl(2(\cos\t A_\mu - Z_\mu^{0u}\sin\t)
q_{-\a} \,\frac{\g_1\a_2-\g_2\a_1}{\g\cdot\g}
+ (Z_\mu^{0u}\cos\t - \sin\t A_\mu)q_{-\a}\langle\a,\g\rangle \Bigr)u^\mu \nn\\
\hskip100pt{}+\frac{v+H(x)}{\sqrt2}\,Q_{-\g}^u \Bigl(u^5 - \frac{i\sin\psi u^6}2 \Bigr)
+\frac i2\,\cos\psi q_{-\a}(2+\langle\a,\b\rangle )=0 \lb C.47
\end{gather}
Simultaneously we get
\beq C.48
q_\b=q_{-\b}=0.
\e

In this way our \e s are simpler, e.g. $h(Q^u,e^*(H_{56}))=h(Q^u,e^*(H_{65}))
=0$. Let us notice that $q_\a$ and $q_{-\a}$ charges are not influenced by
the gauge \tf\ $U(x)$. The electric charge~$q$ does not feel any movement of
additional charges.

Thus one gets eventually
\begin{gather}
\frac{\wt{\ov D}u^\mu}{d\tau} - \frac{\wt Q{}^{iu}}{m_0}\wt g{}^\(\mu\d)
\d_{ij}u^\b W_{\b\d}^{iu} - \frac{Q^{0u}}{m_0}\,\wt g{}^\(\mu\d)u^\b Z_{\b\d}^{0u}
-\frac q{m_0}\,\wt g{}^\(\mu\d)u^\b F_{\b\d}\hskip100pt \nn \\
{}+\frac1{\sqrt2\,m_0}\,\wt g{}^\(\mu\d)\biggl(u^5\Bigl(\frac1{\a\cdot\a}
(q_{-\a}W_\d^{-u} {-} q_\a W_\d^{+u}) + \frac1{\b\cdot\b}\bigl(
h(x_\a,x_\a)q_\a W_\d^{-u}
+h(x_{-\a},x_{-\a})q_{-\a}W_\d^{+u}\bigr) \Bigr) \nn \\
\kern-19pt{}+u^6 \sin\psi\Bigl(\frac{v{+}H(x)}{\a\cdot\a}(q_\a W_\d^{+u}{-}q_{-\a}W_\d^{-u}
)-\frac1{2(\b\cdot\b)}\bigl(h(x_\a,x_\a)q_{-\a}(v{+}H(x))\bigr)W_\d^{+u}\Bigr)
\biggr)=0\kern-3pt
\lb C.49 \\
\frac{\wt Du^5}{d\tau} - \frac1{r^2}\,\frac{u^\b}{\sqrt2\,m_0} \Bigl(
\frac1{\a\cdot\a}(q_{-\a}W_\b^{-u} - q_\a W_\b^{+u})\hskip80pt \nn \\
\hskip80pt {}+\frac1{\b\cdot\b}\bigl(h(x_\a,x_\a)q_\a W_\b^{-u} + h(x_{-\a},x_{-\a})
q_{-\a}W_\b^{+u}\bigr)\Bigr)=0 \lb C.50 \\
\frac{\wt Du^6}{d\tau} - \frac 1{r^2}\,\frac{u^\b}{\sqrt2\,m_0\sin\psi}\Bigl(
\frac{v+H(x)}{\a\cdot\a}(q_\a W_\b^{+u} - q_{-\a}W_\b^{-u})\hskip80pt \nn \\
\hskip80pt {}+\frac1{2(\b\cdot\b)}\bigl(-h(x_\a,x_\a)q_\a W_\b^{-u} + h(x_{-\a},x_{-\a})
(v+H(x))W_\b^{+u}\bigr)\Bigr)=0. \lb C.51
\end{gather}

Let us suppose that $H=G2$. In this case one gets
\beq C.52
\bga
|\b|=|\a|=\sqrt2, \q |\g|=\sqrt6,\\
\a\cdot\a=\b\cdot\b=2, \q \g\cdot\g=6,\\
\langle \g,\a \rangle=3, \q \langle \g,\b \rangle=\langle \a,\b \rangle=-1,\\
\frac{\g_1\a_2-\g_2\a_1}{\g\cdot\g}=\frac{\sqrt3}6,\\
\t=30^\circ, \q \cos\t=\frac{\sqrt3}2\,, \q \sin\t=\frac12\,.
\ega
\e
Thus one gets
\begin{gather}
\frac{dQ_\g^u}{d\tau} - i\Bigl(\frac12(Z_\g^{0u}\sqrt3 - A_\mu)Q_\g^u
-W_\g^{-u}\wt q\Bigr)u^\mu \hskip100pt \nn \\
\hskip100pt {}- \frac{3(v+H(x))}{\sqrt2}\,q_\a
\Bigl(u^5+i\,\frac{\sin\psi}2\,u^6\Bigr) - i\cos\psi Q_\g^u=0 \lb C.53 \\
\frac{dQ_{-\g}^u}{d\tau} - iu^\mu \Bigl(W_\mu^{+u}\wt Q{}^u
-\frac12(\sqrt3\,Z_\mu^{0u}-A_\mu)Q_{-\g}^u\Bigr)\hskip100pt \nn \\
\hskip100pt {}+\frac{3u^5}{\sqrt2}(v+H(x))q_{-\a}\Bigl(u^5-i\,\frac{\sin\psi}
2\, u^6\Bigr) + iu^6\cos\psi Q_{-\g}^u=0 \lb C.54 \\
\frac{dq}{d\tau}=0 \lb C.55 \\
\frac{d\wt Q{}^u}{d\tau} - \frac12 (W_\mu^{+u}Q_\g^u - W_\mu^{-u}Q_{-\g}^u)=0
\lb C.56 \\
\frac{dq_\a}{d\tau} + \frac12\,i\Bigl(\frac32(\sqrt3\,A_\mu - Z_\mu^{0u})
-\frac{\sqrt3}6 (\sqrt3\,Z_\mu^{0u}-A_\mu)\Bigr)q_\a u^\mu \hskip100pt \nn \\
\hskip100pt {}+\frac{u^5(v+H(x))}{\sqrt2}\,Q_\g^u
\Bigl( u^5 + \frac{i\sin\psi}{2\sqrt2}\,u^6\Bigr)
-\frac i2\,u^6\cos\psi q_\a=0 \lb C.57 \\
\frac{dq_{-\a}}{d\tau} + \frac i2\Bigl(\frac{\sqrt3}6(\sqrt3\,A_\mu -
Z_\mu^{0u}) + \frac32(\sqrt3\,Z_\mu^{0u} - A_\mu)\Bigr) u^\mu q_{-\a}
\hskip100pt \nn \\
\hskip100pt {}+\frac{v+H(x)}{\sqrt2}\,Q_{-\g}^u \Bigl(u^5 - \frac{i\sin\psi}2
\,u^6\Bigr)+\frac i2 \cos\psi q_{-\a}=0. \lb C.58
\end{gather}

Eqs \er{C.49}--\er{C.51} and \er{C.53}--\er{C.58} are generalized \KWK\ \e s
in GSW model.

At the end of this appendix we consider a \co ical \ct\ in GSW model. In this
case we have from Eq.~\er{4.30}
$$
\la_c=\Bigl(\frac{\a_s^2}{\ell\pl^2}\,\wt R(\wt\G)+\frac1{r^2}\,\wt{\ul P}
\Bigr).
$$
Moreover, now we have also an additional term for a Higgs' \pt\ $V(0)\ne0$.
One gets
\beq C.12
\la_c'=\la_c-\frac{2\pi(1-2\z^2)}{\sqrt{1+\z^2}\,r^2}
\e
and eventually
\beq C.13
\la_c'=\frac{\a_s^2}{\ell\pl^2}\,\wt R(\wt\G)+\frac1{r^2}\Bigl(\wt{\ul P}
-\frac{2\pi(1-2\z^2)}{\sqrt{1+\z^2}\,r^2}\Bigr)
\e
where $\wt{\ul P}$ is given by the formula \er{4.32}.

Moreover, we should add to \co ical \ct\ term also $V_2(0)$ (see Appendix~D).
The term $\wt P$ has been calculated in Refs \cite{1,5} for $S^2$. $\wt R(\wt
\G)$ is equal to $\wt R_{G2}$ ($G=G2$, see Eq.~\er{A.33}).

One gets
\bml C.67
\la_c' = \frac{\a_s^2}{\ell\pl^2}\,\wt R_{G2} + \frac1{r^2}\biggl(\Bigl(
\frac{16|\z|^3}{3(2\z^2+1)(1+\z^2)^{5/2}} \Bigl(\z^2
E\Bigl(\frac{|\z|}{\sqrt{\z^2+1}}\Bigr) - 2(\z^2+1)
K\Bigl(\frac{|\z|}{\sqrt{1+\z^2}}\Bigr)\Bigr)\\
{}+8\ln\Bigl(|\z|\sqrt{\z^2+1} + \frac{4(1+9\z^2-8\z^4)|\z|^3}
{3(1+\z^2)^{3/2}}\Bigr)\Bigr)\frac{1}{2|\z|\sqrt{1+\z^2}}\\
{}-\frac \pi{\sqrt{1+\z^2}}\Bigl(2(1-2\z^2)+\frac{\xi^2}{2\sqrt{1+\z^2}}\,
\ov K(h_\a+h_\b,h_\a+h_\b)\Bigr)\biggr)
\e
where
\bea C.68
K(k)=\int_0^{\pi/2} \frac{d\t}{\sqrt{1-k^2\sin^2\t}} \\
E(k)=\int_0^{\pi/2} \sqrt{1-k^2\sin^2\t}\,d\t \lb C.69
\e
are elliptic integrals of the first and second order.

$\wt R_{G2}$ strongly depends on $k_{ab}$ and $\xi$.
It seems that we can tune $\la_c'$ to the desired value known from
observational data. $\ov K(\cdot,\cdot)$ is defined by Eq.~\er{D.10a}.

\section*{Appendix D}
\def\theequation{D.\arabic{equation}}
\setcounter{equation}0
In this appendix  we give details of an \ia\ of the Higgs' field and \nos\
(Hermitian) gravity. One gets from Eq.~\er{3.87}
\bml D.1
\cL_{\rm kin}(\gv\n \F)=\frac1{2\pi r^2(1+\z^2)}\Biggl\{
\gv{\n_\mu}\F_5^k \gv{\n_\o}\F_5^d\Bigl(2(\xi\z^2(\z-1)-1-\xi)k_{kd}
+\bigl(\xi\z(2\xi^2-\z-2\xi\z-2)\\
{}+(\pi-2)\xi\bigr)\gd k,b,d, k_{bk}
-2h_{kd}+\xi^3\z(2\xi^2-\pi)k^{nb}k_{nk}k_{bd}\Bigr)g^{\o\mu}
+\gv{\n_\mu}\F_5^d \gv{\n_\g}\F_5^n \Bigl(-2\xi^2h_{nd}\wt g{}^\(\g\b)
g_\[\b\o]\\
-2\xi^3\wt g{}^\(\g\b)g_\[\b\o]k_{nd}
+ 2\wt g^\(\a\nu)\wt g^\(\g\rho)g_\[\nu\o]g_\[\rho\a]k_{nd}
+2\xi^2 k_{nd}\gd k,n,d,\wt g{}^\(\a\nu)\wt g{}^\(\g\rho)
g_\[\nu\o]g_\[\rho\a]\Bigr)g^{\o\mu}\\
+\Bigl(-2\z \gv{\n_\mu}\F_5^k \gv{\n_\o}\wh \F{}_6^n
+2\z\gv{\n_\mu}\wh\F{}_6^k \gv{\n_\o}\F_5^n
+2\z(\z+1)\gd k,n,d,\gv{\n_\mu}\F_5^k \gv{\n_\o}\wh\F{}_6^d\\
{}+2\xi^2(\z\xi+1)\gd k,n,d,\gv{\n_\mu}\F_5^k \gv{\n_\o}\wh\F{}_6^d
+2\xi^2\z k^{nb}k_{bd}\gv{\n_\mu}\wh\F{}_6^k\gv{\n_\o}\F_5^d\Bigr)g^{\o\mu}
\ell_{nk}\\
{}+\ell_{nk}g^{\o\mu}\gd k,n,d,\bigl(\pi+4\z+4\z\xi^2+2\pi\xi\z^2+4\z\xi\bigr)
g^{\a\eta}g_\[\eta\o]\gv{\n_\mu}\F_5^n \gv{\n_\a}\wh\F{}_6^d\\
{}+2\xi\z\ell_{nk}g^{\o\mu}\wt g{}^\(\a\nu)\wt g{}^\(\g\rho)g_\[\nu\o]
g_\[\rho\a]\bigl(\gv{\n_\mu}\F_5^k \gv{\n_\g}\wh\F{}_6^d
+\z^2\gv{\n_\mu}\wh\F{}_6^k\gv{\n_\g}\F_5^d
+2\gv{\n_\mu}\wh\F{}_6^k\gv{\n_\g}\F_5^d \bigr) \\
{}+\ell_{nk}g^{\o\mu}\Bigl[\Bigl(-2\gv{\n_\mu}\wh\F{}_6^k\gv{\n_\o}\wh\F{}_6^n
-2\xi(\z-1)\gd k,n,d,\gv{\n_\mu}\wh\F{}_6^k\gv{\n_\o}\wh\F_6^d
+2\xi^2\z k^{nb}k_{bd}\gv{\n_\mu}\wh\F{}_6^k\gv{\n_\o}\wh\F{}_6^d\Bigr)\\
{}+\wt g^\(\a\eta)g_\[\eta\o]\bigl(2\gv{\n_\a}\wh\F{}_6^d\gv{\n_\mu}\wh\F{}_6^k
-2\xi^2\gv{\n_\mu}\wh\F{}_6^k\gv{\n_\a}\wh\F{}_6^d
-4\z^2\xi \gd k,n,d, \gv{\n_\mu}\wh\F{}_6^k\gv{\n_\a}\wh\F{}_6^d\bigr)
\Bigr]\Biggr\}
\e
where $\gv{\n_\mu}\F_5$ is given by the formula \er{C.4a} before an
electro-weak \s y breaking and by the formula \er{C.27} after an electro-weak
\s y breaking (in a gauge~$U$),
\beq D.2
\gv{\n_\mu}\wh\F{}_6= \frac1{\sin\psi}\gv{\n_\mu}\F_6
\e
and $\gv{\n_\mu}\F_6$ is given by the formula \er{C.5} before an electro-weak
\s y breaking and by the formula \er{C.28} after an electro-weak
\s y breaking (in a gauge~$U$). $\z$ is pure imaginary. We should do
a rescaling \er{B.17} in \er{C.4a} and \er{C.27}.

One can derive $\cL_{\rm int}$ (see Eq.~\er{3.81}) and gets
\beq D.3
\cL_{\rm int}=\frac{i\z\sqrt6}{16\pi r^2(1+\z^2)}\bigl(3F_\m^3 g^\m
(\vf_1\vf_1^* - \vf_2\vf_2^*) - \vf_1\vf_2^*(F_\m^-g^\m+F_\m^+g^\m)\bigr).
\e
One eventually gets in a gauge $U$
\beq D.3a
\cL_{\rm int}=-\frac{i\z 3\sqrt6\,(v+H(x))^2}{64\pi r^2(1+\z^2)}
(Z_\m^{0u}+\sqrt3\,F_\m)g^\m.
\e

Let us notice the following fact. In the formulas \er{D.1} and \er{D.3} an
\ia\ between Higgs' and ``gauge'' fields is \ci, however non-minimal. The
\ia\ between gravity and Higgs' fields has a non-classical kinetic term. The
real significance of this \ia\ demands more investigations.

They are ``interference effects'' between Higgs' field from GSW model and
a \nos\ gravity being an effect of a \un. $\cL_{\rm int}$ is an effect of a
\un\ as well.

Let us consider $\cL_{\rm kin}(\gv\n \F)$ in a Minkowski space $g_\m=\eta_m$
and let us suppose that electroweak \s y breaking took place. One gets
\bml D.4a
\!\!\cL_{\rm kin}(\gv\n\F)=\frac1{2\pi r^2(1+\z^2)}\,\eta^{\o\mu}\Bigl[
\bigl((\xi\z(2\xi^2-\z-2\xi\z-2)+(\pi-2)\z)\gd k,b,d,k_{bk}-2h_{dk}\bigr)
\gv{\n_\o}\F_5^d \gv{\n_\mu}\F_6^k\\
{}+\bigl(\z(2(\z+1)+\xi^2)h_{nk} + 2\xi(\xi(\z\xi+1)-\z)\gd k,n,d,k_{nk}\bigr)
\gv{\n_\o}\F_5^k \gv{\n_\o}\wh\F{}_6^d\\
{}+(h_{nk}-2\xi^2(2\z-1)\gd k,p,n,k_{pk})\gv{\n_\mu}\wh\F{}_6^k \gv{\n_\o}
\wh\F{}_6^n\Bigr]
\e
where
\begin{gather}
\gv{\n_\mu}\F_5 = \frac1{2\sqrt2}\Bigl[
-\frac12(v+H(x))(W_\mu^{-u}x_\a + W_\mu^{+u}x_{-\a})\hskip100pt \nn \\
{}+\Bigl(\pa_\mu H+\frac12\,i(v+H(x))\Bigl(W_\mu^{+u}+\frac12(\sqrt3\,Z_\mu^{0u}
-A_\mu)\Bigr)x_\b\Bigr)\\
\hskip100pt{}-\Bigl(\pa_\mu H+\frac i{\sqrt3}(v+H(x))Z_\mu^{0u}\Bigr)
x_{-\b}\Bigr] \lb  D.5a \\
\gv{\n_\mu}\wh\F_6= \frac1{2\sqrt2}\Bigl(\frac12\,i(v+H(x))(W_\mu^{-u}x_\a
+W_\mu^{+u}x_{-\a})\Bigr) \nn \hskip100pt \\
{}-\Bigl(\pa_\mu H+ \frac12\,i(v+H(x))\Bigl(W_\mu^{+u}+\frac12(\sqrt3\,Z_\mu^{0u}
-A_\mu)\Bigr)x_\b\Bigr) \nn \\
\hskip100pt{}+\Bigl(\pa_\mu H+\frac i{\sqrt3}(v+H(x))Z_\mu^{0u}\Bigr)
x_{-\b}. \lb D.6a
\end{gather}

\goodbreak
Let us come back to the Higgs' \pt. One can write
\bml D.4
V(\vf_1,\vf_2)=\frac1{V_2}\,\frac{2\pi^2}{\sqrt{1+\z^2}}\,\k\Bigl(
\frac12\,i(2-|\vf_1|^2)h_\a+\frac12\,i(2-|\vf_2|^2)h_\b
-\frac32\,i\vf_1\vf_2^* x_\g - \frac32\,i\vf_1\vf_2^*x_{-\g},\\
{}\frac12\,i(2-|\vf_1|^2)h_\a +\frac12\,i(2-|\vf_2|^2)h_\b
-\frac32\,i\vf_1\vf_2^* x_\g - \frac32\,i\vf_1\vf_2^*x_{-\g}\Bigr)
\e
where
\bg D.5
\k(x,y)=\k_{ad}x^ay^d\\
\k_{ad}=(1-2\z^2)h_{ad}+\xi^2\gd k,c,d,k_{ce} \lb D.6
\e
and $\z$ is pure imaginary.

This is of course in the case of Hermitian Kaluza--Klein Theory. This
generalized \pt\ is much more complicated than in the GWS model and can go to
some complicated Higgs' sector structure. Moreover, in the simplest case for
$\xi=0$ it can predict a good agreement with an experiment for a pattern of
masses for $W,Z^0$ bosons and Higgs' boson.

The Higgs' \pt\ can be written in the following form
\bg D.7
V(\vf_1,\vf_2)=V_1(\vf_1,\vf_2)+V_2(\vf_1,\vf_2)\\
V_1(\vf_1,\vf_2)=-\frac{\pi(1-2\z^2)}{2(1+\z^2)r^2}
\Bigl[\frac9{\g\cdot\g}\, |\vf_1|^2|\vf_2|^2
-\frac1{\a\cdot\a}\,(2-|\vf_1|^2)^2\hskip100pt \nn \\
\hskip100pt {}-\frac1{\b\cdot\b}\,(2-|\vf_2|^2)^2
-\frac{2(\a\cdot\b)}{(\a\cdot\a)(\b\cdot\b)}\,(2-|\vf_1|^2)(2-|\vf_2|^2)
\Bigr] \lb D.8 \\
V_2(\vf_1,\vf_2)=\frac{\pi\xi^2}{2(1+\z^2)r^2}\biggl[
-\frac14(2-|\vf_1|^2)^2 \ov K(h_\a,h_\a)
-\frac12(2-|\vf_1|^2)(2-|\vf_2|^2)\ov K(h_\a,h_\b)\hskip30pt\nn \\
+\frac32(2-|\vf_1|^2)\vf_1\vf_2^* \ov K(h_\a,x_\g)
+\frac32(2-|\vf_1|^2)\vf_1\vf_2^* \ov K(h_\a,x_{-\g}) \nn \\
{}-\frac14(2-|\vf_2|^2)^2 \ov K(h_\b,h_\b)
+\frac32(2-|\vf_2|^2)\vf_1\vf_2^* \ov K(h_\b,x_\g)
+\frac32(2-|\vf_2|^2)\vf_1\vf_2^* \ov K(h_\b,x_{-\g}) \nn \\
\hskip50pt{}-\frac94(\vf_1\vf_2^*)^2 \ov K(x_\g,x_\g)
-\frac94(\vf_1^*\vf_2)^2 \ov K(x_{-\g},x_{-\g})
-\frac94|\vf_1|^2|\vf_2|^2\ov K(x_{-\g},x_{-\g})\biggr] \lb D.9
\e
where
\beq D.10a
\bga
\ov K(x,y)=\ov k_{de}x^dy^e = \gd k,c,d, k_{ce}x^dy^e,\\
\gd k,c,d,=h^{cb}k_{bd}, \q \ov k_{dc}=\ov k_{cd}, \q k_{ab}=-k_{ba}
\ega
\e
(a right \iv t tensor on $H=G2$).

$V_1(\vf_1,\vf_2)$ is a part of Higgs' \pt\ known in Manton model corrected
by a \ct\ from \NK\ (Hermitian version).
$V_2(\vf_1,\vf_2)$ is an additional term from Hermitian Kaluza--Klein Theory
and can give additional Higgs' phenomena. Moreover, according to an
experiment we do not see new phenomena. Moreover, we should do a rescaling
\er{B.17} and use primed fields.

One gets
\bml D.16a
V'_1(\vf'_1,\vf'_2) = V_1\Bigl(g\Bigl(\frac{\a\cdot\a}{\g\cdot\g}\Bigr)^{1/2}
\vf'_1, g\Bigl(\frac{\a\cdot\a}{\g\cdot\g}\Bigr)^{1/2}\vf'_2\Bigr)\\
{}=-\frac{\pi(1-2\z^2)}{(1+\z^2)r^2}\biggl[
\frac{9g^2(\a\cdot\a)}{(\g\cdot\g)^3}\,|\vf'_1|^2|\vf'_2|^2
-\frac1{(\a\cdot\a)}\Bigl(2-\frac{g^2(\a\cdot\a)}{(\g\cdot\g)}\,|\vf'_1|^2\Bigr)^2
-\frac1{(\b\cdot\b)}\Bigl(2-\frac{g^2(\a\cdot\a)}{(\g\cdot\g)}\,|\vf'_2|^2\Bigr)^2\\
{}-\frac{2(\a\cdot\b)}{(\a\cdot\a)(\b\cdot\b)}\Bigl(2-\frac{g^2(\a\cdot\a)}{(\g\cdot\g)}
|\vf'_1|^2\Bigr)\Bigl(2-\frac{g^2(\a\cdot\a)}{(\g\cdot\g)}|\vf'_2|^2\Bigr)\biggr]
\e
\bml D.17
V'_2(\vf'_1,\vf'_2) = V_2\Bigl(g\Bigl(\frac{\a\cdot\a}{\g\cdot\g}\Bigr)^{1/2}
\vf'_1, g\Bigl(\frac{\a\cdot\a}{\g\cdot\g}\Bigr)^{1/2}\vf'_2\Bigr)\\
{}=\frac{\pi\xi^2}{2(1+\z^2)r^2}\biggl[
-\frac14\Bigl(2-\frac{g^2(\a\cdot\a)}{(\g\cdot\g)}|\vf'_1|^2\Bigr)^2\ov K(h_\a,h_\a)\\
{}-\frac12\Bigl(2-\frac{g^2(\a\cdot\a)}{(\g\cdot\g)}|\vf'_1|^2\Bigr)
\Bigl(2-\frac{g^2(\a\cdot\a)}{(\g\cdot\g)}|\vf'_2|^2\Bigr)\cdot \ov K(h_\a,h_\b)\\
{}+\frac{3 g^2(\a\cdot\a)}{2(\g\cdot\g)}\Bigl(2-\frac{g^2(\a\cdot\a)}{(\g\cdot\g)}|\vf'_1|^2\Bigr)
\vf_1'\vf'_2{}^* \bigl(\ov K(h_\a,x_\g)+\ov K(h_\a,x_{-\g})\bigr)\\
{}-\frac14\Bigl(2-\frac{g^2(\a\cdot\a)}{(\g\cdot\g)}|\vf'_2|^2\Bigr)^2 \ov K(h_\b,h_\b)
+\frac{3g^2(\a\cdot \a)}{2(\g\cdot\g)}\Bigl(2-\frac{g^2(\a\cdot\a)}{(\g\cdot\g)}|\vf'_2|^2\Bigr)
\bigl(\ov K(h_\b,x_\g)+\ov K(h_\b,x_{-\g})\bigr)\\
{}-\frac{9g^4(\a\cdot\a)^2}{4(\g\cdot\g)^2}\bigl((\vf'_1\vf'_2{}^*)^2\ov K(x_\g,x_\g)
+(\vf'_1{}^*\vf'_2)^2\ov K(x_{-\g},x_{-\g})\bigr)
-\frac{9g^4(\a\cdot\a)^2}{(\g\cdot\g)^2}\,|\vf'_1|^2|\vf'_2|^2\ov K(x_{-\g},x_{-\g})\biggr]
\e

\def\wg#1#2{\wt g{}^\(#1#2)}
\def\bn{{\bar n}}
\def\bk{{\bar k}}
\def\bc{{\bar c}}
\def\bd{{\bar d}}

Eq. \er{4.33} can be rewritten in the form below (see Eq.~\er{2.38}):
\beq D.10
\bga
\cLY = \frac1{8\pi}\biggl[
-\frac1{\g\cdot\g}\Bigl[\d_{\bn\bk}H^{\bk \o\mu}\gd H,\bn,\o\mu,
-2\d_{\bc\bd}H^\bc H^\bd+ 2\d_{\bn\bk} H^{\bk\o\mu}\gd H,\bn,\d\o,
g_\[\a\mu]\wg\a\d \Bigr]\\
{}+\xi \Bigl[2k_{\bn\bk}H^{\bk\o\mu}\gd H,\bn,\d\o,\wg\d\a g_\[\a\mu]
-2k_{\bk\bd}H^{\bk\o\mu}\gd H,\bd,\d\a, \wg\d\b \wg\a\rho g_\[\b\o]g_\[\rho\mu]\\
{}-k_{\bk\bd}H^{\bk\o\mu}\gd H,\bd,\eta\o, \wg\eta\b \wg\a\rho g_\[\mu\a]
g_\[\b\rho]
+k_{\bk\bd}H^{\bk\o\mu}\gd H,\bd,\eta\o, \wg\eta\d \wg\a\rho g_\[\d\b]
r_\[\o\d]\Bigr] \\
{}+\xi^2\Bigl[k_{n\bk}\gd k,n,\bd, H^{\bk\o\mu}\gd H,\bd,\eta\mu, \wg\rho\b
\wg\eta\a g_\[\o\b]g_\[\a\rho]
-2k_{n\bk}\gd k,n,\bd,H^{\bk\o\mu}\gd H,\bd,\d\a, \wg\d\eta \wg\a\rho
g_\[\eta\o]g_\[\rho\mu]\\
{}-k_{n\bk}\gd k,n,\bd, H^{\bk\o\mu}\gd H,\bd,\eta\o, \wg\rho\a \wg\eta\b
g_\[\mu\a]g_\[\b\rho]
+\dg k,\bk,b, k_{b\bd}H^{\bk\o\mu}\gd H,\bd,\a\o, \wg\a\b g_\[\mu\a]\\
{}-\dg k,\bk,b, k_{b\bd}H^{\bk\o\mu}\gd H,\bd,\a\mu, \wg\a\b g_\[\o\b]
+\gd k,p,\bn, k_{p\bk}H^{\bk\o\mu}\gd H,\bn,\o\mu,\Bigr] \\
{}+\xi^3\Bigl[ k_{n\bk}k^{nb}k_{b\bd}H^{\bk\o\mu}\gd H,\bd,\a\o,
\wg\a\b g_\[\mu\b]
-k_{n\bk}k^{nb}k_{b\bd}H^{\bk\o\mu}\gd H,\bd,\a\mu,\wg\a\b g_\[\o\b]\Bigr]
\biggr]
\ega
\e
where $H^\bc=\gd H,\bc,\m,g^\[\m]$, $\bn,\bk,\bc,\bd=1,2,3,4$ in such a way that
\beq D.11
e^*\o_E = A_\mu \ov\t{}^\mu + B_\mu \ov\t{}^\mu = (\dg A,\mu,a, t_a+
B_\mu y)\ov\t{}^\mu
\e
($a=1,2,3$), $t_1=\frac12\,i(x_\g+x_{-\g})$, $t_2=\frac12\,(x_\g-x_{-\g})$,
$t_3=\frac12\,ih_\g=\frac i{\g\cdot\g}(\g_1H_1+\g_2H_2)$,
$t_4=y=\frac12\,ih=\frac i{\g\cdot\g}(\g_1H_2-\g_2H_1)$,
\beq D.12
e^* \O_E = e^*\Bigl(\frac12\,H^\bk \t^\mu \wedge \t^\mu\Bigr)
=\frac12(F_\m^a t_a + B_\m y)\ov\t{}^\mu \wedge \ov\t{}^\mu,
\e
$a,b,c=7,8,\dots,20$ ($\fh=G2$) in such a way that
$$
c=\a,-\a,\b,-\b,\g,-\g,\a',-\a',\b',-\b',\g',-\g',19,20,
$$
where $\a,\b$ etc.\ correspond to 12 roots of the algebra $G2$ and to
generators $x_\a,x_\b$ etc., $19,20$ correspond to the generators
$H_1,H_2$---\el s of Cartan subalgebra of $G2$.

One gets
\beq D.13
\ell_{\bn\bd}=-\frac1{\g\cdot\g}\,\d_{\bn\bd} + \xi \a_\bn^c \a_\bd^d k_{cd}
\e
where
\beq D.14
\a_\bn^c x_c = t_\bn,
\e
$\bn=1,2,3$, $t_\bn=t_a$, $\bn=4$, $t_4=y$.

In this way we have
\beq D.15
\bal
\gd H,\bk,\o\mu,&\to F_{\o\mu}^a, \q \bk=1,2,3,\\
\gd H,4,\o\mu, &\to B_{\o\mu}.
\eal
\e

In a gauge $U$ one gets
\beq D.16
\bal
\gd F,a,\o\mu,&\to W_{\o\mu}^{au}, \q a=1,2,\\
\gd F,3,\o\mu,&\to \gd F,3u,\o\mu,=\frac12(Z_{\o\mu}^{0u}+\sqrt3\,F_{\o\mu})\\
B_\m &=\frac12(\sqrt3\, Z_\m^{0u}-F_\m).
\eal
\e

One should remember that we have to  do with Hermitian version of \NK\
(Hermitian Kaluza--Klein Theory). Moreover, we consider
Hypercomplex--Hermitian version which is effectively equivalent
to a real version of a theory.

\section*{Appendix E}
\def\theequation{E.\arabic{equation}}
\setcounter{equation}0
In this appendix we consider our $\d$-deviation from $\t_W=\frac\pi6$ in
a deeper level. It means we consider a $\D r$ theory up to the second order
known in the literature (see Ref.~\cite{x2} and references therein, see also
\cite{w2}, \cite{v2}). We have
\bg E.1
\D r=\Bigl(1-\frac{\a_{\rm em}(0)}{\a_{\rm em}(M_{Z^0}^2)}\Bigr)
\Bigl(1-\frac{C_W^2}{S_W^2}\,\D\rho\Bigr)+\D r_{\rm rem}, \nonumber\\
C_W^2=\cos^2\t_W,\q S_W^2=\sin^2\t_W,
\e
where
\bg E.2
\D\rho = 3x_t\bigl(1+x_t \rho^{(2)}(z)+ \d\rho_{QCD}\bigr) \\
x_t=\frac{G_F m_t^2}{8\pi^2\sqrt2} \lb E.3 \\
\d\rho_{QCD} = -\frac{\a_s(\mu)}\pi \,c_1 + \Bigl(\frac{\a_s(\mu)}\pi
\Bigr)^2 c_2(\mu) \lb E.4 \\
c_1= \frac23 \Bigl(\frac{\pi^2}3 +1 \Bigr) \lb E.5 \\
c_2=-14.59 \lb E.6 \\
\rho^{(2)}(z)=\frac{49}4 +\pi^2 + \frac{27}2\log z +\frac32\log^2 z
+\frac z3\bigl(2-12\pi^2+12\log z-27\log^2z\bigr)\hskip50pt \nonumber \\
\hskip50pt {}+\frac{z^2}{48}\bigl(1613 - 240\pi^2 - 1500\log z - 720\log^2 z\bigr)
\lb E.7 \\
z=\frac{m_t^2}{m_H^2} \lb E.8 \\
\D r_{\rm rem} = \frac{\sqrt2\,G_F M_W^2}{16\pi^2} \cdot \frac{11}3
\Bigl(\ln \Bigl(\frac{M_H^2}{M_W^2}\Bigr)-\frac56\Bigr). \lb E.9
\e
From the \e
\beq E.10
\sin^2\t_W = \sin^2(\tfrac \pi6 +\d)=4(1-\D r)
\e
one gets
\beq E.11
\d=-\frac{\sqrt3 \bigl(1-\frac{\a_{\rm em}(0)}{\a_{\rm em}(M_{Z^0}^2)}
-3\D\rho\bigl(1+\frac{\a_{\rm em}(0)}{\a_{\rm em}(M_{Z^0}^2)}\bigr)+
\D r_{\rm rem}\bigr)}
{6\bigl(1-4\D\rho\bigl(1-\frac{\a_{\rm em}(0)}{\a_{\rm em}(M_{Z^0}^2)}\bigr)
\bigr)}.
\e
Taking
\bg E.12
\a_s(M_{Z^0}^2) = 0.1185 \\
G_F = G_\mu = 1.66378 \tm 10^{-5}\,({\rm GeV})^{-2} \lb E.13 \\
m_t = 173.21 \,{\rm GeV} \lb E.14 \\
M_H = 125.7\,{\rm GeV} \lb E.15 \\
M_W = 80.385\,{\rm GeV} \lb E.16 \\
M_{Z^0} = 91.18\,{\rm GeV} \lb E.17 \\
\a_{\rm em}(0) \simeq \a_{\rm em}(m_e^2) = \frac1{137.035} \lb E.18 \\
\a_{\rm em}(M_{Z^0}^2) \simeq \a(M_W^2)=\frac1{128} \lb E.19 \\
\D\rho = -0.000044702565 \lb E.20
\e
and eventually
\bg E.21
\d = -0.01652297 \\
\hbox{or} \q \d=-56'48.108'' \lb E.22 \\
\sin^2\t_W = \sin^2\bigl(\tfrac\pi 6+\d\bigr) = 0.23583 \lb E.23 \\
\hbox{and}\q M_W= 79.7067 \lb E.24
\e
which is almost correct value of a mass of $W^\pm$ bosons,
\beq E.25
\t_W = 29^\circ 3' 11.898''.
\e
It seems that this is self-consistent.

\section*{Conclusions and prospects for further research}
In the paper we consider a color \cfn\ in the \nos\ \nA\ Kaluza--Klein
Theory. We derive a condition for a dielectric \cfn\ in the theory. We remind
to the reader some notions of the \NK\ and a new version of the \KWK \e\ in
this theory. We solve constraints in \E\nos\ (\E\nA) Kaluza--Klein Theory
and also constraints in the \NK\ with \sn\ \s y breaking and \Hm.

In our geometrical unification we consider all \ia s unified by one \cn\
defined on many \di al manifold (see Refs \cite{1,*3}).
We consider also \sn\ \s y breaking and \Hm\ in the \NK\ in a general scheme
applicable in a general version for a \un\ of the \nos\ gravity (NGT) with a
grand unified model of gauge field \ia s in a bosonic sector. We combine in
this case a \di al reduction model with Kaluza--Klein Theory. In this
approach Higgs' field is a part of a \YM' field on an extended \spt\ with a
\s y. A~base manifold $V=E\tm M$, where $M=G/G_0$ is a vacuum state manifold
(classical vacuum). This approach has been suggested for the first time in
Ref.~\cite{*5}. We derive a \gn\ of the \KWK \e\ for a case with a Higgs'
field presence.

Due to a geometrical origin of these \e s we get a new kind of a ``charge''
which couples to Higgs' field as an electric charge couples to an \elm c
field in a Lorentz force term. This charge is a \gn\ of a color (isotopic)
charge which couples to \YM' field.

We consider also a Manton model of electro-weak \ia s in the framework of
\NK. In this way we unify \elm c and weak \ia\ (a~bosonic sector) with a
\nos\ gravity (NGT). (This is a 6-\di al Manton model with $G2$ group.) We
get a possibility to obtain a realistic mass spectrum for $W^\pm$, $Z_0$
bosons and a recently discovered Higgs' boson. Our \un\ is justified by the
fact that a small correction to $\t_W=\frac\pi6$ (Weinberg angle) obtained in
the theory can be got by  renormalization procedure known in the literature
($\D r$~theory).

Let us give the following remark. The classical 5-\di al Kaluza--Klein Theory
(formulated as a metrized \elm c fiber bundle) gives the exact results of
Maxwell electrodynamics with a Lorentz force term and Einstein General
Relativity, on a unified geometrical basis. This theory can be considered as
a quintessence of classical physics, even it does not give any ``interference
effects'' between gravity and \elm c theory. Our 20-\di al \un\ of Hermitian
gravity and GSW model (a~bosonic part) in a framework of
Hermitian Kaluza--Klein Theory
with \sn\ \s y breaking can be treated as a prequantum geometrical \un\ of
gravity and electro-weak \ia s. Our 12-\di al \un\ of Hermitian gravity and
Nonabelian \YM' field for $G=\SU(3)$ into Hermitian Nonabelian Kaluza--Klein
Theory can be treated as prequantum geometrical \un\ of gravity and strong
\ia s (a bosonic part of~QCD).
Both \un s give ``interference effects'' between gravity, electro-weak
\ia s and strong \ia s.

There are some further prospects for a research. First of all it is necessary
to incorporate fermions in the theory.

The beautiful theories as Kaluza--Klein theory (a~Kaluza miracle) and its
descendents should pass the \fw\ test if they are treated as real unified
theories. They should incorporate chiral fermions. Since the \fn\ scale in
the theory is a \Pl's mass, fermions should be massless up to the moment of
\sn\ \s y breaking. Thus they should be zero modes. In our approach they can
obtain masses on a \di al reduction scale. Thus they are zero modes in
$(4+n_1)$-\di al case. In this way $(n_1+4)$-\di al fermions are not chiral
(according to the very well known Witten's argument on an index of a Dirac
operator). Moreover, they are not zero modes after a \di al reduction, i.e.\
in 4-\di al case. It means we can get chiral fermions under some assumptions.

We should look for some possibilities of Grand Unified Models (see
Ref.~\cite{31}). First of all we should look for a group~$G$ \st
$$
\SU(3)_c \ot \SU(2)_L \ot \U(1)_Y \subset G.
$$
There are a lot of possibilities. One of most promising is $G=\SO(10)$.
Moreover, we need also a group $G_0$ \st $M=G/G_0$ (see Refs \cite{20,21}).
In our world $G_0=\SU(3)_c \ot \U(1)_{\rm em}$. The group $H$, $G=\SO(10)$ and
$G_0=\U(1)_{\rm em}\ot \SU(3)_c$ should be \st
$$
\SO(10)\ot (\SU(3)_c\ot \U(1)_{\rm em})\subset H.
$$
The simplest choice is $H=\SO(16)$. Why? First of all $G2\subset \SO(16)$ and
$\SO(10)\ot \SO(6)\subset \SO(16)$. Moreover, $\SO(6)\simeq \SU(4)$ and
$\SU(3)\ot \U(1)\subset\SU(4)$. Thus if we identify $\U(1)$ with $U(1)_{\rm em}$ and
$\SU(3)$ with $\SU(3)_c$ we get what we want. In this way
$$
M=\SO(10)/\SU(3)\ot \U(1),
$$
$S^2\subset M$, $\dim\SO(16)=120$, $\dim\SO(10)=45$, $n_1=\dim M=36$. There
is also a possibility to consider a different possibility
$$
M'=\SO(10)/\SU(3)
$$
and $\SO(10)\ot \SU(3)\subset H$ for a $\U(1)$ is an Abelian factor, which is
a little in a spirit of the Manton approach for GSW model (see
Ref.~\cite{21}). Coming back to the problem of fermions we can try to couple
multi\di al spinors in a minimal coupling scheme to multi\di al \cn s
describing a unified field theory. In this way we can get chiral fermions
coupled to gravity, \YM' and Higgs' fields.

Thus a Yukawa mechanism is possible in our approach. The Yukawa sector in the
theory can be obtained due to a minimal coupling to a total \ci\ \dv\
(``gauge'' and \wrt a \LC \cn\ generated by a \s ic part on a multi\di al
metric on a total $(4+n+n_1)$-\di al manifold at one) to many-\di al spinor
in $2^N$-\di al space, where $N={\rm Ent}(\frac{4+n+n_1}2)$. In this way we
can write a Lagrangian of the spinor field in the form
$\frac12i\hbar c(\ov\Ps \G^M \gv{\wt\n_M}\Ps + \gv{\wt\n_M}\ov\Ps \G^M\Ps)$,
where $\gv{\wt\n_M}$ is a \dv\ mentioned above. $\G^M$ are $2^N$-\di al
\gn\ of Dirac matrices and $\ov\Ps=\G^4\Ps^+$. Due to a \di al reduction
procedure, taking only zero-modes for $2^N$-\di al spinor we can get
4-\di al spinors defined on a \spt~$E$. We can try to get chiral spinors
(a~Witten argument of an index of a Dirac operator does not work on a
4-\di al space) and also to arrange many-\di al ($2^N$-\di al) spinor as a
collection of 4-\di al spinors to get fermions (known from an experiment).
Due to a coupling to many-\di al \YM' field (after a \di al reduction
decomposed into 4-\di al \YM' field and a multiple of scalar (four-\di al)
fields---Higgs' fields) we get a Yukawa-type terms for 4-\di al spinors.
Thus due to a \Hm\ (geometrized in our theory) we get a pattern of masses
for 4-\di al fermions. The scale of mass for such fermions is given by a
parameter~$r$---a~radius of a manifold $M=G/G_0$. Very heavy fermions with
masses of order of a Planck's mass are removed from the theory by conditions
of zero-mode for~$\Ps$. The bare masses obtained here can interact according
to the Newton law.

Generalized Dirac matrices are defined by the relations
$$
\{\G^A,\G^B\}=2\eta^{AB} \qh{or} \{\G^{\tA},\G^{\tB}\}=2\eta^{\tA\tB}
$$
where
$$
\bal \eta^{AB}&=\mathop{\rm diag}\{-1,-1,-1,1,\underbrace{-1,\dots,-1}_n\}\\
\eta^{\tA\tB}&=\mathop{\rm diag}\{-1,-1,-1,1,\underbrace{-1,\dots,-1}_{n_1},\underbrace{-1,\dots,-1}_n\}.
\eal
$$
For $(n+4)$ or $(n+n_1+4)$ equal to $2l+2$ (the even case) we define
$$
\bal \G^{4\pm}&=\tfrac12(\pm\G^4+\G^1),\\
\G^{\bar A\pm}&=\tfrac12(\G^{2\bar A}\pm i\G^{2\bar A+1}), \q \ov A=1,\dots,l.
\eal
$$
It is easy to show
$$
\bga
\{\G^{\bar A\pm},\G^{\bar B\pm}\}=-\d^{\bar A\bar B}\\
\{\G^{\bar A+},\G^{\bar B+}\}=\{\G^{\bar A-},\G^{\bar B-}\}=0.
\ega
$$
In particular
$$
(\G^{\bar A+})^2=(\G^{\bar A-})^2=0.
$$
In this way we always have a spinor $\Ps_0$ \st
$$
\G^{\bar A-}\Ps_0=0
$$
for all $\ov A$. We get all possible spinors acting on $\Ps_0$ by $\G^{\bar A+}$.
We get $2^{l+1}$ such spinors (a~full \rp ation). $\G^A$ or $\G^\tA$ can be
derived in such a base by using iterative method.

In the case of $2l+3$ (an odd case) we should have
$$
\G^{2l+3}=i^{-(l+1)}\G^1 \cdots \G^{2l+2}
$$
\st
$$
(\G^{2l+3})^2=-1, \q \{\G^{2l+3},\G^{\bar{\bar A}}\}=0, \q \ov{\ov A}=1,\dots,2l+2.
$$
It is easy to define a basis of spinors for both cases. Let $\z=(\z_1,\dots,
\z_l)$, $\z_{\bar A}=\pm\frac12$,
$$
\Ps_\z = \Bigl(\prod_{\bar A=0}^l (\G^{(l+\bar A)})^{\z_{(l+\bar A)}+1/2} \Bigr)\Ps_0.
$$
$\G^{2l+3}$ in the even case distinguishes between two classes of spinors
$$
\bal \G^{2l+3}\Ps_\z &= +\Ps_\z \qh{($2^l$-\di---first \rp ation)}\\
\G^{2l+3}\Ps_\z &= -\Ps_\z \qh{($2^l$-\di---second \rp ation)}
\eal
$$
In the odd case we have only one \rp ation of $2^{l+1}$-\di.

We can introduce also generators of $\SO(1,3+n)$ or $\SO(1,3+n_1+n)$ algebra
$$
\bga
\wh\si{}^{AB} \qh{or} \wh\si{}^{\tA\tB}\\
\wh\si{}^{AB}=\frac i4[\G^A,\G^B]\\
\wh\si{}^{\tA\tB}=\frac i4[\G^\tA,\G^\tB].
\ega
$$
We have of course
$$
\bga
{}[\wh\si{}^{MN},\wh\si{}^{RS}] = -i [\eta^{NS}\wh\si{}^{MR}+ \eta^{RN}\wh\si{}^{SM}
+\eta^{MR}\wh\si{}^{NS}+\eta^{SM}\wh\si{}^{RS}]\\
[\wh\si{}^{\td M\td N},\wh\si{}^{\td R\td S}] = -i [\eta^{\td N\td S}\wh\si{}^{\td M\td R}+ \eta^{\td R\td N}\wh\si{}^{\td S\td M}
+\eta^{\td M\td R}\wh\si{}^{\td N\td S}+\eta^{\td S\td M}\wh\si{}^{\td R\td S}].
\ega
$$
Our spinors transform as
$$
\bal
\PS&\to \exp\bigl(\tfrac12 \a_{AB}\wh\si{}^{AB})\Ps \\
\hbox{or}\q \PS&\to \exp\bigl(\tfrac12 \a_{\tA\tB}\wh\si{}^{\tA\tB})\Ps \\
&\a_{AB}=-\a_{BA}\\
&\a_{\tA\tB}=-\a_{\tB\tA}.
\eal
$$
We also have
$$
\bal
(\wh\si{}^{AB})^+ \G^4&=\G^4 \wh\si{}^{AB}\\
(\wh\si{}^{\tA\tB})^+ \G^4&=\G^4 \wh\si{}^{\tA\tB}.
\eal
$$

In our particular cases with or without \sn\ \s y breaking we get our
matrices using ordinary Dirac matrices and their tensor products with some
special matrices. One gets for \ci\ \dv s
$$
\bal
\wt D\Ps&=d\Ps + \tfrac12\wt\o_{AB}\wh \si{}^{AB}\Ps\\
\wt D\Ps&=d\Ps + \tfrac12\wt\o_{\tA\tB}\wh \si{}^{\tA\tB}\Ps.
\eal
$$
Moreover, we use as before (see Ref.~\cite8)
$$
\bal
\gv{\wt D}\Ps &= \hor \wt D\Ps = \gv{d}\Ps+\tfrac12\hor(\wt\o_{AB})\wh\si{}^{AB}\Ps\\
\gv{\wt D}\Ps &= \hor \wt D\Ps = \gv{d}\Ps+\tfrac12\hor(\wt\o_{\tA\tB})\wh\si{}^{\tA\tB}\Ps
\eal
$$
and also
$$
\bal
\wt D\ov\Ps &= d\ov\Ps - \tfrac12\wt\o_{AB}\ov\Ps \wh\si{}^{AB}\\
\hbox{or} \q \wt D\ov\Ps &= d\ov\Ps
- \tfrac12\wt\o_{\tA\tB}\ov\Ps \wh\si{}^{\tA\tB}
\eal
$$
where
$$
\ov\Ps = \Ps^+ \G^4
$$
and similarly
$$
\bal
\gv{\wt D}\ov\Ps &= \gv{d}\ov\Ps - \tfrac12\hor(\wt\o_{AB})\ov\Ps
\wh\si{}^{AB}\\
\hbox{or} \q \gv{\wt D}\ov\Ps &= \gv{d}\ov\Ps
- \tfrac12\hor(\wt\o_{\tA\tB})\ov\Ps \wh\si{}^{\tA\tB}.
\eal
$$
$\wt\o_{AB}$ and $\wt\o_{\tA\tB}$ are \LC \cn s defined on~$P$ \wrt a \s ic
part of metrics $\g_\(AB)$ and $\g_\(\tA\tB)$.

How does an iterative method for a construction of $\G$ matrices work? Let us
suppose we have ordinary Dirac matrices $\g^\mu$ and let us define
$$
\bal
\G^\mu &= \g^\mu \otimes \left(\begin{matrix} -1 &\ & 0\\ 0 && 1
\end{matrix}\right), \q \mu=1,2,3,4, \\
\G^5 &= I_4 \otimes \left(\begin{matrix} 0 &\ & 1 \\ 1 && 0
\end{matrix}\right), \\
\G^6 &= I_4 \otimes \left(\begin{matrix} 0 &\ & -i \\ i && 0
\end{matrix}\right) , \q I_4 \hbox{ an identity matrix, }4\tm4.
\eal
$$
Next step
$$
\bal
\ov\G{}^A &= \G^A \otimes \left(\begin{matrix} -1 &\ & 0 \\ 0 && 1
\end{matrix}\right) , \q A=1,2,3,4,5,6,\\
\ov\G{}^7 &= I_6 \otimes I_6 \otimes \left(\begin{matrix} 0 &\ & 1\\ 1 && 0
\end{matrix}\right),\\
\ov\G{}^8 &= I_6 \otimes I_6 \otimes \left(\begin{matrix} 0 &\ & -i \\ i && 0
\end{matrix}\right), \q I_6 \hbox{ an identity matrix, }6\tm6.
\eal
$$
The Lagrangian for out spinor field (multi\di al) looks like
$$
\cL(\Ps,\ov\Ps,\gv{\wt D}) = i\,\frac{\hbar c}2\, \bigl(\ov\Ps\ell
\land \gv{\wt D}\Ps + \gv{\wt D}\ov\Ps \land \ell\Ps)
$$
where
$$
\ell = \G^\mu\ov\eta_\mu
$$
and $\ov\eta_\mu$ is a dual Cartan basis on $E$.

We also write new type of \ci\ \dv\ $\gv{\wt D}$ as $\cD\Ps$ and $\cD\ov\PS$
(see Ref.~\cite8).

The interesting problem is to find exact \so s of field \e s in the case of
GSW-model of Hermitian Kaluza--Klein  Theory with \sn\ \s y breaking and also
for the \E\nos\ Nonabelian (real or Hermitian) Kaluza--Klein Theory with
$G=\SU(3)$. We expect some nonsingular, particle-like stationary \so s in the case
of spherical \s y. Axially \s ic stationary \so s in both cases seem to be
very interesting from more general point of view and we will seek for them.
These \so s can be considered with and without fermion sources. We also look
for some wave-like \so s: a \nA\ plane wave, spherical and cylindrical waves.
The waves can be considered as gravito-\YM' waves.

Let us give some comments. There are two versions of the \NK: real and
Hermitian. Both versions work very well in the case of 5-\di al (\elm c) and
in the case of \E\nA\ \YM' field. We get charge and color \cfn\ and
nonsingular \so s. However, if we want to apply the theory for GSW model
(a~bosonic part of this model), only Hermitian version works getting pattern
of masses of~$Z^0$ and $W^\pm$ bosons and Higgs' boson agreed with an
experiment. It seems that an experiment chooses the Hermitian version.
In this way an idea of deriving a unified field theory from higher-\di al
gravity is maintained, together with much of the appealing simplicity and
unity of the theory.

Hermitian Kaluza--Klein Theory seems to be closer to quantum theory even it
is a classical field theory. According to A.~Einstein Hermitian version of
Unified Field Theory would be prequantum gravity.

Let us express $E_c$ and $H^e$ in terms of $D^n$ and $B_d$ in the \NK\ (\elm c
case):
$$
E_c=\D^r_c\cdot \ov L_{rn}D^n + \D^r_c \ov K_{re}\ov C{}^{ed}B_d
$$
where $\ov L_{rn},\ov K_{re}$ and $\ov C{}^{ed}$ are given resp.\ by Eqs \er{2.123n},
\er{2.122n}, and by Eq.~(1.73) of Ref.~\cite8 and $\D^r_c$ is an inverse tensor of
$\d^c_r - \ov K_{re}\ov A{}^{ec}$, i.e.
$$
(\d^c_r - \ov K_{re}\ov A{}^{ec})\D^r_d = \d^c_d
$$
where $\ov A{}^{ec}$ is given by Eq.~(1.70) of Ref.~\cite8 and
$$
\bga
\det(\d^c_r - \ov K_{re}\ov A{}^{ec})\ne 0\\
H^e = \Xi_a^e \ov A{}^{ac} \ov L_{cn} D^n + \Xi_a^e \ov C{}^{ad}B_d
\ega
$$
and $\Xi^e_a$ is an inverse tensor of $\d^a_e - \ov A{}^{ec}\ov K_{ce}$, i.e.
$$
(\d^a_e - \ov A{}^{ec}\ov K_{ce})\Xi^e_f=\d^a_f, \q
\det(\d^a_e - \ov A{}^{ec}\ov K_{ce})\ne0.
$$
In the case of Hermitian Kaluza--Klein Theory we define vectors
$$
\mathcal F_c=\frac1{\sqrt2}(D_c+ iB_c)
$$
and
$$
\mathcal G_c=\frac1{\sqrt2}(E_c+iH_c).
$$
One gets
$$
\mathcal G_c=\frac1{\sqrt2}\bigl[(\D^r_c \dg\ov L,r,n, + i\Xi_{ca}\ov A{}^{af}
\dg\ov L,f,n,)D_n
+ (\D^r_c \ov K_{re} \ov C{}^{ed} + i\Xi_{ca}\ov C{}^{ad})B_d\bigr].
$$
We remind to the reader that Latin indices (3-\di al space indices)
are keeping in up or down position
only for convenience. Thus we exchange
$$
\bal
\gd \Xi,c,a, &\to \Xi_{ca}\\
D^n &\to D_n\\
\ov L_{fn} &\to \dg\ov L,f,n,\\
\ov L_{rn} &\to \dg\ov L,r,n,\\
H^c &\to H_c.
\eal
$$
In this way we describe the Riemann--Silberstein vector in the Hermitian
Kaluza--Klein Theory in an \elm c case (see Refs \cite{43}, \cite{44},
\cite{45}, \cite{46}) $\mathcal F_c$ and the second vector $\mathcal G_c$.

For a vector $\mathcal F_c$ is considered as a wave \f\ of a photon we are
closer to the quantum theory. This really is a prequantum theory.

We can do the same in the case of \YM' field getting
$$
\mathcal F^e_{\bar c}=\frac1{\sqrt2}(D^e_{\bar c}+ iB^e_{\bar c})
$$
and
$$
\mathcal G^e_{\bar c}=\frac1{\sqrt2}(E^e_{\bar c}+iH^e_{\bar c}),
$$
which can be calculated in the Hermitian-Nonabelian Kaluza--Klein Theory
using formulas \er{2.44}--\er{2.49} and \er{2.111n}--\er{2.112n}.

How to quantize the \NK? First of all we can quantize it using
Ashtekar--Lewandowski formalism considering it as GR with additional sources,
i.e.\ $g_\[\m]$, gauge fields, Higgs' fields (see Ref.~\cite{17}). This will be
done elsewhere. The second approach is to consider our theory also as~GR with
additional geometrized sources (see Appendix~E of Ref.~\cite8) and develop it
into a nonlocal theory. There are several approaches of quantization of
nonlocal theories (see Refs \cite{A1},~\cite{A2}). In this case we can avoid
infinities appearing in perturbation calculus, getting a theory which is
renormalizable, super-renormalizable and even finite.

Nonlocal theories, roughly speaking, are equivalent to theories with higher
\dv s up to an infinite order. An integral \tf\ is equivalent to a differential
operator of an infinite order.

Moreover, introducing nonlocality or a differential operator of an infinite order
can be considered as a special type of a regularization procedure to remove
infinities from Feynman diagrams calculations in perturbation calculus. It is
possible to consider such a procedure as a generalized (\ci\ and gauge \iv t)
Pauli--Villars regularization procedure. Simultaneously we can quantize the
theory (as an ordinary field theory) using Faddeev--Popov prescription in
path-integral formalism for gravity and \YM' field. A~divergence of a one loop
in the case of gravity can be removed using \di al regularization-renormalization
procedure. In order to avoid massive ghosts we should carefully design higher
\dv\ corrections to gravity, \YM's fields, Higgs' fields using differential
operator of infinite order. According to Ref. \cite{B1,B2} we should add
$$
\ell_{ab}L^{a\m}h_1\Bigl(-\frac{\D}{\La^2}\Bigr)\gd H,b,\m,
$$
where $\D$ is a Laplace operator, a gauge \ci\ and a \ci\ \wrt a \LC \cn\ generated by $g_\(\a\b)$,
$\La$ is a scale:
$$
\D=\wt g{}^\(\a\b)\gv{\wt{\ov \n}_\a}\,\gv{\wt{\ov \n}_\b}.
$$
$h_1$ is an entire \f\ (non-polynomial) which should be carefully chosen.

We should also add
$$
\bga
h_{ab}H^a h_2\Bigl(-\frac\D{\La^2}\Bigr)H^b\\
\ov R(\ov W)h_3\Bigl(-\frac{\n^2}{\La^2}\Bigr)\ov R(W)\\
\ov R_\m(\ov W)h_4\Bigl(-\frac{\n^2}{\La^2}\Bigr)\ov R{}^\m(\ov W)\\
\n^2=\wt g{}^\(\a\b)\wt{\ov \n}_\a\wt{\ov \n}_\b.
\ega
$$
$\ov R(\ov W)$ and $\ov R_\m(\ov W)$ should be expressed by $\wt{\ov R}$,
$\wt{\ov R}_\m$ and additional fields, i.e. $g_\[\m]$ and $\ov W_\mu$.

In the case where we have to do with Higgs' fields and \sn\ \s y breaking
we add also the terms
\begin{gather*}
\frac1{V_2}\int_M \sqrt{|\wt g|}\,d^mx \Bigl(\ell_{ab}\wt g^{\td b\td n}
\gd L,a\mu,b, h_5\Bigl(-\frac\D{\La^2}\Bigr)\gv{\n_\mu}\gd\F,b,\td n,\Bigr)\\
\hbox{and}\q
h_{ab}\gd\mu,a,i, \wt H{}^i h_6\Bigl(-\frac\D{\La^2}\Bigr)g^\[\td a\td b]
\bigl(\gd C,b,cd,\F^c_{\td a}\F^d_{\td b} - \gd \mu,b,\hat\imath, \gd
f,\hat\imath,\td a\td b, - \gd\F,b,\td d, \gd f,\td d,\td a\td b,\bigr)
\end{gather*}
where $h_3,h_4,h_5$ and $h_6$ are entire transcendental \f s of a complex
variable.

The problem which arises now is as follows: Is it possible to choose
$h_1,h_2,h_3,h_4,h_5,h_6$ in such a way that no physical poles are introduced
while the theory will be (super-) renormalizable and unitary. It seems that
such entire transcendental \f s can be defined (see also Refs \cite{C1,C2}).
This will be examined elsewhere.

Let us consider the following Lagrangian in the theory
\begin{multline*}
L=\ov R(W)+\ov R(\ov W)h_3\Bigl(-\frac{\n^2}{\La^2}\Bigr)\ov R(\ov W)
+\ov R_\m(\ov W)h_4\Bigl(-\frac{\n^2}{\La^2}\Bigr)\ov R{}^\m(\ov W)\\
{}+\ell_{ab}L^{a\m}h_1\Bigl(-\frac{\D}{\La^2}\Bigr)\gd H,b,\m,
+h_{ab}H^ah_2\Bigl(-\frac{\D}{\La^2}\Bigr)H^b+\ov R(\ov\G),
\end{multline*}
i.e.\ a Lagrangian \er{2.66} plus higher order in \dv s terms (this Lagrangian
can be extended to the case with Higgs' fields).

Let us apply a path-integral method to quantize \gr al and \YM' field. We write
rather formally (see Ref.~\cite{P1})
$$
\cZ=\int e^{iS[A,g,\bar W]} \cD A\cD g \cD \ov W,
$$
where $S$ is a classical action
\begin{align*}
\cD A&=\prod_\mu \cD A_\mu=\prod_{x,\mu} dA_\mu(x)\\
\cD g&=\prod_{\substack{\a,\b\\\a\le\b}}Dg_\(\a\b)\prod_{\substack{\a,\b\\\a<\b}}
Dg_\[\a\b]=\prod_{\substack{x,\a,\b\\\a\le\b}} dg_\(\a\b)(x)
\prod_{\substack{x,\a,\b\\\a<\b}} dg_\[\a\b](x)\\
\cD \ov W&=\prod_\mu \cD \ov W_\mu=\prod_{x,\mu} d \ov W_\mu(x)
\end{align*}
mean \f al (nonexiting) measure for gauge field and gravity.

According to Ref.~\cite{B1} we add gauge-fixing terms
$$
L_g = -\frac{\eta^\m}{2\b_1} \, f_\nu[g]W_{\rm g}\Bigl(-\frac\Box{\La^2}\Bigr)
f_\mu[g]
-\frac1{2\b_2}\,f_a[A]W_{\rm YM}\Bigl(-\frac\Box{\La^2}\Bigr)h^{ab}f_b[A]
-\frac1{2\b_3}\,h[\ov W]\wt W\Bigl(-\frac\Box{\La^2}\Bigr)h[\ov W],
$$
where $\b_1,\b_2,\b_3$ are \ct s and $\Box$ is an ordinary d'Alembert operator
in a Minkowski space. $f^\mu[g]=f^\mu[g_{\a\b}]$ is a gauge fixing \f\ for
a \gr al field, $f_a[A]$
for a gauge field, $W_{\rm g}$ is a gravity gauge-fixing weighting \f,
$W_{\rm YM}$ is a gauge-term weight for a \YM' field. We add also a gauge-fixing
\f\ for $\ov W_\mu$ field with its weight~$\wt w$.

We can also add gauge-fixing terms for Higgs' fields. Sometimes it is possible
to consider a gauge condition which involves gauge and Higgs' fields together.
In this way we can get also additional Faddeev--Popov ghosts. But this does not
threaten us. These ghosts are easily exorcized. The most important problems
in this theory are possible massive ghosts which could appear if \f s
$h_i$ are not properly chosen.

The FP  (Faddeev--Popov) ghosts are not dangerous as we mention above
from quantum field theory
point of view. They are also exorcized from geometrical point of view, i.e.\
they can be geometrized (see Ref.~\cite{P2}). According to Refs \cite{P2,P3}
a gauge field (in specific fixed gauge, i.e.\ in a section of a principal
bundle) plus a ghost field is a globally defined \cn\ on the principal
bundle (see also Eq.~\er{2.55}). The anticommuting property of a ghost field can be easily derived and
a nilpotent BRST charge obtained as a differential operator. In order to
proceed a \f al integration we apply a well-known Faddeev--Popov trick in
order to do an integration over those configurations which \sf y a gauge
fixing conditions. In this approach ghosts fields appear in the Lagrangian. We have
two kinds of ghosts---gauge field ghosts and gravity field ghosts. Thus we
have a ghost field Lagrangian
$$
L_{gh}=\ov c_a M_{ab}c_b + \ov c{}^\mu N_\m C^\nu + \ov cMc
$$
coming from an exponentiation of a Faddeev--Popov determinant (an infinite analogue
of a ``Jacobian'') \st
\begin{align*}
M_{ab}c_b = \d_c f_a[A,x] \q &\hbox{(scalars)}\\
N_\m c_\mu = \d_\a f_v[g,x] \q &\hbox{(vector)}\\
Mc = \d h[\ov W,x] \q &\hbox{(scalar)}
\end{align*}
where $\d_c f_a$ is the infinitesimal \tf\ of $f_a$ with gauge parameters
$c_b$, $\d_\a f_v$ is the infinitesimal \tf\ of $f_v$ with a changing of a
frame with parameter $c_\mu$ and $\d h$ is an analogue of a gauge changing
of~$\ov W_\nu$. They do not depend on weighting \f s. The Faddeev--Popov
ghosts  are ghost fields in this sense that they do not have a right
statistics. In order to get ghost Lagrangian we should integrate using
anticommuting fields. In this way, they are anticommuting bosons. Thus one
gets
$$
Z=\int e^{iS[A,g,\bar W,c_a,c_\mu,c]}\,DF\,Dh=V\cdot
\int e^{iS[A,g,\bar W,c_a,c_\mu,c]}\,DF
$$
where
$$
\bga
DF=\prod_x dA^{\rm fix}(x) \prod_x dg^{\rm fix}(x) \prod_x d\ov W{}^{\rm fix}
(x) \prod_x(dc_a(x)\,d\ov c_a(x)) \prod_x(dc_\mu(x)\,d\ov c_\mu(x))
\prod_x(dc(x)\,d\ov c(x))\\
Dh=\prod_x dh(x)\q \hbox{(integration over a gauge group), }
\ega
$$
$\ov c_a,\ov c_\mu,\ov c$ mean antighost
fields. $V$ is an ``infinite volume'' of a local gauge group, $A^{\rm fix}(x)$
means that the gauge has been fixed. The same for $g^{\rm fix}(x)$ and
$\ov W^{\rm fix}(x)$. (In a more geometrical language we say that an
integration is over an orbits space of a local gauge group.)
$$
S[A,g,\ov W,c_a,c_\mu,c]= \int d^4x\,\sqrt{-g}(L+L_g+L_{gh}).
$$
The above formulae are starting points for a path-integral quantization of
the \NK\ after a careful choice of entire \f s $h_i$, $i=1,2,3,4$. We hope
to find them to get (super-) renormalizable or even finite theory unifying
\nos\ gravity and other \fn\ \ia s (in a bosonic section) with ``interference
effects'' obtained on a level of a classical field theory. After an inclusion
of fermion fields this program accomplishes the Einstein idea of a Unified
Field Theory of all \ia s, which is geometrical (geometrization of physical
\ia s), nonlinear (nonlinear field \e s) and also non-local.
This nonlocality should of course be causal and
this depends on \f s $h_i$. Such \f s are entire transcendental \f s. They
are not polynomials, i.e.
$$
h(z)=\sum_{n=0}^\iy a_nz^n \qh{and} \lim_{n\to\iy} \root n \of {|a_n|}=0.
$$
It means they are defined on a whole open complex plane and according to Liouville
theorem they have a pole or an essential singularity at infinity.

The construction of such \f s can be done according to Refs \cite{B1,B2,C1,C2}.
In any cases we can write ($\wt g$ is a coupling \ct)
$$
h(z)=1+\wt g{}^2\exp\Bigl(\int_0^{p_\g(z)}\frac{1-\z(w)}w\,dw - 1\Bigr)
$$
where $p_\g(z)$ is a real polynomial of degree $\g$ and $p_\g(0)=0$, $\z(z)$
is an entire \f\ and real on the real axis and $\z(0)=1$,
$$
|\z(z)|\to\iy \qh{for} |z|\to\iy,\ z\in\C.
$$
There are several propositions for such \f s in some applications for GR and
\YM' fields. We can perform a perturbation calculus using Feynman diagrams
for S-matrix which is unitary.
The full program will be developed elsewhere.

We should look for K\"ahler structure on $M=G/G_0$ (on a homogeneous space)
and also on compact Lie groups $G$ and~$H$ in order to get a more sound
mathematical constructions.

Let us notice that all the conclusions from Appendix~D of Ref.~\cite8 can be
applied also here.

The important problem in this theory is a problem of ghosts and tachions.
This problem should be solved on the level of a classical field theory
before a quantization, on a prequantum level.
The problem is connected with the existence of the skew-\s ic tensor
$g_\[\m]$---a~skewon. This particle in a linear \ap ion is massive getting
mass term from a \co ical \ct\ which appears in the theory. According to modern
ideas a \co ical \ct\ is not zero and we are (in principle) able to tune our
\co ical term to the observation data. In this way we can predict a value
of a mass of a skewon. Skewon has a spin zero and has a positive energy in
the case of a pure real or hypercomplex  Hermitian Theory (see Refs
\cite{o}, \cite{oo}, \cite{ooo}, \cite{oooo}).

Thus our 20-\di al unification should be considered in the case of
hypercomplex Hermitian gravity with K\"ahlerian \sc e on~$S^2$ combined to
Hermitian (complex or hypercomplex) Kaluza--Klein Theory.

Our 12-\di al \un\ should be considered in the case of hypercomplex
Hermitian gravity combined to Hermitian (complex or hypercomplex)
Kaluza--Klein Theory.

Our 5-\di al \un\ from Ref.~\cite8 should be considered in the case of
hypercomplex Hermitian gravity combined to Hermitian (complex or hypercomplex)
Kaluza--Klein Theory.

We do not exhaust a possible research in this direction. First of all we can
consider a Quaternionic (or Split-Quaternion) Hermitian gravity (if we find
some applications of additional degrees of freedom) to extend to Hermitian
(quaternionic or split-quaternion) Hermitian Kaluza--Klein Theory.
The second direction is to consider in place of a Cartesian product
$V=E\tm M=E\tm G/G_0$ or $V=E\tm S^2$ a nontrivial principal fiber bundle
over $E$ with a fiber $M=G/G_0$ or~$S^2$.
Moreover, we need application of additional degrees of freedom, i.e.\ a \cn\
on the bundle.

In Ref.~\cite{53} the authors write in a very pessimistic way. We quote:
``\textit{Unfortunately, although our understanding of gauge theories has
continued to develop, we have made very little progress in understanding the
origin of \sn\ \s y breakdown. For the most part, the \Hm\ continues to be
described by the ad hoc introduction into the Lagrangian of \el ary, weakly
self-coupled scalar fields. In the minimal model, a complex $\SU(2)$ doublet
is used, providing three Goldstone bosons (longitudinal $W$ and $Z$ bosons) and
one physical massive scalar.}''

According to our research this pessimistic view of Higgs' sector is not
longer true. Higgs' fields are part of gauge fields (\di al reduction
procedure). A~full bosonic sector of GSW model can be incorporated as a part
of the Hermitian (\E\nos) Kaluza--Klein Theory getting masses of~$W$ and~$Z$
bosons, Higgs' boson and Weinberg angle agreed with an experiment.
All mentioned \pc s have been discovered and some additional phenomena can be
predicted due to the existence of anti\s ic tensors in the theory.

Let us give some historical remarks. The \di al reduction and \iv t \cn s
which lead to the interpretation of the Higgs-like scalar multiplets as
a part of \YM' field in higher \di al bundle over a quotient space has been
also introduced in Refs \cite{78, 79, 80}. Moreover, in our approach we
follow Refs \cite{18}--\cite{*2}. They are by any means published earlier.
Let us notice the following fact. In~1977 A.~Trautman communicated to me
that K.~A.~Olive found a possibility to get a kinetic term for a scalar field
from the fifth \di\ which was similar to my observation.

The idea to interpret the \nA\ gauge field as a torsion appeared in Refs
\cite{81, 82}. The quartic \pt\ of generalized Higgs' field can be obtained in
the framework of non-commutative geometry, see Refs \cite{83,84}. Actually
we do not follow this approach.

We mention above on our plans to consider chiral fermions in our approach.
Probably we use some ideas from Ref.~\cite{84A}.

In Refs \cite{85,86} an idea was considered to use supergroups in order to
unify physical \ia s. In this way a \nos\ tensor on a supergroup appears
naturally as a part of generalized \KC tensor connected anticommuting
generators. We mention this possibility in Ref.~\cite1. Moreover, in Refs
\cite{85,86} this idea is not connected to any \nos\ geometry as in
Ref.~\cite1. Let us notice the idea  to geometrized BRST \s y has been
developed in Ref.~\cite{87}. Let us notice that in Refs \cite{85,86}
supergroups are considered as global \s ies. Our suggestions from Ref.~\cite1
considered a supergroup as a local \s y combined via Einstein geometry with
\nos\ gravity or even supergravity. This idea can be considered as future
prospects for further research. Let us notice that we give some historical
remarks on Einstein Unified Field Theory in the last section of Ref.~\cite8.
It is interesting that A.~Einstein in Ref.~\cite{14} came back to his first
ideas in Unified Field Theory (from 1920--30) and we develop them further
in a new context. Let us mention on Ref.~\cite{88} where A.~Crumeyrolle
developed a program of geometrization and unification using a manifold with
\hc\ \cd s close to our prospects with quaternionic metric.

Finally, let us give some remarks. We do not consider in our (even do not
touch) approaches in which weak \ia s and gravity are cross-correlated. In
particular, the possibility to see \gr al \ia s as emerging from the long-range
behavior of the Higgs' field (see Ref.~\cite{89}), the possibility that
\gr\ morphs into weak \ia s at the Fermi scale (see Ref.~\cite{90}) and the
relationship between weak \ia s chirality and gravity (see Refs \cite{91,92}).
All mentioned approaches are very interesting in \pp. However, they have not
any application in our approach---geometrization and unification of \fn\
\ia s. Only in Ref.~\cite{90} we see some possibility to extend it to an
Einstein--Cartan-like theory in order to get current-current \ia\ known in
an old weak \ia\ theory. However, this approach even interesting from
conceptual point of view cannot be maintained because we have now GSW-model
employing \YM' and Higgs' fields. The relationship between weak \ia s
chirality and gravity (see Refs \cite{91,92}) is not applicable in our \NK\
because we can get chiral fermions in a completely different setting
(mentioned above).

We should look for a flavor-chiral fermion \rp ation in our approach as we
described above. However this is a still unresolved problem to be considered
in future works. Moreover, we can claim that we deal with GSW-model and
QCD-model in the \NK. Our theory is a real candidate of TOE. Moreover, it
does not make any development to ``modern'' Kaluza--Klein Theory for the
reasons given in Conclusions of Ref.~\cite8. Now we are waiting for results
from new LHC and future accelerators.

The \nos\ metric considered here is a crucial point and it has real physical
motivation described in the paper (dielectric model of a \cfn\ and a correct
pattern of $W^\pm,Z^0$ and Higgs bosons in GSW-model). The \gr al influence
on GSW-model (Higgs kinetic energy) can be of course testable in an
experiment as a skewon-Higgs' \ia\ which may be discovered in LHC even before
graviton-Higgs' \ia.

\section*{Acknowledgement}
I would like to thank Professor B. Lesyng for the opportunity to carry out
computations using Mathematica\TM~9\footnote{Mathematica\TM\ is the
registered mark of Wolfram Co.} in the Centre of Excellence
BioExploratorium, Faculty of Physics, University of Warsaw, Poland.

\end{document}